\documentclass[11pt,a4paper,fleqn]{article}
\usepackage[utf8]{inputenc}
\usepackage{authblk}
\usepackage{geometry}
\usepackage{xcolor}
\usepackage{natbib}
\usepackage{graphicx}
\usepackage{pdflscape}
\usepackage{appendix}
\usepackage{setspace}
\usepackage{makecell}

\usepackage{colortbl}
\providecommand{\shadeRow}{\cellcolor[rgb]{0.9, 0.9, 0.9}}
\providecommand{\shadeBench}{\cellcolor[rgb]{0.95, 0.3, 0.3}}

\usepackage{threeparttable}
\usepackage{booktabs}

\usepackage{amsmath}
\usepackage{amssymb}
\usepackage{bm}

\usepackage{xcolor}

\usepackage{enumitem}
\newlist{steps}{enumerate}{1}
\setlist[steps, 1]{label = Step \arabic*:}

\usepackage{dcolumn}
\newcolumntype{d}[1]{D{.}{.}{#1}}



\usepackage{caption}
\usepackage{subcaption}
\definecolor{nblue}{HTML}{000660}
\usepackage[colorlinks=true,urlcolor=nblue,linkcolor=nblue,citecolor=nblue]{hyperref}

\title{\LARGE \textbf{Real-time Inflation Forecasting Using Non-linear Dimension Reduction Techniques}}
\author[1, 2]{\MakeUppercase{Niko Hauzenberger}}
\author[1]{\MakeUppercase{Florian Huber}}
\author[1]{\MakeUppercase{Karin Klieber}\thanks{Corresponding author: Karin Klieber. Department of Economics, University of Salzburg. \textit{Address}: M\"{o}nchsberg 2a, 5020 Salzburg, Austria. \textit{Email}: \href{mailto:karin.klieber@sbg.ac.at}{karin.klieber@sbg.ac.at}. We would like to thank the
participants of the 2021 Annual Meeting
of the Austrian Economic Association (NOeG, June 2021), a joint internal seminar of the University of Salzburg and the University of Strathclyde and the KOF Research Seminar at the ETH Zurich, two anonymous referees as well as Gary Koop, Michael Pfarrhofer, Aristeidis Raftapostolos, and Anna Stelzer for valuable comments and suggestions. The authors gratefully acknowledge financial support from the Austrian Science Fund (FWF, grant no. ZK 35) and the Oesterreichische Nationalbank (OeNB, Anniversary Fund, project no. 18127).} }
\affil[1]{\textit{University of Salzburg}}
\affil[2]{\textit{Vienna University of Economics and Business}}

\date{\today}

\begin{document}

\maketitle\thispagestyle{empty}\normalsize\vspace*{-2em}\small

\begin{center}
\begin{minipage}{0.8\textwidth}
\noindent\small In this paper, we assess whether using non-linear dimension reduction techniques pays off for forecasting inflation in real-time. Several recent methods from the machine learning literature are adopted to map a large dimensional dataset into a lower dimensional set of latent factors. We model the relationship between inflation and the latent factors using constant and time-varying parameter (TVP) regressions with shrinkage priors. Our models are then used to forecast monthly US inflation in real-time. The results suggest that sophisticated dimension reduction methods yield inflation forecasts that are highly competitive to linear approaches based on principal components. Among the techniques considered, the Autoencoder and squared principal components yield factors that have high predictive power for one-month- and one-quarter-ahead inflation. Zooming into model performance over time reveals that controlling for non-linear relations in the data is of particular importance during recessionary episodes of the business cycle or the current COVID-19 pandemic.\\\\ 
\textbf{JEL}: C11, C32, C40, C53, E31\\
\textbf{Keywords}: Non-linear principal components, machine learning, time-varying parameter regression, density forecasting, real-time data\\
\end{minipage}
\end{center}

\onehalfspacing\normalsize\renewcommand{\thepage}{\arabic{page}}

\newpage
\section{Introduction}\label{sec:intro}

Inflation expectations are used as crucial inputs for economic decision making in central banks such as the European Central Bank (ECB) and the US Federal Reserve (Fed). Given current and expected inflation, economic agents decide on how much to consume, save and invest. In addition, measures of inflation expectations are often employed to estimate the slope of the Phillips curve, infer the output gap or  the natural rate of interest. Hence, being able to accurately predict inflation is key for designing and implementing appropriate monetary policies in a forward looking manner. 

Although the literature on modeling inflation is voluminous and the efforts invested considerable, predicting inflation remains a difficult task and simple univariate models are still difficult to beat \citep{sw07inflation}. The recent literature, however, has shown that using large datasets \citep{stock2002macroeconomic} and/or sophisticated models \citep[see][]{koop2007estimation, kk2012, giannone2013, koop2013var, clark2015macroeconomic, CCK2018, JaroLenza2018} has the potential to improve upon simpler benchmarks. 

These studies often exploit information from huge datasets. This is commonly achieved by extracting a relatively small number of principal components (PCs) and including them in a second stage regression model \citep[see, e.g.,][]{stock2002macroeconomic}. While this approach performs well empirically and yields consistent estimators for the latent factors, it fails to capture non-linear relations in the dataset. In the presence of non-linearities, using simple PCs potentially reduces predictive accuracy by ignoring important features of the data. Some studies deal with this issue by using flexible factor models which allow for non-linearities in the data.  \cite{Bai-Ng2008} use targeted predictors coupled with quadratic principal components and show that allowing for non-linearities yields non-trivial improvements in predictive accuracy for inflation. This suggests that non-linearities (of a known form) are present in US macroeconomic datasets which are commonly employed for inflation forecasting. More recently, \cite{pelger2021state} propose a flexible state-dependent factor model and apply this method to US bond yields and stock returns. Using this non-linear and non-parametric technique yields results which differ from linear, PC-based models by extracting significantly more information from the data. 

One additional assumption commonly made is that the relationship between inflation and the latent factors is constant. For longer time series which feature multiple structural breaks this assumption is a strong one and may be deleterious for predictive accuracy. Several recent papers deal with this issue by using time-varying parameter (TVP) regressions which, in addition, allow for heteroscedasticity through stochastic volatility (SV) models \citep{koop2007estimation, kk2012, giannone2013, belmonte2014hierarchical, clark2015macroeconomic, JaroLenza2018, korobilis2019high}.

Investigating whether allowing for non-linearities in the compression stage  pays off for inflation forecasting is the key objective of the present paper. Building on recent advances in machine learning \citep[see][]{gallant-white1992, McNelis2004Forecasting, exterkate2016nonlinear, Chakraborty2017ML, heaton2017, Mullainathan2017ML, Polson2018, Coulombe2019ML, Kelly2018AE, medeiros2019forecasting}, we adopt several non-linear dimension reduction techniques. The resulting latent factors are then linked to inflation in a second stage regression. To investigate whether there exists a relationship between non-linear factor estimation and flexible modeling of the predictive inflation equation, we introduce  dynamic regression models that allow for TVPs and SV. Since the inclusion of a relatively large number of latent factors can still imply a considerable number of parameters (and this problem is even more severe in the TVP regression case), we rely on state-of-the-art shrinkage techniques. 

From an empirical standpoint it is necessary to investigate how these dimension reduction techniques perform over  time and during different business cycle phases.  We show this by carrying out a thorough real-time forecasting experiment for the US. Our forecasting application uses monthly real-time datasets \citep[i.e., the FRED-MD database proposed in][]{mccracken2016fred} and includes a battery of well established models commonly used in central banks and other policy institutions to forecast inflation. These include simple benchmarks as well as more elaborate models such as the specification proposed in \cite{stock2002macroeconomic}.

Our results show  that non-linear dimension reduction techniques yield forecasts that are highly competitive to (and in fact often better than)  the ones obtained from using linear methods based on PCs. In terms of one-month-ahead  forecasts we find that models based on the Autoencoder yield point and density forecasts which are more precise than the ones obtained from other sophisticated non-linear dimension reduction techniques as well as traditional methods based on  PCs. When the focus is on one-quarter-ahead forecasts we find that non-linear variants of PCs perform best. This performance, however, is not homogeneous over time and some of the models do better than others during different stages of the business cycle. In a brief discussion, we also analyze how our set of models performs during the COVID-19 pandemic.


These findings give rise to the second contribution of our paper. Since we observe that more sophisticated non-linear dimension reduction methods outperform simpler techniques during recessions, we combine the different models using dynamic model averaging \citep[see][]{RafteryDMA, koop2013var}. We show that combining our proposed set of models with a variety of standard forecasting models yields predictive densities which are very close to the single best performing model in overall terms. Since the set of models we consider is huge, this indicates that using model and forecast averaging successfully controls for model uncertainty.

The remainder of this paper is structured as follows. Section \ref{sec:dim} discusses our proposed set of dimension reduction techniques. Section \ref{sec: regression} introduces the econometric modeling environment that we use to forecast inflation. Section \ref{sec:resultsA} first provides some in-sample features, then discusses the results of the forecasting horse race and finally presents our findings based on forecast averaging. The last section summarizes and concludes the paper. The Online Appendix provides further details on the econometric techniques as well as the data and additional empirical results.

\section{Linear and non-linear dimension reduction techniques}\label{sec:dim}
Suppose that we are interested in predicting inflation using a large number of $K$ regressors that we store in a $T \times K$ matrix $\bm X = (\bm x_1, \dots, \bm x_T)'$, where $\bm x_t$ denotes a $K$-dimensional vector of observations at time $t$.  If $K$ is large relative to $T$, estimation of an unrestricted model that uses all columns in $\bm X$ quickly becomes cumbersome and overfitting issues arise. As a solution, dimension reduction techniques are commonly employed \citep[see, e.g.,][]{stock2002macroeconomic, bernanke2005measuring}. These methods  strike a balance between model fit and parsimony. At a very general level, the key idea is to introduce a function $f$ that takes the matrix $\bm X$ as input and yields a lower dimensional representation $\bm Z = f(\bm X) = (\bm z_1, \dots, \bm z_T)'$, which is of dimension $T \times q$, as output. The critical assumption to achieve parsimony is that $q \ll K$.  The latent factors in $\bm Z$ are then linked to inflation through a dynamic regression model (see Section \ref{sec: regression}).

The function $f: \mathbb{R}^{T \times K} \to \mathbb{R}^{T \times q}$ is typically assumed to be linear with the most prominent example being PCs. In this paper, we will consider several choices of $f$ that range from linear to highly non-linear (such as manifold learning as well as deep learning) specifications. We subsequently analyze how these different specifications impact inflation forecasting accuracy. In the following sub-sections, we briefly discuss the different techniques and refer to the original papers for additional information.

\subsection{Principal component analysis}
We start our discussion by considering principal component analysis (PCA).
Minor alterations of the standard PCA approach allow for introducing non-linearities in two ways. First, we can introduce a non-linear function $g$ that maps the covariates onto a matrix $\bm W = g(\bm X)$. Second, we could alter the sample covariance matrix (the kernel) with a function $h$: $\bm \kappa = h(\bm W' \bm W)$. Both $\bm W$ and $\bm \kappa$ form the two main ingredients of a general PCA reducing the dimension to $q$, as outlined below \citep[for details, see][]{Schoelkopf1998}. 

Independent of the functional form of $g$ and $h$, we obtain PCs by performing a truncated singular value decomposition (SVD) of the transformed sample covariance matrix  $\bm \kappa$. Conditional on the first $q$ eigenvalues, the resulting factor matrix $\bm Z$ is of dimension $T \times q$. These PCs, for appropriate $q$, explain the vast majority of variation in $\bm X$. In the following, the relationship between the PCs and $\bm X$ is:
\begin{equation}
\bm Z = f(\bm X) = g(\bm X) \bm \Lambda(\bm \kappa) = \bm W \bm \Lambda(\bm \kappa), \label{eq:PCA}
\end{equation}
with $\bm \Lambda(\bm \kappa)$ being the truncated $K \times q$ eigenvector matrix of $\bm \kappa$ \citep{stock2002macroeconomic}. Notice that this is always conditional on deciding on a suitable number $q$ of PCs. The number of factors is a crucial parameter that strongly influences predictive accuracy and inference \citep{Bai-Ng-2002}. In our empirical work, we consider a small ($q=5$), moderate ($q=15$), and large ($q=30$) number of PCs. 

By varying the functional form of $g$ and $h$ we are now able to discuss the first set of linear and non-linear dimension reduction techniques belonging to the class of PCA: 
\begin{enumerate}
\item \textbf{Linear PCs}

The simplest way is to define both $g$ and $h$ as the unity function, resulting in $\bm W = \bm X$ and $\bm \kappa = \bm X' \bm X$.  Due to the linear link between the PCs and the data, PCA is very easy to implement and yields consistent estimators for the latent factors if $K$ and $T$ go to infinity \citep{stock2002macroeconomic, Bai-Ng2008}. Even if there is some time-variation in the factor loadings (and $K$ is large), \cite{stock2002forecasting} show that principal components asymptotically (i.e., $T \rightarrow \infty)$ remain a consistent estimator for the factors and also that the resulting forecast is efficient.\footnote{Note that this result holds only asymptotically. With relatively small $T$ and large $K$, however, forecast efficiency may be improved by better capturing important non-linear features of the dataset.}

\item \textbf{Quadratic and squared PCs}

The literature suggests several ways to overcome the linearity restriction of PCs. \cite{Bai-Ng2008}, for example, apply a quadratic link function between the latent factors and the regressors, yielding a more flexible factor structure. While squared PC considers just squaring the elements of $\bm X$ resulting in
\begin{equation*}
\bm W = \bm X^2 \quad \text{and} \quad \bm \kappa = (\bm X^2)'(\bm X^2),
\end{equation*}
with $\bm X^2 = (\bm X \odot \bm X)$ and $\odot$ denoting element-wise multiplication, quadratic PC is defined as 
\begin{equation*}
\bm W = (\bm X, \bm X^2) \quad \text{and} \quad \bm \kappa = \bm W'\bm W.
\end{equation*}
Both variants also focus on the second moments of the covariate matrix and allow for a non-linear relationship between the principal components and the predictors. \cite{Bai-Ng2008} show that quadratic variables can have substantial predictive power as they provide additional information on the underlying time series. Intuitively speaking, given that we transform our data to stationarity in the empirical work, this transformation strongly overweights situations characterized by sharp movements in the columns of $\bm X$ (such as during a recession). By contrast, periods characterized by little variation in our macroeconomic panel are transformed to mildly fluctuate around zero (and thus carry little predictive content for inflation). In our empirical model, our regressions always feature lagged inflation and this transformation thus effectively implies that in tranquil periods, the model is close to an autoregressive model whereas in crisis periods, more information is introduced.

\item \textbf{Kernel PCs}

Another approach for non-linear PCs is the kernel principal component analysis (KPCA). KPCA dates back to \cite{Schoelkopf1998}, who proposed using integral operator kernel functions to compute PCs in a non-linear manner. In essence, this amounts to implicitly applying a non-linear transformation of the data through a kernel function and then applying PCA on this transformed dataset. Such an approach has been used for forecasting in \cite{giovannelli2012} and \cite{exterkate2016nonlinear}.

We allow for non-linearities in the kernel function between the data and the factors by defining $h$ to be a Gaussian or a polynomial kernel  $\bm \kappa$  (which is of dimension $K \times K$) with the $(i,j)$th element given by
\begin{equation*}
 \kappa_{ij} = \exp\left(-\frac{||\bm x_{\bullet i}-\bm x_{\bullet j}||}{2c_1^2}\right) 
\end{equation*}
for a Gaussian kernel and 
\begin{equation*}
\kappa_{ij} = \left(\frac{\bm x'_{\bullet i} \bm x_{\bullet j}}{c_0^2} + 1\right)^2 \\ 
\end{equation*}
for a polynomial kernel.

Here, $\bm W = \bm X$ (i.e., $g$ is the unity function), $\bm x_{\bullet i}$ and $\bm x_{\bullet j}$ $(i,j = 1, \dots, K)$ denote two columns of $\bm X$ while $c_0$ and $c_1$ are scaling parameters. As suggested by \cite{exterkate2016nonlinear} we set $c_0 = \sqrt{(K+2)/2}$ and $c_1 = \sqrt{c_K}/ \pi$ with $c_K$ being the $95$th percentile of the $\chi^2$ distribution with $K$ degrees of freedom.
\end{enumerate}

\subsection{Diffusion maps}
Diffusion maps, originally proposed in \cite{coifman2005diff} and \cite{coifman2006diffusion}, are another set of non-linear dimension reduction techniques that retain local interactions between data points in the presence of substantial non-linearities in the data.\footnote{For an application to astronomical spectra, see \cite{richards2009diff}.}  The local interactions are preserved by introducing a random walk process. 

The random walk captures the notion that moving between similar data points is more likely than moving to points which are less similar. We  assume that the weight function which determines the strength of the relationship between $\bm x_{\bullet i}$ to $\bm x_{\bullet j}$ is given by
\begin{equation*}
{w}(\bm x_{\bullet i}, \bm x_{\bullet j}) = \exp\left(\frac{||\bm x_{\bullet i} - \bm x_{\bullet j}||^2}{c_2} \right),
\end{equation*}
where $||\bm x_{\bullet i} - \bm x_{\bullet j}||$ denotes the Euclidean distance between $\bm x_{\bullet i}$ and $\bm x_{\bullet j}$ and $c_2$ is a tuning parameter set such that ${w}(\bm x_{\bullet i}, \bm x_{\bullet j})$ is close to zero except for $\bm x_{\bullet i} \approx \bm x_{\bullet j}$. Here, $c_2$ is determined by the median distance of the $k$-nearest neighbors of $\bm x_{\bullet i}$ as suggested by \cite{Zelnik2005diff}. The number of $k$ is approximated using the algorithm suggested by \cite{angerer2016destiny}. 

The probability of moving from $\bm x_{\bullet i}$ to $\bm x_{\bullet j}$ is then simply obtained by normalizing:
\begin{equation*}
p_{i \to j} = \text{Prob}(\bm x_{\bullet i} \to \bm x_{\bullet j}) = \frac{{w}(\bm x_{\bullet i}, \bm x_{\bullet j})}{\sum_j {w}(\bm x_{\bullet i}, \bm x_{\bullet j})}.
\end{equation*}
This probability tends to be small except for the situation where $\bm x_{\bullet i}$ and $\bm x_{\bullet j}$ are similar to each other. As a result, the probability that the random walk moves from $\bm x_{\bullet i}$ to $\bm x_{\bullet j}$ will be large if they are equal but rather small if both covariates differ strongly. 

Let $\bm P$ denote a transition matrix of dimension $K \times K$ with $(i,j)$th element given by $p_{i \to j}$. The probability of moving from $\bm x_{\bullet i}$ to $\bm x_{\bullet j}$ in $n=1,2, \dots$ steps is then simply the matrix power of $\bm P^n$, with typical element denoted by $p_{i \to j}^n$.
Using a biorthogonal spectral decomposition of $\bm P^n$ yields:
\begin{equation*}
p^n_{i \to j} = \sum_{s \ge 0} \lambda_s^n \psi_s(\bm x_{\bullet i}) \phi_s (\bm x_{\bullet j}),
\end{equation*}
with $\psi_s$ and $\phi_s$ denoting left and right eigenvectors of $\bm P$, respectively. The corresponding eigenvalues are given by $\lambda_s$.

We then proceed by computing the so-called diffusion distance as follows:
\begin{equation*}
\xi^2_n(\bm x_{\bullet i}, \bm x_{\bullet j}) = \sum_j \frac{(p_{i \to j}^n - p_{s \to j}^n)^2}{p_0(\bm x_{\bullet j})}, \label{eq: diffusion_distance}
\end{equation*}
with $p_0$ being a normalizing factor that measures the proportion the random walk spends at $\bm x_{\bullet j}$. This measure turns out to be robust with respect to noise and outliers.
\cite{coifman2006diffusion} show that 
\begin{equation*}
\xi^2_n(\bm x_{\bullet i}, \bm x_{\bullet j}) = \sum_{s=1}^\infty \lambda_s^{2n} (\psi_s(\bm x_{\bullet i}) - \psi_s (\bm x_{\bullet j}))^2.
\end{equation*}
This allows us to introduce the family of diffusion maps from $\mathbb{R}^{K} \to \mathbb{R}^{q}$ given by:
\begin{equation*}
\bm \Xi_n(\bm x_{\bullet i}) = [\lambda_1^n \psi_1(\bm x_{\bullet i}), \dots, \lambda_q^n \psi_q(\bm x_{\bullet i})].
\end{equation*}
The distance matrix can then be approximated as:
\begin{equation*}
\xi^2_n(\bm x_{\bullet i}, \bm x_{\bullet j}) \approx \sum_{s=1}^q \lambda_s^{2n} (\psi_s(\bm x_{\bullet i}) - \psi_s (\bm x_{\bullet j}))^2 = ||\bm \Xi_n(\bm x_{\bullet i}) - \bm \Xi_n (\bm x_{\bullet j})||^2.
\end{equation*}
Intuitively, this equation states that we now approximate diffusion distances in $\mathbb{R}^K$ through the Euclidian distance between $\bm \Xi_n(\bm x_{\bullet i})$ and  $\bm \Xi_n (\bm x_{\bullet j})$. This discussion implies that we have to choose $n$ and $q$ and we do this by setting $q=\{5,15,30\}$ according to our approach with either a small, moderate or large number of factors and $n=T$, the number of time periods. The algorithm in our application is implemented using the \texttt{R} packages \texttt{diffusionMap} and \texttt{destiny} \citep{diffusionMap, angerer2016destiny}.

\subsection{Local linear embeddings}
Locally linear embeddings (LLE) have been introduced by \cite{roweis2000lle}. Intuitively, the LLE algorithm maps a high dimensional input dataset $\bm X$ into a lower dimensional space while preserving the neighborhood structure. This implies that points which are close to each other in the original space are also close to each other in the transformed space. 

The LLE algorithm is based on the assumption that each $\bm x_{\bullet i}$ is sampled from some underlying manifold. If this manifold is well defined, each $\bm x_{\bullet i}$ and its neighbors $\bm x_{\bullet j}$ are located close to a locally linear patch of this manifold. One consequence is that each $\bm x_{\bullet i}$ can be reconstructed from its neighbors $\bm x_{\bullet j}$ with $j \neq i$, conditional on suitably chosen linear coefficients. This reconstruction, however, will be corrupted by measurement errors. \cite{roweis2000lle} introduce a cost function to quantify these errors:
\begin{equation*}
C(\bm \Omega) = \sum_{i}( \bm x_{\bullet i} - \sum_{j} \omega_{ij} \bm x_{\bullet j})^2,
\end{equation*}
with $\omega_{ij}$ denoting the $(i,j)$th element of a weight matrix $\bm \Omega$. This cost function is then minimized subject to the constraint that each $\bm x_{\bullet i}$ is reconstructed only from its neighbors. This implies that $\omega_{ij} = 0$ if $\bm x_{\bullet j}$ is not a neighbor of $\bm x_{\bullet i}$. The second constraint is that the matrix $\bm \Omega$ is row-stochastic, i.e., the rows sum to one. Conditional on these two restrictions, the cost function can be minimized by solving a least squares problem.

To make this algorithm operational we  need to define our notion of neighbors. In the following, we will use  the $k$-nearest neighbors in terms of the Euclidean distance. We choose the number of neighbors by applying the algorithm proposed by \cite{kayo2006}, which automatically determines the optimal number for $k$. The $q$ latent factors in $\bm Z$, with typical $i$th column $\bm z_{\bullet i}$, are then obtained by minimizing:
\begin{equation*}
\Phi (\bm Z) = \sum_i | \bm z_{\bullet i} - \sum_j \Omega_{ij} \bm z_{\bullet j}|^2,
\end{equation*}
which implies a quadratic form in $\bm z_t$. Subject to suitable constraints, this problem can be easily solved by computing:
\begin{equation*}
\bm M  = (\bm I_T - \bm \Omega)' (\bm I_T - \bm \Omega),
\end{equation*}
and finding the $q+1$ eigenvectors of $\bm M$ associated with the $q+1$ smallest eigenvalues. The bottom eigenvector is then discarded to arrive at $q$ factors. For our application, we use the \texttt{R} package \texttt{lle} \citep{LLE}.

\subsection{Isometric feature mapping}

Isometric Feature Mapping (ISOMAP) is one of the earliest methods developed in the category of manifold learning algorithms. Introduced by \cite{tenenbaum2000}, the ISOMAP algorithm determines the geodesic distance on the manifold and uses multidimensional scaling to come up with a low number of factors describing the underlying dataset. Originally, ISOMAP was constructed for applications in visual perception and image recognition. In economics and finance, some recent papers highlight its usefulness \citep[see, e.g.,][]{Ribeiro2008, lin2011isomap, orsenigo2013isomap, Zime2014}.

The algorithm consists of three steps. In the first step, a dissimilarity index that measures the distance between data points is computed. These distances are then used to identify neighboring points on the manifold. In the second step, the algorithm estimates the geodesic distance between the data points as shortest path distances. In the third step, metric scaling is performed by applying classical multidimensional scaling (MDS) to the matrix of distances. For the dissimilarity transformation, we determine the distance between point $i$ and $j$ by the Manhattan index $d_{ij} = \sum_k |x_{ki} - x_{kj}|$ and collect those points where $i$ is one of the $k$-nearest neighbors of $j$ in a dissimilarity matrix. For our empirical application, we again choose the number of neighbors by applying the algorithm proposed by \cite{kayo2006} and use the implementation in the \texttt{R} package \texttt{vegan} \citep{vegan}.

The described non-linear transformation of the dataset enables the identification of a non-linear structure hidden in a high-dimensional dataset and maps it to a lower dimension. Instead of pairwise Euclidean distances, ISOMAP uses the geodesic distances on the manifold and compresses information under consideration of the global structure. 

\subsection{Non-linear compression with deep learning}
Deep learning algorithms are characterized by not only non-linearly converting input to output but also representing the input itself in a transformed way. This is called representation learning in the sense that representations of the data are expressed in terms of other, simpler representations before mapping the data input to output values. 

One tool which performs representation of itself as well as representation to output is the Autoencoder (AE). The first step is accomplished by the encoder function, which maps an input to an internal representation. The second part, which maps the encoded (transformed) data to the output, is called the decoder function. Their ability to extract factors, which explain a large fraction of the variability in the observed data, in a non-linear manner makes deep learners a powerful tool complementing the range of commonly used dimension reduction techniques \citep{Goodfellow-et-al-2016}. \cite{andreini2020deep}, for example, embed a dynamic Autoencoder structure in a dynamic factor model and show that it yields a good now- and forecasting performance for US GDP. In their paper, they allow for additional flexibility by simultaneously estimating the non-linear latent factors and the parameters. In empirical finance, \cite{heaton2017}, \cite{Polson2018} and \cite{Kelly2018AE} find that the application of these methods is beneficial to predict asset returns.

Based on deep learning techniques, we propose obtaining hierarchical predictors $\bm Z$ by applying a number of $l \in \{1, \dots ,L\}$ non-linear transformations to $\bm X$. These transformations are called hidden layers with $L$ giving the depth of our architecture and $f$ denoting an univariate activation function.\footnote{In principle, $f$ can vary over the different layers.} More specifically, in each layer,  activation functions (non-linearly) transform the inputs (which are the outputs of the previous layer). A common choice, which we adopt, is the hyperbolic tangent (tanh) given by
\begin{equation*}
f(\bm X) = \frac{\exp(\bm X) - \exp(-\bm X)}{\exp(\bm X) + \exp(-\bm X)}.  
\end{equation*}
We apply  this function element-wise to the entries of $\bm X$.  Using tanh activation functions is justified by its strong empirical properties identified in recent studies  such as \cite{Saxe2019deep} and \cite{andreini2020deep}. 

The structure of our deep learning algorithm can be represented in form of a composition of univariate semi-affine functions given by
\begin{align*}
\hat{\bm X}^{(l)} = f \left( \hat{\bm X}^{(l-1)} \bm W^{(l-1)} +  \bm \iota_T \otimes \bm b'_{l-1} \right)\bm W^{(l)} + \bm \iota_T \otimes \bm b'_{l}, \quad \text{ for } 1 \leq l \leq L,
\end{align*}
and $\hat{\bm X}^{(0)} = \bm X$ for $l=0$. Here, $\bm W^{(l)}$ denotes a weighting matrix of dimension $N_{l-1} \times N_{l}$ (with $N_l$ being the number of neurons in layer $l$), $\bm b_{l}$ is a $N_l \times 1$ bias vector and $\bm \iota_T$ is a $T \times 1$ vector of ones. 

The output of the network is then obtained by setting:
\begin{equation*}
    \bm Z = \hat{\bm X}^{(L)} = f \left( \hat{\bm X}^{(L-1)} \bm W^{(L-1)} + \bm \iota_T \otimes \bm b_{L-1} \right) \bm W^{(L)} + \bm \iota_T \otimes \bm b_L.
\end{equation*}
Notice that if we set $N_L=q (\ll K)$, we achieve dimension reduction and the output of the network is a (non-linearily) compressed version of the input dataset. In principle, what we have just described constitutes the encoding part of the Autoencoder. If we are interested in recovering the original dataset $\bm X$ we simply have to add additional layers characterized by increasing numbers of neurons until we reach $N_{L+j} = K$ for $j=1,2,\dots$.



The optimal sets of $\hat{\bm W} = (\hat{\bm W}^{(1)},\dots,\hat{\bm W}^{(L)})$ and $\hat{\bm b} = (\hat{\bm b}_1,\dots,\hat{\bm b}_L)$ are obtained by computing a loss function, most commonly the mean squared error of the in-sample fit. The complexity of the neural network is determined by choosing the number of hidden layers $L$ and the number of neurons in each layer $N_l$. 
We perform our forecasting exercise with different sets of tuning parameters and choose one, three, five, and eight hidden layers with the number of neurons evenly being downsized to the desired number of factors. 

For the loss function and the optimization algorithm we stick to common choices in the literature and use the mean squared error loss function and the Adaptive Moment Estimation (ADAM). We repeat the optimization procedure in $100$ epochs on at least $84$ batches which corresponds to the average duration of a business cycle in the US.\footnote{The average duration of a business cycle was determined using data provided by The National Bureau of Economic Research on business cycle expansions and recessions.} This implies that we train the algorithm in each epoch with a partition of the original data set of at least the length of one business cycle. To capture the dynamics of the different cycles present in the data the optimization procedure needs to be repeated in a reasonably high number of epochs. We find that the algorithm converges quickly and setting the number of epochs to $100$ is sufficient.

We employ the \texttt{R} interface to \texttt{keras} \citep{keras}, a high-level neural networks API and widely used package for implementing deep learning models.

\section{A TVP regression for forecasting inflation} \label{sec: regression}
In the following, we introduce the predictive regression that links our target variable, inflation in consumer prices, to $\bm Z$ and other observed factors. Following \cite{stock1999forecasting}, inflation is specified such that:
\begin{equation}\label{eq:inflation}
y_{t+h} = \log \left( \frac{\text{CPI}_{t+h}}{\text{CPI}_{t}} \right) - \log \left( \frac{\text{CPI}_{t}}{\text{CPI}_{t-1}} \right),
\end{equation}
with  $\text{CPI}_{t+h}$ denoting the consumer price index in period $t+h$. 

In the empirical application we set $h \in \{1, 3\}$. $y_{t+h}$ is then modeled using a dynamic regression model:
\begin{equation}
y_{t+h} = \bm d_{t}' {\bm  \beta}_{t+h} + \epsilon_{t+h}, \quad \epsilon_{t+h} \sim \mathcal{N}(0, \sigma^2_{t+h}), \label{eq: TVP_regression}
\end{equation}
where $\bm \beta_{t+h}$ is a vector of TVPs associated with $M (=q + p)$ covariates denoted by $\bm d_t$ and $\sigma^2_{t+h}$ is a time-varying error variance.  $\bm d_t$ might include the latent factors extracted from the various methods discussed in the previous sub-section, lags of inflation, an intercept term or other covariates which are not compressed.

Following much of the literature \citep{Taylor1982stochvol, bkk, kg2014, kastner2014ancillarity, stock2016core, chan2017stochastic, huber2020} we assume that the TVPs and the error variances evolve according to independent stochastic processes:
\begin{align}
\begin{pmatrix}
\bm \beta_{t+h} \\
\log \sigma^2_{t+h}
\end{pmatrix}
\sim \mathcal{N}\left( \begin{pmatrix}
\bm \beta_{t+h-1} \\
\mu_h + \rho_h \log \sigma^2_{t+h-1}
\end{pmatrix}, \begin{pmatrix}
\bm V & 0 \\
0 & \vartheta_h^2
\end{pmatrix}
\right), \label{eq: state}
\end{align}
with $\mu_h$ denoting the conditional mean of the log-volatility, $\rho_h$ its persistence parameter and $\vartheta_h^2$ the error variance of $\log \sigma_{t+h}^2$. The matrix $\bm V$ is an $M \times M$-dimensional variance-covariance matrix with $\bm V = \text{diag}(v_1^2, \dots, v_M^2)$ and $v_j^2$ being the process innovation variance that determines the amount of time-variation in $\bm \beta_{t+h}$. This setup implies that the TVPs are assumed to follow a random walk process while the log-volatilities evolve according to an AR(1) process.

The model described by Eq. (\ref{eq: TVP_regression}) and Eq. (\ref{eq: state}) is a flexible state space model that encompasses a wide range of models commonly used for forecasting inflation. For instance, if we set $\bm V = \bm 0_M$ and $\vartheta^2=0$, we obtain a constant parameter model with homoscedastic errors. If $\bm V$ is instead a full $M \times M$ matrix but of reduced-rank, we obtain the model proposed in \cite{chan2020reducing}. If $\bm d_t$ includes the lags of inflation and (lagged) PCs, we obtain a model closely related to the one used in \cite{stock2002macroeconomic}. If we set $d_t =1$ and allow for TVPs, we obtain a specification similar to the unobserved components stochastic volatility model successfully adopted in \cite{stock1999forecasting}. A plethora of other models can be identified by appropriately choosing $\bm d_t$, $\bm V$ and $\vartheta^2$. This flexibility, however, calls for model selection. We select appropriate submodels by using Bayesian methods for estimation and forecasting. These techniques are further discussed in Section \ref{sec:App B} of the Online Appendix and allow for data-based shrinkage towards simpler nested alternatives.


\section{Forecasting US inflation}\label{sec:resultsA}
\subsection{Data overview, design of the forecasting exercise and competitors}\label{sec:descresults}

In our  empirical application we consider the popular FRED-MD database. This dataset is publicly accessible and available in real-time. The monthly data vintages ensure that we only use information that would have been available at the time a given forecast is being produced. A detailed description of the databases can be found in \cite{mccracken2016fred}. To achieve approximate stationarity we transform the dataset as outlined in Section \ref{sec:App C} of the Online Appendix. Furthermore, each time series is standardized to have sample mean zero and unit sample variance prior to using the non-linear dimension reduction techniques. 

Our US dataset includes 105 monthly variables that span the period from 1963:01 to 2021:01. The forecasting design relies on a rolling window, as justified in \cite{clark2011}, that initially ranges from 1980:01 to 1999:12. For each month of the hold-out sample, which starts in 2000:01 and ends in 2019:12, we compute the $h$-month-ahead predictive distribution for each model (for $h \in \{1, 3\}$), keeping the length of the estimation sample fixed at $240$ observations (i.e., a rolling window of $20$ years).\footnote{In addition to our baseline sample ending in 2019:12, we present the results of our forecasting exercise including observations covering the COVID-19 pandemic (2020:01 to 2020:08) in Sub-section \ref{sec:pandemic}. Since the pandemic caused severe outliers in our dataset, including those periods helps to test the forecasting performance of our models during turbulent times.} 
For these periods we contrast each forecast with the realization of inflation in the vintage  one-quarter-ahead, following the evaluation approach of \cite{chan2017stochastic}. As most data revisions take place in the first quarter while afterwards the vintages remain relatively unchanged \citep[see, e.g.,][]{croushore2011frontiers, pfarrhofer2020forecasts}, we make sure that realized inflation is not subject to revisions anymore.

One key limitation is that all methods are specified conditionally on $\bm d_t$ and thus implicitly on the specific function $f$ used to move from $\bm X$ to $\bm Z$. Another key objective of this paper is to control for uncertainty with respect to $f$ by using dynamic model averaging techniques. For obtaining predictive combinations, we use the first $24$ observations of our hold-out sample. The remaining periods (i.e., ranging from 2002:01 to 2019:12) then constitute our evaluation sample and the respective predictions are again contrasted to the one-quarter-ahead vintage of inflation. 

In terms of competing models we can classify the specifications along two dimensions: 
\begin{enumerate}
\item \textbf{How $\bm d_t$ is constructed.} First, let $\bm s_t$ denote a $K_0$-dimensional vector of covariates except for $y_t$. $\bm x_t =  (\bm s'_t, \dots , \bm s'_{t-p+1})'$ is then composed of $p$ lags of $s_t$ with $K=p K_0$. In our empirical work we set $p=12$ and include all variables in the dataset (except for the transformed CPI series, i.e., $K_0 = 104$). We then use the different dimension reduction techniques outlined in Section \ref{sec:dim} to estimate $\bm z_t$. Moreover, we include $p$ lags of $y_t$ as additional observed factors to $\bm d_t$. This serves to investigate how different dimension reduction techniques perform when interest centers on predicting inflation. We also consider simple AR($12$) models as well as a small- and a large-scale AR specification augmented with (observed) exogenous covariates (henceforth labeled ARX) as additional competitors. For the small-scale variants we include five exogenous regressors, while for the large-scale ARX model we use $20$ additional covariates.
Since the macroeconomic forecasting literature is quite inconclusive about variable inclusion in such predictive ARX models for inflation \citep[see, e.g.,][]{de2008forecasting, SW08, kk2012, hauzenberger2019fast}, we use a semi-automatic approach which handles this issue rather agnostically. We discuss this in more detail  in Sub-section \ref{sec:varselect}.

\item \textbf{The relationship between $\bm d_t$ and $y_{t+h}$.} The second dimension along which our models differ is the specific relationship described by Eq. (\ref{eq: TVP_regression}). To investigate whether non-linear dimension reduction techniques are sufficient to control for unknown forms of non-linearities, we benchmark all our models that feature TVPs with their respective constant parameter counterparts. To perform model selection we consider two priors. The first one is the horseshoe \citep[HS,][]{carvalho2010horseshoe} prior and the second one is an adaptive Minnesota \citep[MIN, see][]{carriero2015bayesian, giannone2015prior} prior (for further details see Section \ref{sec:App B} of the Online Appendix). 
\end{enumerate}

\subsection{Properties of the factors}\label{sec:varselect}
In this sub-section we analyze bivariate correlations between the factors, obtained from using different dimension reduction techniques, and the variables in our dataset as well as inflation. These correlations provide some information on the specific factor dynamics and (with caution) on how to interpret the factors in $\bm Z$ from a structural perspective.\footnote{The estimates of the factor are considerably more difficult to interpret. Nevertheless, to provide some intuition on how the factors for the best performing specifications evolve over time, see \autoref{fig:factors} in the Online Appendix.} The recent literature \citep{crawford2018approx, crawford2019approx, joseph2019parametric} advocates using linear approximations or Shapley values to improve interpretability of these highly non-linear models. In this paper, we opt for a simple correlation-based approach  given the large amount of competing dimension reduction techniques and the fact that for some of these the different techniques work better than for other methods.

\begin{figure}[!htbp]
\caption{Correlations between the variables in the input dataset and the factors obtained from the different dimension reduction techniques. \label{fig:corrfac}}
\centering
\includegraphics[scale=.55]{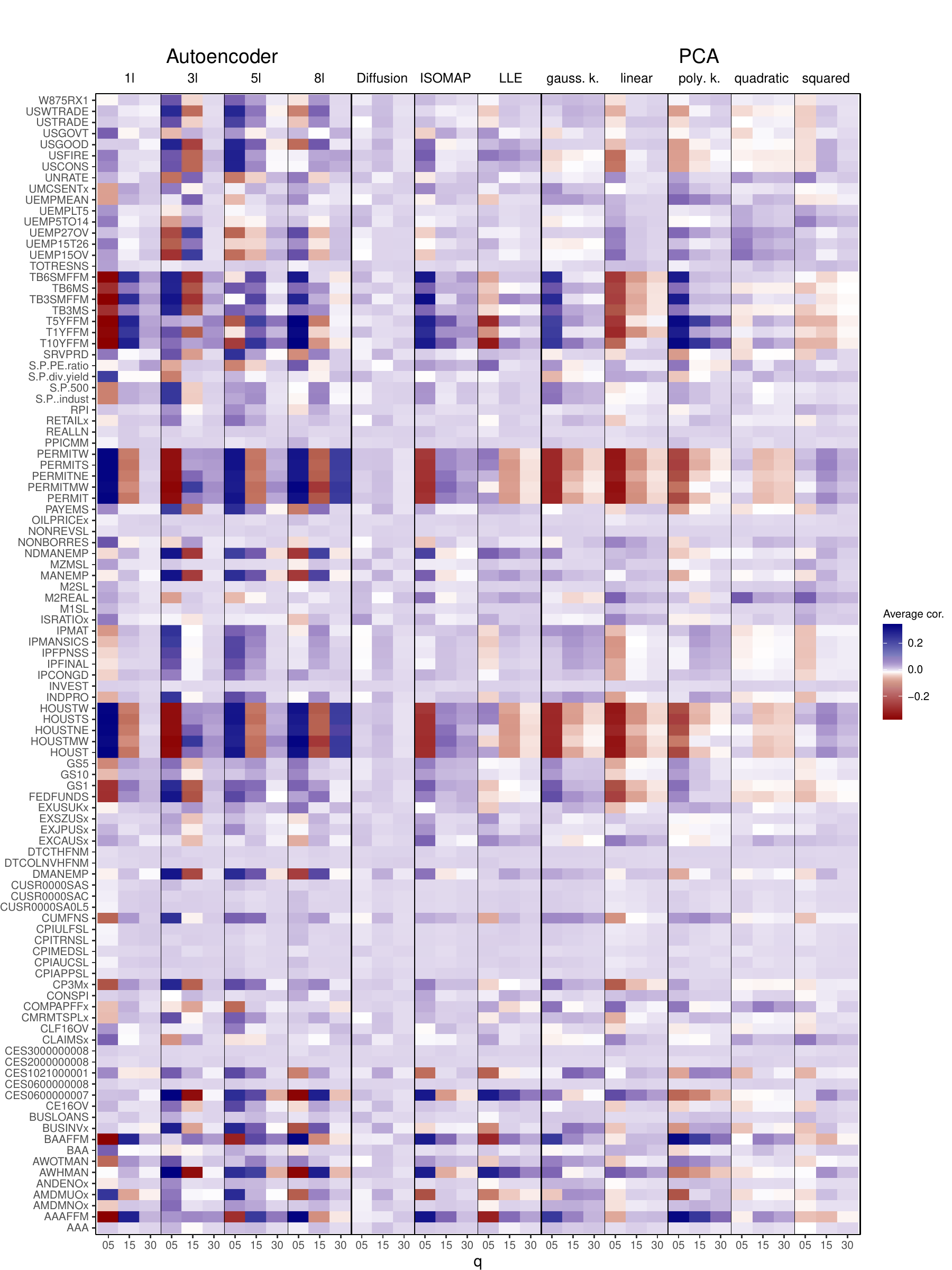}
\end{figure}

\autoref{fig:corrfac} is a heatmap of the correlations with rows denoting the different covariates in $\bm X$ and columns representing the different dimension reduction techniques. These correlations are averages across the factors (in case that $q>1$) and, since we include several lags of the input dataset, are also averaged across the lags. 

The figure suggests for most dimension reduction techniques that the factors are correlated with housing quantities (PERMIT and HOUST alongside their sub-components) as well as interest rate spreads. Some variables which measure real activity (such as industrial production and several of its components) also display comparatively large correlations with the factors. In some cases, these correlations are positive whereas in other cases, correlations are negative. In both instances, however, the absolute magnitudes are similar. The three exceptions from this rather general pattern are diffusion maps as well as PCA quadratic and squared. In this case, the corresponding columns indicate lower correlations.  

\begin{figure}[!htbp]
\caption{Correlations between the top-five variables in the input dataset and the factors obtained from the Autoencoder with one layer. \label{fig:corrAE}}
\centering
\includegraphics[scale=.55]{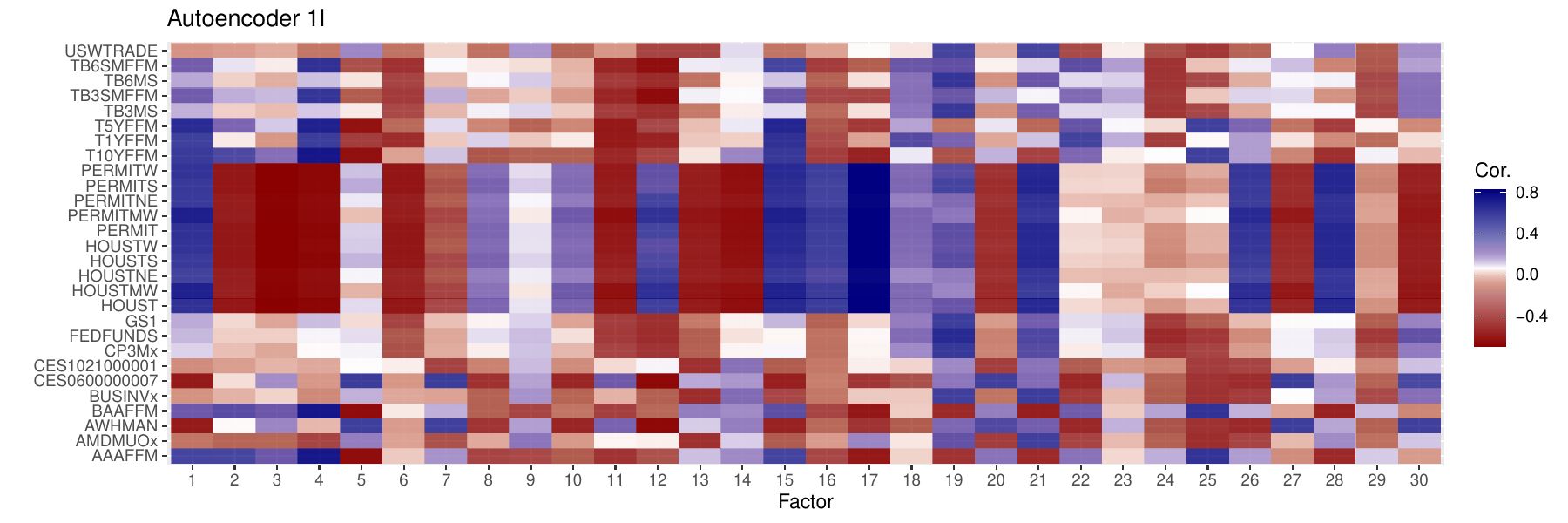}
\end{figure}

Averaging over the factors, as done in \autoref{fig:corrfac}, potentially masks important features of individual factors.  Next we ask whether there are relevant  differences by analyzing the correlations between each $\bm z_j~(j=1, \dots, q)$ and each column of $\bm X$. For brevity, we focus on a specific model that performs extraordinarily well in terms of density forecasts: the Autoencoder with a single hidden layer and $30$ factors. \autoref{fig:corrAE} shows, for each factor, the five variables which display the largest absolute correlation. The variables in the rows are a union over the sets of top-five variables for each factor. This figure shows that several factors display quite similar correlation patterns. For all of them, housing quantities are either positively or negatively correlated (with similar magnitudes). Apart from that, and in consistence with the findings discussed above, we observe that financial market variables (such as interest rate spreads) show up frequently for several factors. Only very few factors depart from this overall pattern. In the case of factors 9, 22, 23 and 24 we find low correlations with housing  and much stronger correlations with financial markets. In fact, factor 9 is closely tracking the credit (BAAFFM) and term spreads (e.g., T10YFFM).

These two heatmaps provide a rough overview on what variables drive the factors. Next, we ask whether we can construct models based on including variables which display the strongest correlations with the factors. This approach can be interpreted as a simple selection device which takes non-linearities in the input dataset implicitly into account. Since the heatmap is based on full-sample results and we are interested in using these small-scale models for out-of-sample forecasting we compute the correlation for each point in our hold-out period. In summary, the variables which frequently show up across hold-out periods and dimension reduction techniques are:
\begin{itemize}
\item \textbf{Real activity and housing:} Variables on industrial production (INDPRO, IPMANSICS), capacity utilization (CUMFNS) and private housing starts (HOUST) and permits (PERMIT),  
\item \textbf{Labor market:} Variables on (un-)employment (MANEMP, USGOOD) and average hours worked (CES0600000007, AWHMAN),
\item \textbf{Prices:} Sub-indicators of consumer prices (CUSR0000SA0L5),
\item \textbf{Interest rates and other stock market variables:} Spreads (to the Fed funds rate) of treasuries (TB3SMFFM, TB6SMFFM, T1YFFM, T10YFFM) and of corporate bonds (AAAFFM, BAAFFM, COMPAPFFx), 
\item \textbf{Money stocks and reserves:} Non-borrowed reserves (NONBORRES) and adjusted monetary base (AMBSL).
\end{itemize}
These variables are also the ones which display high correlations to the factors in \autoref{fig:corrfac} and are included in the large-scale ARX model. Here, it is worth stressing that there seems to be appreciable heterogeneity with respect to dimension reduction methods.  Most of them generate factors that are highly correlated with real activity and housing measures as well as interest rates and other stock market variables. Interestingly, when we focus on the second group we observe that the factors arising from using PCA squared (and to a somewhat lesser extent PCA quadratic) are heavily related to labor market measures. Average correlations with prices (i.e., CUSR0000SA0L5) are small for most techniques (with PCA quadratic yielding the largest correlations  of around $0.3-0.4$). Some methods also yield factors that are strongly correlated to money stocks and reserves (e.g., diffusion maps). \autoref{tab:corr_variables} of the Online Appendix provides a much more detailed picture on the precise variables used to build the small-scale models. 

\begin{table}[t]
\caption{Average correlation with inflation}\label{tab:correlation}
\small{
\begin{center}
\scalebox{0.6}{
\begin{tabular}{clclcccccccccc}
\toprule
\multicolumn{1}{l}{}&\multicolumn{1}{c}{\makecell{Business \\Cycle}}&\multicolumn{1}{c}{}&\multicolumn{1}{c}{\makecell{No. of \\factors}}&\multicolumn{1}{c}{}&\multicolumn{1}{c}{\makecell{PCA\\linear}}&\multicolumn{1}{c}{\makecell{PCA\\quadratic}}&\multicolumn{1}{c}{\makecell{PCA\\squared}}&\multicolumn{1}{c}{\makecell{PCA\\gauss. kernel}}&\multicolumn{1}{c}{\makecell{PCA\\poly. kernel}}&\multicolumn{1}{c}{ISOMAP}&\multicolumn{1}{c}{\makecell{Diffusion\\Maps}}&\multicolumn{1}{c}{LLE}&\multicolumn{1}{c}{\makecell{Autoen-\\coder 1l}}\tabularnewline
\midrule
   ~~&   Full Sample&   &   q = 05&   &   0.017&   \textbf{0.114}&   0.113&   0.013&   0.016&   0.007&   0.061&   0.009&   0.008\tabularnewline
   ~~&   &   &   q = 05&   &   (0.003,0.043)&   (0.006,0.251)&   (0.006,0.250)&   (0.004,0.024)&   (0.003,0.035)&   (0.002,0.010)&   (0.002,0.163)&   (0.003,0.017)&   (0.002,0.018)\tabularnewline
   ~~&   &   &   q = 15&   &   \textbf{0.086}&   0.069&   0.068&   0.030&   0.069&   0.034&   0.052&   0.013&   0.035\tabularnewline
   ~~&   &   &   q = 15&   &   (0.003,0.198)&   (0.003,0.251)&   (0.005,0.250)&   (0.004,0.133)&   (0.003,0.206)&   (0.003,0.095)&   (0.001,0.163)&   (0.001,0.035)&   (0.004,0.090)\tabularnewline
   ~~&   &   &   q = 30&   &   \textbf{0.108}&   0.049&   0.050&   0.052&   0.106&   0.043&   0.063&   0.036&   0.036\tabularnewline
   ~~&   &   &   q = 30&   &   (0.002,0.259)&   (0.003,0.251)&   (0.005,0.250)&   (0.004,0.274)&   (0.003,0.292)&   (0.001,0.143)&   (0.001,0.185)&   (0.001,0.168)&   (0.001,0.163)\tabularnewline
\midrule
   ~~&   Expansion&   &   q = 05&   &   0.027&   \textbf{0.069}&   0.055&   0.033&   0.030&   0.021&   0.045&   0.010&   0.023\tabularnewline
   ~~&   &   &   q = 05&   &   (0.016,0.049)&   (0.003,0.151)&   (0.003,0.110)&   (0.013,0.059)&   (0.019,0.050)&   (0.009,0.050)&   (0.022,0.098)&   (0.003,0.02)&   (0.006,0.043)\tabularnewline
   ~~&   &   &   q = 15&   &   \textbf{0.073}&   0.057&   0.055&   0.030&   0.061&   0.035&   0.049&   0.020&   0.039\tabularnewline
   ~~&   &   &   q = 15&   &   (0.001,0.192)&   (0.003,0.151)&   (0.003,0.110)&   (0.003,0.087)&   (0.009,0.194)&   (0.003,0.118)&   (0.005,0.098)&   (0.003,0.068)&   (0.003,0.115)\tabularnewline
   ~~&   &   &   q = 30&   &   \textbf{0.103}&   0.047&   0.045&   0.049&   0.098&   0.040&   0.053&   0.043&   0.041\tabularnewline
   ~~&   &   &   q = 30&   &   (0.001,0.290)&   (0.001,0.151)&   (0.001,0.110)&   (0.001,0.276)&   (0.009,0.285)&   (0.001,0.134)&   (0.005,0.116)&   (0.003,0.209)&   (0.001,0.196)\tabularnewline
\midrule
   ~~&   Recession&   &   q = 05&   &   0.098&   \textbf{0.250}&   0.248&   0.121&   0.116&   0.042&   0.180&   0.065&   0.071\tabularnewline
   ~~&   &   &   q = 05&   &   (0.040,0.156)&   (0.123,0.442)&   (0.123,0.442)&   (0.049,0.149)&   (0.051,0.163)&   (0.014,0.091)&   (0.122,0.269)&   (0.007,0.114)&   (0.012,0.190)\tabularnewline
   ~~&   &   &   q = 15&   &   \textbf{0.152}&   0.137&   0.136&   0.097&   0.143&   0.090&   0.133&   0.067&   0.068\tabularnewline
   ~~&   &   &   q = 15&   &   (0.004,0.286)&   (0.012,0.442)&   (0.016,0.442)&   (0.003,0.321)&   (0.022,0.326)&   (0.015,0.205)&   (0.019,0.269)&   (0.003,0.172)&   (0.010,0.188)\tabularnewline
   ~~&   &   &   q = 30&   &   0.152&   0.110&   0.114&   0.118&   \textbf{0.158}&   0.104&   0.132&   0.081&   0.077\tabularnewline
   ~~&   &   &   q = 30&   &   (0.004,0.297)&   (0.009,0.442)&   (0.011,0.442)&   (0.003,0.321)&   (0.022,0.424)&   (0.005,0.366)&   (0.003,0.306)&   (0.003,0.180)&   (0.003,0.297)\tabularnewline
\midrule
   ~~&   Pandemic&   &   q = 05&   &   0.320&   0.285&   0.287&   0.124&   0.231&   0.199&   \textbf{0.367}&   0.303&   0.087\tabularnewline
   ~~&   &   &   q = 05&   &   (0.059,0.631)&   (0.188,0.424)&   (0.216,0.423)&   (0.020,0.210)&   (0.017,0.642)&   (0.101,0.303)&   (0.127,0.610)&   (0.087,0.506)&   (0.046,0.131)\tabularnewline
   ~~&   &   &   q = 15&   &   0.292&   \textbf{0.322}&   0.314&   0.292&   0.260&   0.173&   0.308&   0.228&   0.126\tabularnewline
   ~~&   &   &   q = 15&   &   (0.059,0.631)&   (0.015,0.641)&   (0.012,0.642)&   (0.020,0.734)&   (0.008,0.642)&   (0.022,0.559)&   (0.053,0.610)&   (0.088,0.402)&   (0.017,0.365)\tabularnewline
   ~~&   &   &   q = 30&   &   0.269&   0.348&   0.320&   0.260&   0.262&   0.261&   0.313&   \textbf{0.358}&   0.160\tabularnewline
   ~~&   &   &   q = 30&   &   (0.029,0.631)&   (0.015,0.641)&   (0.010,0.642)&   (0.020,0.796)&   (0.008,0.646)&   (0.001,0.764)&   (0.001,0.886)&   (0.057,0.821)&   (0.007,0.475)\tabularnewline
\bottomrule
\end{tabular}
}
\begin{minipage}{\textwidth}
      \tiny
      \item \textit{Note:} The values are averaged across the number of factors stated in the second column with minimum and maximum values in parentheses. All correlation values are absolute values. The periods are divided into business cycle phases according to the NBER (\url{https://www.nber.org/research/business-cycle-dating}).
\end{minipage}
\end{center}}
\end{table}
Next, we ask whether the factors are correlated to inflation. \autoref{tab:correlation} shows the correlation with inflation averaged across the number of factors for each dimension reduction techniques as well as the minimum and maximum value (across these factors) in parentheses. To assess whether these correlations differ over time,  we divide our sample into expansionary and recessionary periods.\footnote{Recessions are defined by using the the business cycle classification of the National Bureau of Economic Research (NBER).} Since the COVID-19 pandemic marks an extraordinary period in our sample, we also compute the correlations for 2020 only and include it at the bottom of \autoref{tab:correlation}.

For the full sample as well as during expansions, we find that the factors obtained from using the linear variants of PCA display comparatively higher correlations relative to the other dimension reduction techniques (with some of the factors featuring a correlation of close to $0.2$). In recessions and the pandemic, these correlations increase substantially to reach average correlations close to $0.3$ (with the factor displaying the maximum correlation being strongly related to inflation, with values of around $0.6$).  The non-linear dimension reduction techniques yield strong correlations during turbulent times (i.e., recessions and the pandemic). This is not surprising since these methods tend to work well if there are strong deviations from linearity (which mostly occurs in recessions). Such a feature can be easily demonstrated by considering PCA squared. In normal times, the factors will be centered around zero and typically display little variation. But in recessions the link function implies that larger changes will dominate the shape of the factors and imply pronounced movements which could be helpful for predicting turning points in inflation.

\subsection{Density and point forecast performance}

We now consider point and density forecasting performance of the different models and dimension reduction techniques. 
The forecast performance is evaluated through log predictive likelihoods (LPLs) for density forecasts and root mean squared errors (RMSEs) for point forecasts. Superior models are those with high scores in terms of LPL and low values in terms of RMSE. We benchmark all models relative to the autoregressive (AR) model with constant parameters and the Minnesota prior. The first entry in the tables gives the actual value of the LPL (cumulated over the hold-out sample) with actual RMSEs in parentheses (averaged over the hold-out sample) for our benchmark model. The remaining entries are differences in  LPLs with relative RMSEs in parentheses. We mark statistically significant results according to the \cite{diebold1995dmtest} test at the one, five and ten percent significance levels with one, two and three asterisks, respectively.

\begin{table*}[!tbp]
{\tiny
\begin{center}
\caption{One-month-ahead forecast performance. \label{tab:main1}}
\begin{tabular*}{\textwidth}{l @{\extracolsep{\fill}} llcllll}
\toprule
\multicolumn{1}{l}{\bfseries }&\multicolumn{1}{c}{\bfseries Specification}&\multicolumn{1}{c}{\bfseries }&\multicolumn{1}{c}{const. (MIN)}&\multicolumn{1}{c}{const. (HS)}&\multicolumn{1}{c}{TVP (MIN)}&\multicolumn{1}{c}{TVP (HS)}\tabularnewline
\midrule
{\scshape }&&&&&&\tabularnewline
   ~~&   AR&   &  \shadeBench -329.28&     0.23&   -0.99&    -1.04\tabularnewline
  ~~&   &   &  \shadeBench  (1.24)&   (0.98)&   (0.98)&   (0.98)\tabularnewline
   ~~&   Large ARX&   &      2.46&    -9.80***&      &   -8.95***\tabularnewline
   ~~&   &   &   (0.97)&   (1.06***)&   &   (1.04***)\tabularnewline
\midrule
   ~~&   Autoencoder 1l (q = 05)&   &      2.19&    -2.06&    1.12&    -1.43\tabularnewline
   ~~&   &   &   (0.98)&   (0.99)&   (0.97)&   (0.99)\tabularnewline
   ~~&   Autoencoder 1l (q = 15)&   &     16.26***&    12.71***&   21.41***&    13.69***\tabularnewline
   ~~&   &   &   (0.90***)&   (0.91***)&   (0.87***)&   (0.90***)\tabularnewline
   ~~&   Autoencoder 1l (q = 30)&   &   \textbf{38.19***}&   \textbf{29.92***}&   \textbf{35.48***}&   \textbf{31.39***}\tabularnewline
   ~~&   &   &   (\textbf{0.81***})&   (\textbf{0.84***})&   (\textbf{0.80***})&   (\textbf{0.83***})\tabularnewline
   ~~&   Autoencoder 3l (q = 05)&   &      3.55&    -1.59&    1.90&    -2.35\tabularnewline
   ~~&   &   &   (0.97)&   (0.99)&   (0.97)&   (0.99)\tabularnewline
   ~~&   Autoencoder 3l (q = 15)&   &     11.52***&     8.90**&   13.19**&     8.35**\tabularnewline
   ~~&   &   &   (0.94**)&   (0.95*)&   (0.92**)&   (0.95**)\tabularnewline
   ~~&   Autoencoder 3l (q = 30)&   &     23.63***&    16.44***&   18.62*&    12.81*\tabularnewline
   ~~&   &   &   (0.86***)&   (0.88***)&   (0.83***)&   (0.87***)\tabularnewline
   ~~&   Autoencoder 5l (q = 05)&   &      2.50&    -2.07&    1.55&    -2.74\tabularnewline
   ~~&   &   &   (0.97)&   (0.99)&   (0.97)&   (0.99)\tabularnewline
   ~~&   Autoencoder 5l (q = 15)&   &     10.05*&     4.45&   11.65*&     5.62\tabularnewline
   ~~&   &   &   (0.94**)&   (0.95)&   (0.93**)&   (0.94*)\tabularnewline
   ~~&   Autoencoder 5l (q = 30)&   &     26.91***&    22.80***&   26.72***&    23.12***\tabularnewline
   ~~&   &   &   (0.85***)&   (0.86***)&   (0.84***)&   (0.86***)\tabularnewline
   ~~&   Autoencoder 8l (q = 05)&   &      1.79&    -2.58&    3.27&    -3.05\tabularnewline
   ~~&   &   &   (0.97)&   (0.99)&   (0.97)&   (0.99)\tabularnewline
   ~~&   Autoencoder 8l (q = 15)&   &      2.86&    -4.34&    1.49&    -2.52\tabularnewline
   ~~&   &   &   (0.98)&   (1.00)&   (0.97)&   (1.00)\tabularnewline
   ~~&   Autoencoder 8l (q = 30)&   &      5.64&    -0.97&    7.21*&     0.13\tabularnewline
   ~~&   &   &   (0.97)&   (0.99)&   (0.95)&   (0.98)\tabularnewline
\midrule
   ~~&   Diffusion Maps (q = 05)&   &      2.12&    -3.64&    2.97&    -2.40\tabularnewline
   ~~&   &   &   (0.98)&   (1.00)&   (0.98)&   (1.00)\tabularnewline
   ~~&   Diffusion Maps (q = 15)&   &      2.10&    -8.17**&    2.13&    -7.54**\tabularnewline
   ~~&   &   &   (0.98)&   (1.03**)&   (0.99)&   (1.03**)\tabularnewline
   ~~&   Diffusion Maps (q = 30)&   &      2.55&    -9.13**&    3.81&    -6.79*\tabularnewline
   ~~&   &   &   (0.98)&   (1.03**)&   (0.97)&   (1.02**)\tabularnewline
\midrule
   ~~&   ISOMAP (q = 05)&   &      2.55&    -3.36&    2.28&    -1.24\tabularnewline
   ~~&   &   &   (0.98)&   (0.99)&   (0.97)&   (0.99)\tabularnewline
   ~~&   ISOMAP (q = 15)&   &      3.01&    -5.33*&    1.31&    -5.46*\tabularnewline
   ~~&   &   &   (0.97)&   (1.01)&   (0.98)&   (1.01*)\tabularnewline
   ~~&   ISOMAP (q = 30)&   &      1.50&    -6.04**&    2.36&    -4.53\tabularnewline
   ~~&   &   &   (0.97)&   (1.01**)&   (0.97)&   (1.01)\tabularnewline
\midrule
   ~~&   LLE (q = 05)&   &      1.39&    -2.23&    1.93&    -1.84\tabularnewline
   ~~&   &   &   (0.98)&   (0.99)&   (0.98)&   (1.00)\tabularnewline
   ~~&   LLE (q = 15)&   &      2.68&    -5.87**&    0.83&    -5.18*\tabularnewline
   ~~&   &   &   (0.97)&   (1.02*)&   (0.97)&   (1.01)\tabularnewline
   ~~&   LLE (q = 30)&   &      1.86&   -11.38**&    0.45&   -11.49**\tabularnewline
   ~~&   &   &   (0.98)&   (1.02*)&   (0.98)&   (1.01*)\tabularnewline
\midrule
   ~~&   PCA gauss. kernel (q = 05)&   &      2.44&    -1.37&    1.50&     0.07\tabularnewline
   ~~&   &   &   (0.98)&   (0.99)&   (0.97)&   (0.98)\tabularnewline
   ~~&   PCA gauss. kernel (q = 15)&   &      2.97&    -1.30&    1.85&    -3.88\tabularnewline
   ~~&   &   &   (0.98)&   (0.99)&   (0.98)&   (0.99)\tabularnewline
   ~~&   PCA gauss. kernel (q = 30)&   &      2.68&    -2.94&    3.10&    -1.36\tabularnewline
   ~~&   &   &   (0.97)&   (1.00)&   (0.97)&   (0.99)\tabularnewline
\midrule
   ~~&   PCA linear (q = 05)&   &      1.96&    -3.74&    1.27&    -2.46\tabularnewline
   ~~&   &   &   (0.98)&   (1.00)&   (0.98)&   (0.99)\tabularnewline
   ~~&   PCA linear (q = 15)&   &      3.78&    -4.89&    3.35&    -3.76\tabularnewline
   ~~&   &   &   (0.97)&   (1.01)&   (0.97)&   (1.00)\tabularnewline
   ~~&   PCA linear (q = 30)&   &      3.25&    -7.51**&    2.10&    -5.38\tabularnewline
   ~~&   &   &   (0.97)&   (1.02**)&   (0.98)&   (1.02**)\tabularnewline
\midrule
   ~~&   PCA poly. kernel (q = 05)&   &      3.74&    -1.06&    2.50&    -0.76\tabularnewline
   ~~&   &   &   (0.98)&   (0.99)&   (0.98)&   (0.99)\tabularnewline
   ~~&   PCA poly. kernel (q = 15)&   &      1.01&    -1.90&    2.37&    -2.17\tabularnewline
   ~~&   &   &   (0.98)&   (0.99)&   (0.98)&   (0.99)\tabularnewline
   ~~&   PCA poly. kernel (q = 30)&   &      4.28&    -1.78&    2.51&    -2.43\tabularnewline
   ~~&   &   &   (0.97)&   (0.99)&   (0.97)&   (0.99)\tabularnewline
\midrule
   ~~&   PCA quadratic (q = 05)&   &     10.56**&     7.15&    9.72*&     8.32\tabularnewline
   ~~&   &   &   (0.91*)&   (0.93)&   (0.90*)&   (0.93)\tabularnewline
   ~~&   PCA quadratic (q = 15)&   &      4.63&     0.01&    6.79&     1.34\tabularnewline
   ~~&   &   &   (1.00)&   (0.99)&   (0.97)&   (0.98)\tabularnewline
   ~~&   PCA quadratic (q = 30)&   &      3.15&    -7.23&    5.11&    -4.72\tabularnewline
   ~~&   &   &   (0.97)&   (1.04**)&   (0.96)&   (1.03*)\tabularnewline
\midrule
   ~~&   PCA squared (q = 05)&   &      9.78*&     8.44&   11.30*&     7.10\tabularnewline
   ~~&   &   &   (0.90*)&   (0.92)&   (0.89*)&   (0.92)\tabularnewline
   ~~&   PCA squared (q = 15)&   &      5.77&     0.70&    7.36&     2.47\tabularnewline
   ~~&   &   &   (0.97)&   (0.98)&   (0.93)&   (0.98)\tabularnewline
   ~~&   PCA squared (q = 30)&   &      4.33&    -1.49&    6.79*&    -3.58\tabularnewline
   ~~&   &   &   (0.97)&   (1.01)&   (0.95)&   (1.02)\tabularnewline
\bottomrule
\end{tabular*}
\begin{tablenotes}[flushleft]
\tiny
\item \textit{Note:} The table shows LPLs with RMSEs in parentheses below. The first (red shaded) entry gives the actual LPL and RMSE scores of our benchmark (an autoregressive (AR) model with constant parameters and a Minnesota prior). Asterisks indicate statistical significance for each model relative to the benchmark at the 1\% (***), 5\% (**) and 10\% (*) significance levels. Since the large ARX model with time-varying parameters would features $273$ period-specific coefficients and is computationally intractable, we assume that the TVPs feature a factor structure (with three factors) to reduce the dimension of the state space \citep[see Section \ref{sec:App B} of the Online Appendix and][]{chan2020reducing}.
\end{tablenotes}
\end{center}}
\end{table*}

Starting with the one-month-ahead horizon, \autoref{tab:main1} depicts the inflation forecasting results. This table suggests that, in terms of density forecasts, using dimension reduction techniques (both linear and non-linear) improves predictions substantially. These improvements arise not only relative to the AR benchmark but also related to the large AR models with additional exogenous regressors. For some models, these improvements are sizable, irrespective of the regression specification (i.e., whether we use a constant parameter or a TVP model). Especially the Autoencoder with one and five layers sharply improves upon the benchmark (and all the remaining competitors) by large margins. Moreover, it yields statistically significant improvements at the one percent level. A similar story emerges when we focus on point forecasts. Non-linear dimension reductions help slightly. Relative RMSEs are smaller but close to one for most models. Again, the Autoencoder works well and yields RMSEs which are, across regression specifications, almost 20 percent lower than the ones from the benchmark. It is worth emphasizing that PCA squared also yields highly competitive point forecasts which are statistically significant according to the \cite{diebold1995dmtest} test.

When we compare model performance across regression specifications and focus on the Minnesota-type priors, we find that constant parameter models  work quite well if non-linear dimension reduction techniques such as the Autoencoder are adopted. With a single exception (diffusion maps), introducing TVPs does not pay off and yields density forecasts which are slightly more imprecise than the ones obtained from their time-invariant counterparts. If we use a horseshoe prior, this result somewhat reverses (with the caveat that the models coupled with the horseshoe sometimes yield weaker inflation forecasts than the benchmark). Here, we observe that introducing TVPs often improves log predictive likelihoods relative to the constant parameter model with the same prior.

Summing up this discussion, we observe that the Autoencoder yields favorable point and density forecasts, irrespective of the prior and regression specification chosen. This strong performance of the Autoencoder, however, depends on the number of layers as well as the number of factors. The  literature \citep[see, e.g.,][]{huang2003learning, heaton2008introduction} suggests that the number of hidden layers should increase with the complexity of the dataset. Our results, however, suggest the opposite. For a typical US macroeconomic dataset the forecast performance of the Autoencoder seems to be strongest when a single hidden layer coupled with a large number of factors is used.

\begin{table*}[!tbp]
{\tiny
\begin{center}
\caption{One-quarter-ahead forecast performance. \label{tab:main3}}
\begin{tabular*}{\textwidth}{l @{\extracolsep{\fill}} llcllll}
\toprule
\multicolumn{1}{l}{\bfseries }&\multicolumn{1}{c}{\bfseries Specification}&\multicolumn{1}{c}{\bfseries }&\multicolumn{1}{c}{const. (MIN)}&\multicolumn{1}{c}{const. (HS)}&\multicolumn{1}{c}{TVP (MIN)}&\multicolumn{1}{c}{TVP (HS)}\tabularnewline
\midrule
{\scshape }&&&&&&\tabularnewline
   ~~&   AR&   &  \shadeBench -322.66&     9.79&    20.02&    18.53\tabularnewline
  ~~&   &   &  \shadeBench  (1.27)&   (0.94*)&   (0.89*)&   (0.89**)\tabularnewline
   ~~&   Large ARX&   &     19.68&    14.44&       &   11.03\tabularnewline
   ~~&   &   &   (0.89*)&   (0.94)&   &   (0.94)\tabularnewline
\midrule
   ~~&   Autoencoder 1l (q = 05)&   &     18.26&    15.86&    14.72&    17.25\tabularnewline
   ~~&   &   &   (0.89*)&   (0.89*)&   (0.89*)&   (0.89*)\tabularnewline
   ~~&   Autoencoder 1l (q = 15)&   &     24.56&    27.39&    22.94&    26.17\tabularnewline
   ~~&   &   &   (0.86**)&   (0.85**)&   (0.86**)&   (0.85**)\tabularnewline
   ~~&   Autoencoder 1l (q = 30)&   &     22.17&    27.55&    24.62&    26.65\tabularnewline
   ~~&   &   &   (0.85**)&   (0.84**)&   (0.85**)&   (0.83**)\tabularnewline
   ~~&   Autoencoder 3l (q = 05)&   &     19.24&    22.33&    21.18&    19.49\tabularnewline
   ~~&   &   &   (0.87**)&   (0.85**)&   (0.87**)&   (0.86**)\tabularnewline
   ~~&   Autoencoder 3l (q = 15)&   &     20.24&    21.49&    20.54&    20.60\tabularnewline
   ~~&   &   &   (0.86**)&   (0.86**)&   (0.86**)&   (0.85**)\tabularnewline
   ~~&   Autoencoder 3l (q = 30)&   &     23.97&    20.29&    21.60&    20.45\tabularnewline
   ~~&   &   &   (0.86**)&   (0.85**)&   (0.86**)&   (0.85**)\tabularnewline
   ~~&   Autoencoder 5l (q = 05)&   &     20.87&    16.60&    18.60&    19.66\tabularnewline
   ~~&   &   &   (0.86**)&   (0.86**)&   (0.87**)&   (0.86**)\tabularnewline
   ~~&   Autoencoder 5l (q = 15)&   &     15.15&    14.34&    15.71&    14.00\tabularnewline
   ~~&   &   &   (0.86**)&   (0.86**)&   (0.87**)&   (0.86**)\tabularnewline
   ~~&   Autoencoder 5l (q = 30)&   &     21.13&    22.40&    19.67&    21.44\tabularnewline
   ~~&   &   &   (0.86**)&   (0.85**)&   (0.86**)&   (0.85**)\tabularnewline
   ~~&   Autoencoder 8l (q = 05)&   &     20.82&    18.27&    16.17&    17.18\tabularnewline
   ~~&   &   &   (0.88*)&   (0.88*)&   (0.89*)&   (0.88*)\tabularnewline
   ~~&   Autoencoder 8l (q = 15)&   &     12.21&     6.51&     8.01&     7.27\tabularnewline
   ~~&   &   &   (0.88*)&   (0.92)&   (0.90)&   (0.91)\tabularnewline
   ~~&   Autoencoder 8l (q = 30)&   &     -3.75&   -16.20&   -11.91&   -14.54\tabularnewline
   ~~&   &   &   (0.89)&   (0.90)&   (0.89)&   (0.90)\tabularnewline
\midrule
   ~~&   Diffusion Maps (q = 05)&   &     16.27&    20.18&    12.09&    18.18\tabularnewline
   ~~&   &   &   (0.86**)&   (0.86**)&   (0.86**)&   (0.86**)\tabularnewline
   ~~&   Diffusion Maps (q = 15)&   &     19.69&    16.58&    17.53&    14.27\tabularnewline
   ~~&   &   &   (0.86**)&   (0.86**)&   (0.86**)&   (0.86**)\tabularnewline
   ~~&   Diffusion Maps (q = 30)&   &     17.02&    13.20&    20.08&    12.27\tabularnewline
   ~~&   &   &   (0.87**)&   (0.89*)&   (0.86**)&   (0.88**)\tabularnewline
\midrule
   ~~&   ISOMAP (q = 05)&   &     18.01&    16.89&    15.62&    18.45\tabularnewline
   ~~&   &   &   (0.86**)&   (0.86**)&   (0.87**)&   (0.86**)\tabularnewline
   ~~&   ISOMAP (q = 15)&   &     19.75&    11.58&    13.79&    11.19\tabularnewline
   ~~&   &   &   (0.86**)&   (0.86**)&   (0.87**)&   (0.86**)\tabularnewline
   ~~&   ISOMAP (q = 30)&   &     12.01&     3.63&     5.41&     2.66\tabularnewline
   ~~&   &   &   (0.86**)&   (0.85**)&   (0.87**)&   (0.86**)\tabularnewline
\midrule
   ~~&   LLE (q = 05)&   &     14.10&     5.25&     9.71&     5.27\tabularnewline
   ~~&   &   &   (0.87**)&   (0.87**)&   (0.87**)&   (0.87**)\tabularnewline
   ~~&   LLE (q = 15)&   &    -11.38&   -13.20&   -18.14&   -10.97\tabularnewline
   ~~&   &   &   (0.88*)&   (0.88)&   (0.89)&   (0.88)\tabularnewline
   ~~&   LLE (q = 30)&   &     -1.39&   -38.94&   -11.37&   -40.18\tabularnewline
   ~~&   &   &   (0.87**)&   (0.87*)&   (0.88*)&   (0.87*)\tabularnewline
\midrule
   ~~&   PCA gauss. kernel (q = 05)&   &     17.21&    17.81&    20.04&    15.67\tabularnewline
   ~~&   &   &   (0.88*)&   (0.88**)&   (0.88*)&   (0.88*)\tabularnewline
   ~~&   PCA gauss. kernel (q = 15)&   &     15.63&    14.29&    16.88&    17.30\tabularnewline
   ~~&   &   &   (0.88*)&   (0.88*)&   (0.88*)&   (0.88*)\tabularnewline
   ~~&   PCA gauss. kernel (q = 30)&   &     19.03&    19.01&    18.04&    18.17\tabularnewline
   ~~&   &   &   (0.88**)&   (0.88**)&   (0.88**)&   (0.88**)\tabularnewline
\midrule
   ~~&   PCA linear (q = 05)&   &     17.05&    15.66&    15.05&    14.73\tabularnewline
   ~~&   &   &   (0.90*)&   (0.90*)&   (0.90)&   (0.90*)\tabularnewline
   ~~&   PCA linear (q = 15)&   &     18.79&    19.39&    16.35&    14.59\tabularnewline
   ~~&   &   &   (0.88*)&   (0.88*)&   (0.89*)&   (0.88*)\tabularnewline
   ~~&   PCA linear (q = 30)&   &     21.39&    18.10&    19.18&    18.21\tabularnewline
   ~~&   &   &   (0.87**)&   (0.87*)&   (0.88*)&   (0.87*)\tabularnewline
\midrule
   ~~&   PCA poly. kernel (q = 05)&   &     18.98&    16.48&    15.09&    14.71\tabularnewline
   ~~&   &   &   (0.89*)&   (0.90*)&   (0.89*)&   (0.89*)\tabularnewline
   ~~&   PCA poly. kernel (q = 15)&   &     18.73&    19.44&    18.52&    20.78\tabularnewline
   ~~&   &   &   (0.88*)&   (0.88**)&   (0.88*)&   (0.88**)\tabularnewline
   ~~&   PCA poly. kernel (q = 30)&   &     22.11&    20.63&    19.63&    21.22\tabularnewline
   ~~&   &   &   (0.87**)&   (0.87**)&   (0.88**)&   (0.87**)\tabularnewline
\midrule
   ~~&   PCA quadratic (q = 05)&   &     40.33***&   \textbf{46.67**}&    41.31***&   \textbf{48.50***}\tabularnewline
   ~~&   &   &   (0.84**)&   (0.79***)&   (0.84**)&   (0.79***)\tabularnewline
   ~~&   PCA quadratic (q = 15)&   &     20.37&    33.39&    17.63&    34.40\tabularnewline
   ~~&   &   &   (0.91)&   (0.86*)&   (0.91)&   (0.86*)\tabularnewline
   ~~&   PCA quadratic (q = 30)&   &     24.90&    26.61&    28.50*&    26.91\tabularnewline
   ~~&   &   &   (0.87**)&   (0.89**)&   (0.87**)&   (0.90**)\tabularnewline
\midrule
   ~~&   PCA squared (q = 05)&   &   \textbf{46.84***}&    46.30**&   \textbf{45.90***}&    47.62**\tabularnewline
   ~~&   &   &   (\textbf{0.80**})&   (\textbf{0.76***})&   (\textbf{0.81**})&   (\textbf{0.76***})\tabularnewline
   ~~&   PCA squared (q = 15)&   &     24.28&    33.35*&    23.02&    32.54*\tabularnewline
   ~~&   &   &   (0.87*)&   (0.87*)&   (0.87*)&   (0.88*)\tabularnewline
   ~~&   PCA squared (q = 30)&   &     28.25*&    27.59&    26.46*&    27.35\tabularnewline
   ~~&   &   &   (0.86**)&   (0.88**)&   (0.86**)&   (0.88**)\tabularnewline
\bottomrule
\end{tabular*}
\begin{tablenotes}[flushleft]
\tiny
\item \textit{Note:} The table shows LPLs with RMSEs in parentheses below. The first (red shaded) entry gives the actual LPL and RMSE scores of our benchmark (an autoregressive (AR) model with constant parameters and a Minnesota prior). Asterisks indicate statistical significance for each model relative to the benchmark at the 1\% (***), 5\% (**) and 10\% (*) significance levels. Since the large ARX model with time-varying parameters would features $273$ period-specific coefficients and is computationally intractable, we assume that the TVPs feature a factor structure (with three factors) to reduce the dimension of the state space \citep[see Section \ref{sec:App B} of the Online Appendix and][]{chan2020reducing}. 
\end{tablenotes}
\end{center}}
\end{table*}

Next, we inspect the longer forecast horizon in greater detail. \autoref{tab:main3} depicts the forecast performance of all competitors for the one-quarter-ahead horizon. The table indicates that several non-linear dimension reduction techniques (most notably the Autoencoder, PCA quadratic and PCA squared) clearly outperform the autoregressive benchmark as well as models based on linear PCs.  The improvements relative to linear PCs are sizable and statistically significant (especially for squared and quadratic PCs).  For this horizon, the large ARX models also exhibit excellent forecasting properties.

Zooming into the different approaches to dimension reduction reveals that PCA quadratic with TVPs yields highest LPLs. Among the different dimension reduction techniques, both PCA quadratic and squared stand out and improve appreciably against all other dimension reduction techniques. The Autoencoder also provides density forecasts which are highly competitive.

When we consider point forecasts a similar picture arises. Here we find that models which do well in terms of LPLs also yield precise point forecasts. The single best performing specification, however, is PCA squared with five factors, constant parameters and a horseshoe prior. This model improves upon the AR benchmark by around 24 percent. Notice, however, that the same model but with TVPs also yields forecasts which are 24 percent more precise than the ones of the benchmark. This suggests that the horseshoe shrinks the TVPs close to zero and the corresponding point estimators are almost identical. Since the LPLs differ, the remaining time-variation in the coefficients mainly impacts the LPLs through the predictive variance.

\autoref{tab:mainARX} provides a summary of the best performing models for the one-month and the one-quarter-ahead forecasts, respectively. Moreover, we also assess how a model which includes the five variables displaying the highest average correlations to the factors performs (see discussion in Sub-section \ref{sec:varselect}). These models are labeled small ARX in the table. The results suggest that for one-month-ahead forecasts, replacing the latent factors arising from the Autoencoder with observed variables that display a high correlation to the factors does not pay off in terms of point and density forecasts. Across all regression specifications, the RMSEs are higher and the LPLs lower. When we consider one-quarter-ahead forecasts, however, this strategy seems to work much better. In this case, smaller models with covariates selected based on their correlations to the factors obtained by using PCA squared and PCA quadratic yield density predictions which are close to the ones obtained from exploiting all available data. Notice, however, that for point forecasts, the performance of the small models is inferior.
\begin{table*}[!tbp]
{\tiny
\begin{center}
\caption{Summary of best performing models and comparision to small-scale unsupervised ARX models.    \label{tab:mainARX}}
\begin{tabular*}{\textwidth}{l @{\extracolsep{\fill}} llcllll}
\toprule
\multicolumn{1}{l}{\bfseries }&\multicolumn{1}{c}{\bfseries Specification}&\multicolumn{1}{c}{\bfseries }&\multicolumn{1}{c}{const. (MIN)}&\multicolumn{1}{c}{const. (HS)}&\multicolumn{1}{c}{TVP (MIN)}&\multicolumn{1}{c}{TVP (HS)}\tabularnewline
\midrule
\multicolumn{7}{c}{\textit{One-month-ahead}}\tabularnewline
\midrule
~~&   AR&   &  \shadeBench -329.28&  &&\tabularnewline
  ~~&   &   &  \shadeBench  (1.24)&  &&\tabularnewline
~~&   Autoencoder 1l (q = 30) &   &   38.19***&   29.92***&   35.48***&   31.39***\tabularnewline
 ~~&   &   &   (0.81***)&   (0.84***)&   (0.80***)&   (0.83***)\tabularnewline  
{\scshape }&&&&&&\tabularnewline  
~~&  Small ARX based on&   &      1.39&   -2.02&   1.97&   -2.27      \tabularnewline
   ~~& Autoencoder 1l (q = 30)  &   &   (0.98)&   (0.99)&   (0.98)&   (0.99)   \tabularnewline
\midrule
\multicolumn{7}{c}{\textit{One-quarter-ahead}}\tabularnewline
\midrule
~~&   AR&   &  \shadeBench -322.66&  &&\tabularnewline
~~&   &   &  \shadeBench  (1.27)&  &&\tabularnewline
~~&   PCA quadratic (q = 05)&   &     40.33***&  46.67**&    41.31***&  48.50***\tabularnewline
~~&   &   &   (0.84**)&   (0.79***)&   (0.84**)&   (0.79***)\tabularnewline
~~&   PCA squared (q = 05)&   &   46.84***&    46.30**&  45.90***&    47.62**\tabularnewline
~~&   &   &   (0.80**)&   (0.76***)&   (0.81**)&   (0.76***)\tabularnewline
{\scshape }&&&&&&\tabularnewline 
~~& Small ARX based on  &   &   38.20**&   43.73**&   39.41**&   43.28**    \tabularnewline
~~&   PCA quadratic (q = 5)&   &   (1.05)&   (0.93)&   (1.04)&   (0.93)  \tabularnewline
~~& Small ARX based on  &   &     36.25**&   42.90**&   38.87**&   44.25**\tabularnewline
~~& PCA squared (q = 5)  &   &   (1.09)&   (0.95)&   (1.09)&   (0.95)    \tabularnewline

\bottomrule
\end{tabular*}
\begin{tablenotes}[flushleft]
\tiny
\item \textit{Note:} The table shows LPLs with RMSEs in parentheses below. While the first entry denotes the benchmark (shaded in red), other entries contrast the best performing model(s) as presented in the Tables \ref{tab:main1} and \ref{tab:main3} with ARX models. 
For these ARX specifications the best performing model solely serves as an unsupervised variable selection device. Conditional on the latent factors of this best performing dimension reduction technique, we always include the top-five correlated variables as covariates alongside the lags of inflation (see Figure \ref{fig:corrfreq} of the Online Appendix).  Asterisks indicate statistical significance for each model relative to the benchmark at the 1\% (***), 5\% (**) and 10\% (*) significance levels. 
\end{tablenotes}
\end{center}}
\end{table*}

Before proceeding to the next sub-section we briefly discuss two important issues. First, it is worth stressing that the factors used in this forecasting exercise are extracted from the full set of variables in $\bm X$. In \autoref{tab:main1_slwfst} and \autoref{tab:main3_slwfst} of the Online Appendix we divide the dataset into slow- and fast-moving variables \citep{bernanke2005measuring} and extract the latent factors from these partitioned datasets exclusively. The main results based on extracting the factors from the full dataset remain in place: for one-month-ahead forecasts we find the Autoencoder to perform particularly well whereas for one-quarter-ahead predictions PCA squared and quadratic yield accurate forecast densities. 

Second, for one of our best performing models (the Autoencoder with one hidden layer) forecasting performance changes sharply when the number of factors is changed. This raises the question on how the relationship between the number of factors and forecast performance is. In \autoref{fig:AE_fac_errors} in the Online Appendix we show two graphs that discuss how point and density forecasting performance change with the number of factors. In this exercise we find that the largest jumps in predictive accuracy is found when increasing the number of factors from 17 to 24 and again from 29 to 30 in terms of LPLs and from 18 to 26 and 29 to 30 in terms of RMSE.

\subsection{Assessing model calibration using probability integral transforms}\label{sec:PITs}
The results based on RMSEs and LPLs provide information on relative forecasting performance. In the next step, we ask whether the different methods and models we propose yield predictive distribution which are better calibrated. To this end, we consider the normalized forecast errors obtained through the probability integral transform (PIT). If a model is correctly specified the PITs are iid uniformly distributed and the respective standardized forecast errors should be iid normally distributed. Departures from the standard Gaussian distribution allow us to inspect along what dimensions the model is poorly calibrated. For instance, if the variance of the normalized forecast error is too small (i.e., below one) this is evidence that the predictive distribution is too wide (i.e., too many predictions are in the tails) while values greater than one indicate that the variance is too tight (i.e., the tails are not adequately represented).

\begin{landscape}
\begin{table*}[ht]
{\tiny
\caption{Test statistics of one-month-ahead probability integral transformations \label{tab:PITs1}}
\begin{center}
\begin{tabular*}{\linewidth}{l @{\extracolsep{\fill}} lclllclllclllclll}
\toprule
\multicolumn{1}{l}{\bfseries }&\multicolumn{1}{c}{\bfseries Specification}&\multicolumn{1}{c}{\bfseries }&\multicolumn{3}{c}{\bfseries const. (MIN)}&\multicolumn{1}{c}{\bfseries }&\multicolumn{3}{c}{\bfseries const. (HS)}&\multicolumn{1}{c}{\bfseries }&\multicolumn{3}{c}{\bfseries TVP (MIN)}&\multicolumn{1}{c}{\bfseries }&\multicolumn{3}{c}{\bfseries TVP (HS)}\tabularnewline
\cmidrule{4-6} \cmidrule{8-10} \cmidrule{12-14} \cmidrule{16-18}
\multicolumn{1}{l}{}&\multicolumn{1}{c}{}&\multicolumn{1}{c}{}&\multicolumn{1}{c}{Mean}&\multicolumn{1}{c}{Variance}&\multicolumn{1}{c}{AR(1) coef.}&\multicolumn{1}{c}{}&\multicolumn{1}{c}{Mean}&\multicolumn{1}{c}{Variance}&\multicolumn{1}{c}{AR(1) coef.}&\multicolumn{1}{c}{}&\multicolumn{1}{c}{Mean}&\multicolumn{1}{c}{Variance}&\multicolumn{1}{c}{AR(1) coef.}&\multicolumn{1}{c}{}&\multicolumn{1}{c}{Mean}&\multicolumn{1}{c}{Variance}&\multicolumn{1}{c}{AR(1) coef.}\tabularnewline
\midrule
{\scshape }&&&&&&&&&&&&&&&&&\tabularnewline
   ~~&   AR&   &   0.018&   1.171&   0.063&   &   0.035&   1.182&   0.067&   &   0.024&   1.165&   0.032&   &   0.038&   1.189&   0.063\tabularnewline
   ~~&   Large ARX&   &   0.036&   1.166&   0.086&   &   0.029&   1.198&   0.061&   &   &   &   &   &   0.026&   1.195&   0.069\tabularnewline
\midrule
   ~~&   Autoencoder 1l (q = 05)&   &   0.028&   1.182&   0.082&   &   0.031&   1.194&   0.068&   &   0.024&   1.188&   0.081&   &   0.034&   1.191&   0.063\tabularnewline
   ~~&   Autoencoder 1l (q = 15)&   &   0.020&   1.176&   0.067&   &   0.029&   1.203&   0.054&   &   0.007&   1.155&   0.066&   &   0.030&   1.199&   0.054\tabularnewline
   ~~&  Autoencoder 1l (q = 30)&   &   0.010&  1.293** & 0.010&   &   0.028&   1.336**&   0.002&   &   0.005&   1.307**&   -0.013&   &   0.023&   1.317**&   -0.004\tabularnewline
   ~~&   Autoencoder 3l (q = 05)&   &   0.030&   1.175&   0.083&   &   0.038&   1.191&   0.063&   &   0.030&   1.183&   0.067&   &   0.035&   1.196&   0.067\tabularnewline
   ~~&   Autoencoder 3l (q = 15)&   &   0.045&   1.187&   0.084&   &   0.044&   1.200&   0.059&   &   0.029&   1.193&   0.062&   &   0.045&   1.203&   0.057\tabularnewline
   ~~&   Autoencoder 3l (q = 30)&   &   -0.005&   1.277*&   0.050&   &   0.009&   1.315**&   0.018&   &   -0.041&   1.343*&   0.061&   &   0.005&   1.357**&   0.027\tabularnewline
   ~~&   Autoencoder 5l (q = 05)&   &   0.034&   1.181&   0.084&   &   0.036&   1.189&   0.065&   &   0.031&   1.183&   0.077&   &   0.039&   1.192&   0.062\tabularnewline
   ~~&   Autoencoder 5l (q = 15)&   &   0.040&   1.191&   0.068&   &   0.057&   1.220&   0.041&   &   0.032&   1.196&   0.056&   &   0.058&   1.218&   0.035\tabularnewline
   ~~&   Autoencoder 5l (q = 30)&   &   0.062&   1.225&   0.074&   &   0.067&   1.230&   0.058&   &   0.039&   1.244&   0.059&   &   0.066&   1.233&   0.046\tabularnewline
   ~~&   Autoencoder 8l (q = 05)&   &   0.034&   1.190&   0.088&   &   0.039&   1.193&   0.068&   &   0.034&   1.167&   0.081&   &   0.036&   1.201&   0.064\tabularnewline
   ~~&   Autoencoder 8l (q = 15)&   &   0.038&   1.174&   0.079&   &   0.033&   1.201&   0.075&   &   0.030&   1.193&   0.090&   &   0.039&   1.189&   0.071\tabularnewline
   ~~&   Autoencoder 8l (q = 30)&   &   0.032&   1.164&   0.070&   &   0.027&   1.187&   0.048&   &   0.025&   1.152&   0.072&   &   0.026&   1.180&   0.045\tabularnewline
\midrule
   ~~&   Diffusion Maps (q = 05)&   &   0.025&   1.193&   0.079&   &   0.023&   1.212&   0.070&   &   0.012&   1.166&   0.070&   &   0.026&   1.198&   0.067\tabularnewline
   ~~&   Diffusion Maps (q = 15)&   &   0.037&   1.192&   0.079&   &   0.031&   1.211&   0.068&   &   0.027&   1.188&   0.075&   &   0.031&   1.210&   0.064\tabularnewline
   ~~&   Diffusion Maps (q = 30)&   &   0.032&   1.194&   0.083&   &   0.034&   1.212&   0.059&   &   0.029&   1.187&   0.070&   &   0.033&   1.190&   0.054\tabularnewline
\midrule
   ~~&   ISOMAP (q = 05)&   &   0.024&   1.170&   0.085&   &   0.026&   1.197&   0.057&   &   0.020&   1.175&   0.082&   &   0.031&   1.178&   0.060\tabularnewline
   ~~&   ISOMAP (q = 15)&   &   0.027&   1.180&   0.086&   &   0.027&   1.189&   0.058&   &   0.024&   1.181&   0.080&   &   0.021&   1.194&   0.059\tabularnewline
   ~~&   ISOMAP (q = 30)&   &   0.031&   1.195&   0.079&   &   0.026&   1.194&   0.068&   &   0.026&   1.193&   0.085&   &   0.031&   1.188&   0.065\tabularnewline
\midrule
   ~~&   LLE (q = 05)&   &   0.029&   1.196&   0.083&   &   0.025&   1.191&   0.071&   &   0.019&   1.177&   0.085&   &   0.026&   1.181&   0.071\tabularnewline
   ~~&   LLE (q = 15)&   &   0.027&   1.189&   0.083&   &   0.025&   1.206&   0.067&   &   0.020&   1.185&   0.084&   &   0.022&   1.196&   0.066\tabularnewline
   ~~&   LLE (q = 30)&   &   0.031&   1.194&   0.080&   &   0.034&   1.261*&   0.045&   &   0.013&   1.194&   0.086&   &   0.033&   1.247*&   0.040\tabularnewline
\midrule
   ~~&   PCA gauss. kernel (q = 05)&   &   0.027&   1.174&   0.081&   &   0.035&   1.190&   0.066&   &   0.025&   1.194&   0.084&   &   0.035&   1.175&   0.072\tabularnewline
   ~~&   PCA gauss. kernel (q = 15)&   &   0.032&   1.178&   0.087&   &   0.041&   1.189&   0.067&   &   0.030&   1.192&   0.080&   &   0.036&   1.205&   0.066\tabularnewline
   ~~&   PCA gauss. kernel (q = 30)&   &   0.032&   1.178&   0.078&   &   0.034&   1.203&   0.069&   &   0.023&   1.170&   0.083&   &   0.038&   1.182&   0.065\tabularnewline
\midrule
   ~~&   PCA linear (q = 05)&   &   0.033&   1.190&   0.083&   &   0.042&   1.208&   0.075&   &   0.023&   1.192&   0.091&   &   0.041&   1.183&   0.063\tabularnewline
   ~~&   PCA linear (q = 15)&   &   0.034&   1.177&   0.083&   &   0.037&   1.192&   0.070&   &   0.026&   1.172&   0.083&   &   0.038&   1.187&   0.064\tabularnewline
   ~~&   PCA linear (q = 30)&   &   0.031&   1.185&   0.081&   &   0.035&   1.210&   0.067&   &   0.025&   1.176&   0.087&   &   0.030&   1.187&   0.063\tabularnewline
\midrule
   ~~&   PCA poly. kernel (q = 05)&   &   0.029&   1.170&   0.081&   &   0.036&   1.193&   0.067&   &   0.029&   1.178&   0.081&   &   0.037&   1.189&   0.061\tabularnewline
   ~~&   PCA poly. kernel (q = 15)&   &   0.028&   1.201&   0.078&   &   0.034&   1.193&   0.061&   &   0.034&   1.187&   0.077&   &   0.031&   1.192&   0.068\tabularnewline
   ~~&   PCA poly. kernel (q = 30)&   &   0.031&   1.166&   0.090&   &   0.043&   1.185&   0.066&   &   0.029&   1.182&   0.080&   &   0.042&   1.199&   0.059\tabularnewline
\midrule
   ~~&   PCA quadratic (q = 05)&   &   0.029&   1.071&   0.044&   &   0.031&   1.070&   0.023&   &   0.034&   1.055&   0.026&   &   0.032&   1.052&   0.016\tabularnewline
   ~~&   PCA quadratic (q = 15)&   &   0.032&   1.178&   0.069&   &   0.008&   1.190&   0.063&   &   0.021&   1.158&   0.056&   &   0.003&   1.181&   0.052\tabularnewline
   ~~&   PCA quadratic (q = 30)&   &   0.028&   1.177&   0.086&   &   -0.003&   1.251*&   0.085&   &   0.032&   1.163&   0.061&   &   -0.004&   1.232&   0.090\tabularnewline
\midrule
   ~~&  PCA squared (q = 05)&   &   0.042&   1.071&   0.035&   &   0.035&   1.056&   0.021&   &   0.040&   1.030&   0.027&   &   0.031&   1.061&   0.013\tabularnewline
   ~~&   PCA squared (q = 15)&   &   0.031&   1.174&   0.070&   &   0.006&   1.173&   0.049&   &   0.033&   1.145&   0.041&   &   0.007&   1.157&   0.043\tabularnewline
   ~~&   PCA squared (q = 30)&   &   0.037&   1.172&   0.079&   &   -0.002&   1.203&   0.069&   &   0.043&   1.148&   0.062&   &   -0.008&   1.224&   0.080\tabularnewline
\bottomrule
\end{tabular*}
\begin{tablenotes}[flushleft]
      \tiny
      \item \textit{Note:} This table summarizes the normalized forecast errors, which are obtained with probability integral transformations (PIT). 
      Similar to \cite{clark2011} we show the mean, the variance and the AR($1$) coefficient of the normalized forecast errors. Given a well-calibrated model (i.e. the null-hypothesis), normalized forecast errors should have zero mean, a variance of one and experience no autocorrelation. These conditions are tested separately: 1) To test for a zero mean we compute the p-values with a Newey–West variance (with five lags). 2) To test for a unit variance we regress the squared normalized forecast errors on an intercept and allow for a Newey–West variance (with three lags). 3) To test for no autocorrelation we obtain the p-values with an AR(1) model that features an unconditional mean and heteroskedasticity-robust standard errors. Asterisks indicate statistical significance for each model at the 1\% (***), 5\% (**) and 10\% (*) significance levels.
    \end{tablenotes}
\end{center}}
\end{table*}

\end{landscape}

\autoref{tab:PITs1} shows the results for the one-month-ahead normalized forecast errors.\footnote{The results for one-quarter-ahead forecasts are provided in Section \ref{sec:App A} of the Online Appendix.} In principle, we observe that the mean across methods is close to zero (with some few exceptions such as PCA quadratic for $q=15$). Nevertheless, these differences are never statistically significantly different from zero. Considering the variances shows that most models yield forecast distributions which seem to be slightly too narrow (with variances exceeding one). The asterisks indicate whether the variances are significantly different from one. For some few models, this is the case (especially if we assume constancy of the parameters) but if we allow for TVPs there are only a handful of cases left. This, however, strongly depends on the shrinkage prior adopted. Turning to the autocorrelation of the normalized shocks reveals that these are mostly close to zero and never statistically significantly different from zero. 

Comparing sophisticated to simple dimension reduction methods suggests no discernible differences in model calibration. In principle, approaches based on linear PCs yield normalized forecast errors with similar statistical properties than the ones obtained from using more sophisticated dimension reduction techniques.


\begin{figure}[htb!]
\caption{Detailed analysis of one-month-ahead predictive distributions for selected models. \label{fig:AE_fac_errors}}
\begin{minipage}{\textwidth}
\centering
(a) Autoencoder 1l (q = 30), const. (MIN)
\end{minipage}
\begin{minipage}{0.32\textwidth}
\centering
\includegraphics[scale=.4]{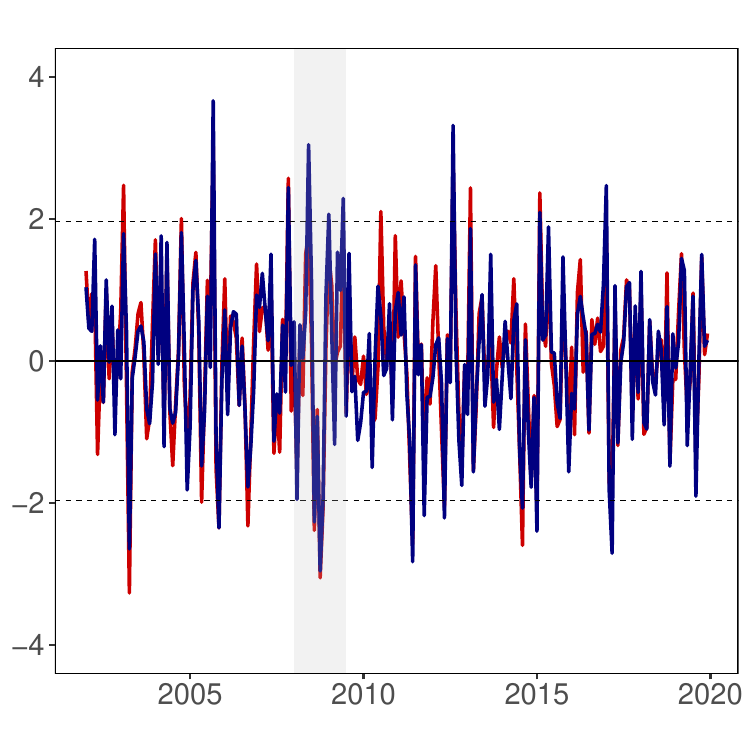}
\end{minipage}
\begin{minipage}{0.32\textwidth}
\centering
\includegraphics[scale=.4]{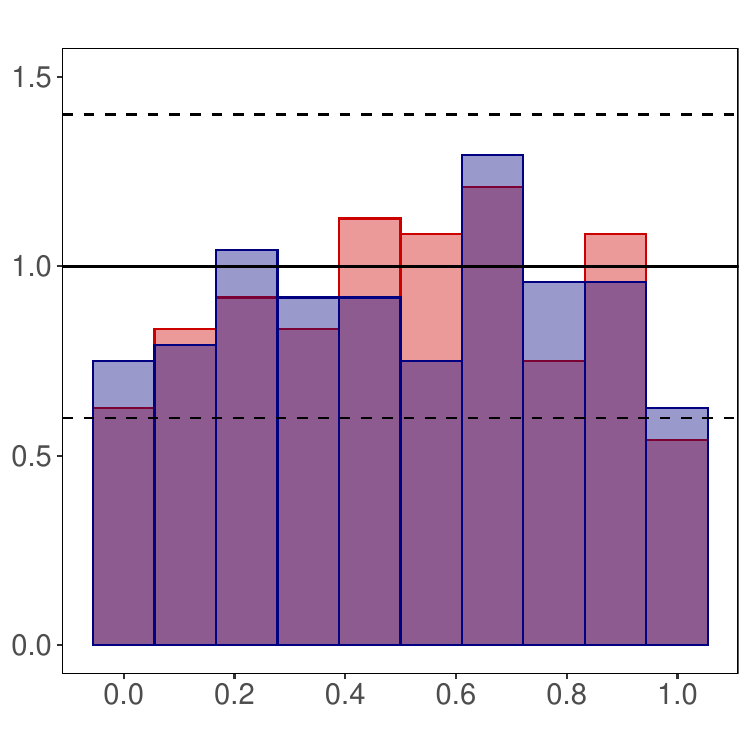}
\end{minipage}
\begin{minipage}{0.32\textwidth}
\centering
\includegraphics[scale=.4]{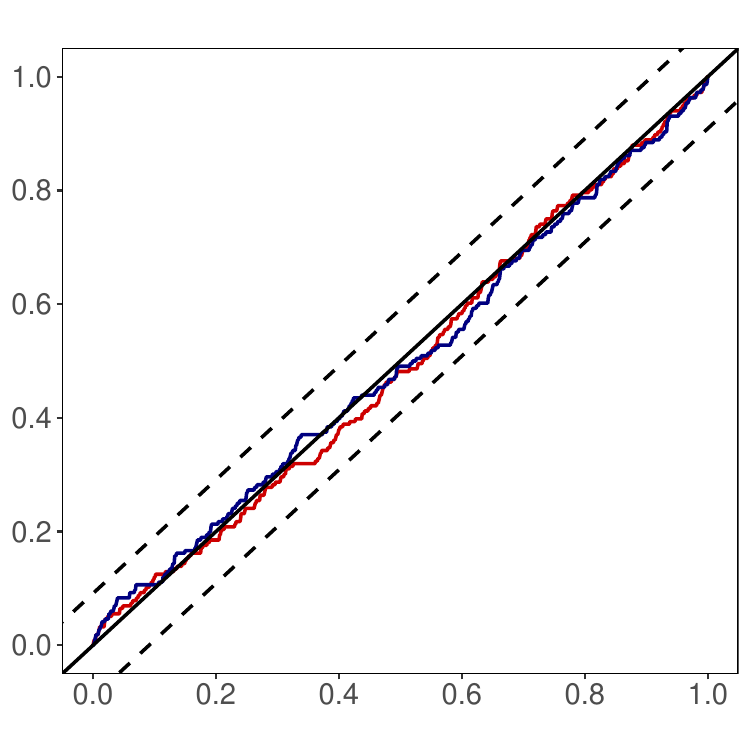}
\end{minipage}
\begin{minipage}{\textwidth}
\centering
(b) PCA squared (q = 05), const. (MIN)
\end{minipage}
\begin{minipage}{0.32\textwidth}
\centering
\includegraphics[scale=.4]{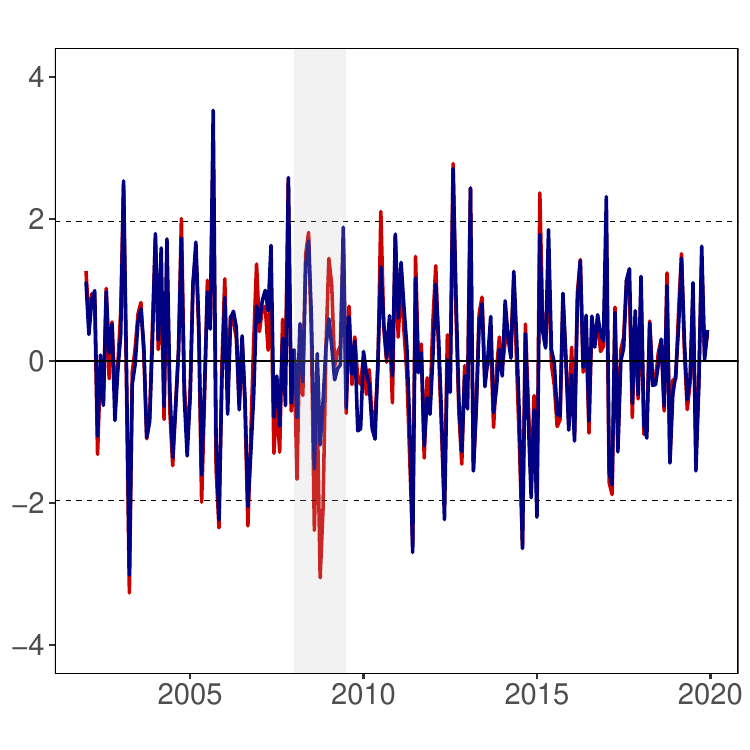}
\end{minipage}
\begin{minipage}{0.32\textwidth}
\centering
\includegraphics[scale=.4]{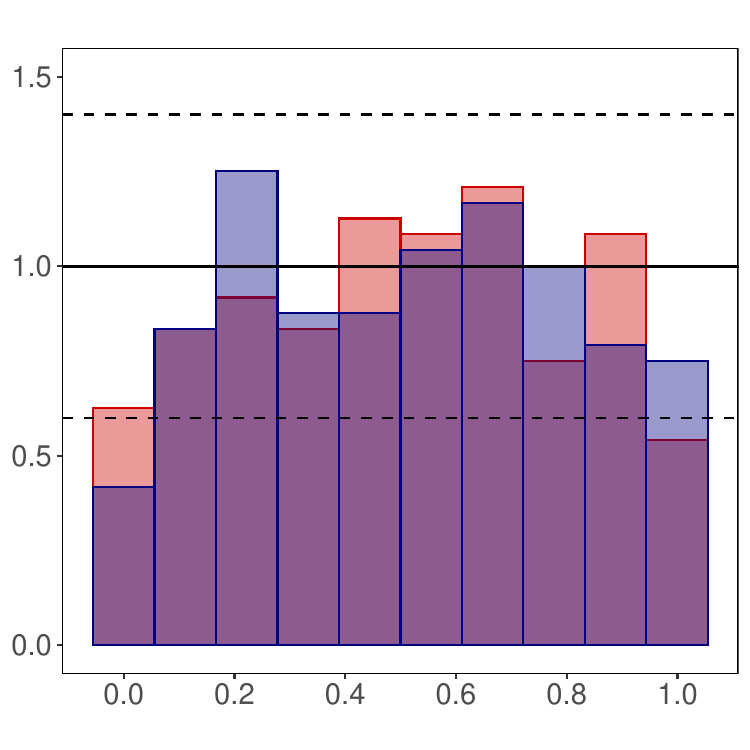}
\end{minipage}
\begin{minipage}{0.32\textwidth}
\centering
\includegraphics[scale=.4]{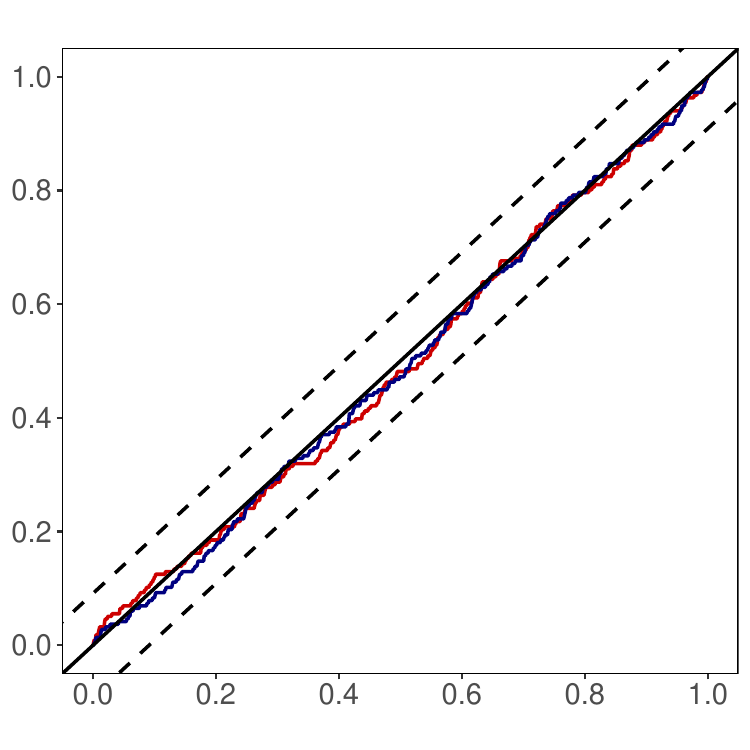}
\end{minipage}
\begin{minipage}{\textwidth}
\scriptsize \textit{Note: This figure shows the evolution of normalized forecast errors for the one-month-ahead horizon in the left panel, the histogram of the PITs in the middle panel and the empirical cumulative density function of the PITs in the right panel. If a model is correctly specified the PITs are standard uniformly distributed and the normalized forecast errors standard normally distributed. This theoretically correct specification is indicated by the black lines, with the dashed lines referring to the respective $95\%$ confidence interval. In red we present the results of the benchmark, whereas in blue we indicate the respective model. The light gray shaded areas refer to the global financial crisis.} 
\end{minipage}
\end{figure}

The discussion above might mask important differences in calibration of different parts of the predictive distribution. We now turn to a deeper analysis of the one-month-ahead predictive distribution of the two best performing models vis-\'{a}-vis the benchmark: the Autoencoder 1l $(q=30)$ and PCA squared $(q=5)$. This analysis is based on visual inspection of the normalized forecast errors (left panel of \autoref{fig:AE_fac_errors}), a histogram of the PITs (middle panel of \autoref{fig:AE_fac_errors}) and the visual diagnostic of the empirical cumulative density function proposed in \cite{rossi2019alternative} (right panel of \autoref{fig:AE_fac_errors}). Recall that, under correct specification, the PITs should be iid uniformly  and the normalized forecast errors should be iid standard normally distributed, respectively.

The left panel of the figure indicates that for both models under consideration, normalized forecast errors are centered on zero, display little serial correlation and a variance close to one (with the Autoencoder generating slightly more spread out normalized forecast errors). In some periods, normalized forecast errors depart significantly from the standard normal distribution (i.e., the corresponding observations lie outside the 95\% confidence intervals). But in general, and for both models (and the benchmark), model calibration seems to be adequate.  Next, we focus on the histogram in the middle panel of \autoref{fig:AE_fac_errors} (which includes 95\% confidence intervals). From this figure, we learn that both models are well calibrated with some tendency to overestimate the upper tail risk. Finally, considering the right panel shows that all models appear to be well calibrated, with most observations being clustered around the 45 degree lines and not a single observation being outside the 95\% confidence intervals.

\subsection{A note on the pandemic}\label{sec:pandemic}
To the detriment of linear modeling techniques, the COVID-19 pandemic caused severe outliers for several of the time series we include in our dataset. Following the recent literature \citep[e.g.,][]{huber2020nowcasting, chkmp2021, coulombe2021covid19} which advocates using non-linear and non-parametric modeling techniques in turbulent times, we briefly investigate whether the non-linear dimension reduction techniques proposed in this paper yield more precise inflation forecasts during the pandemic.

\begin{figure}[!htpb]
\caption{Evolution of one-month-ahead cumulative LPLs against the benchmark for the COVID-19 pandemic.
  \label{fig:lps1_pand}}

\begin{minipage}{\textwidth}
\centering
\vspace{10pt}
\textit{q = 30}
\vspace{5pt}
\end{minipage}

\begin{minipage}{0.24\textwidth}
\centering
\textit{const. (HS)}
\vspace{2pt}
\end{minipage}
\begin{minipage}{0.24\textwidth}
\centering
\textit{const. (MIN)}
\vspace{2pt}
\end{minipage}
\begin{minipage}{0.24\textwidth}
\centering
\textit{TVP (HS)}
\vspace{2pt}
\end{minipage}
\begin{minipage}{0.24\textwidth}
\centering
\textit{TVP (MIN)}
\vspace{2pt}
\end{minipage}

\begin{minipage}{0.24\textwidth}
\centering
\includegraphics[scale=.3]{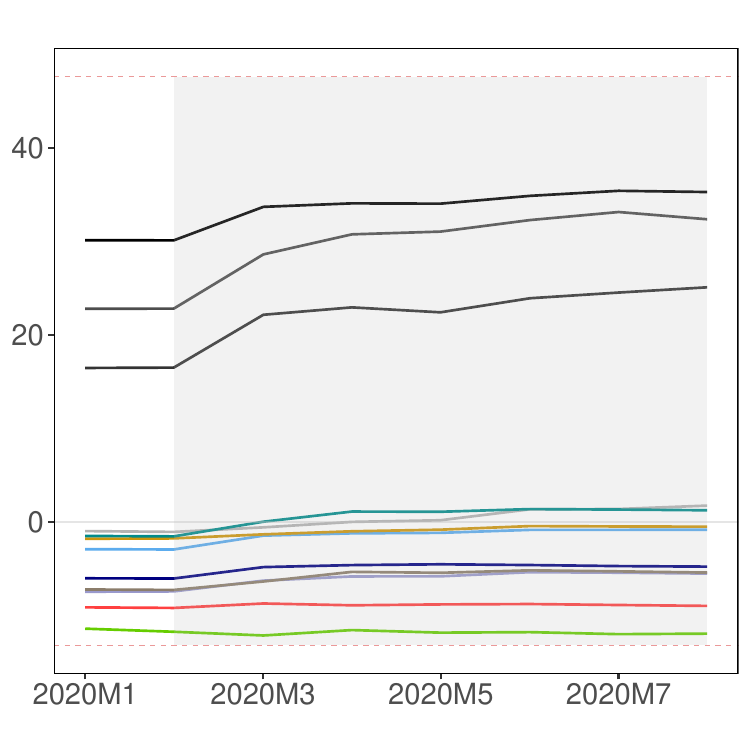}
\end{minipage}
\begin{minipage}{0.24\textwidth}
\centering
\includegraphics[scale=.3]{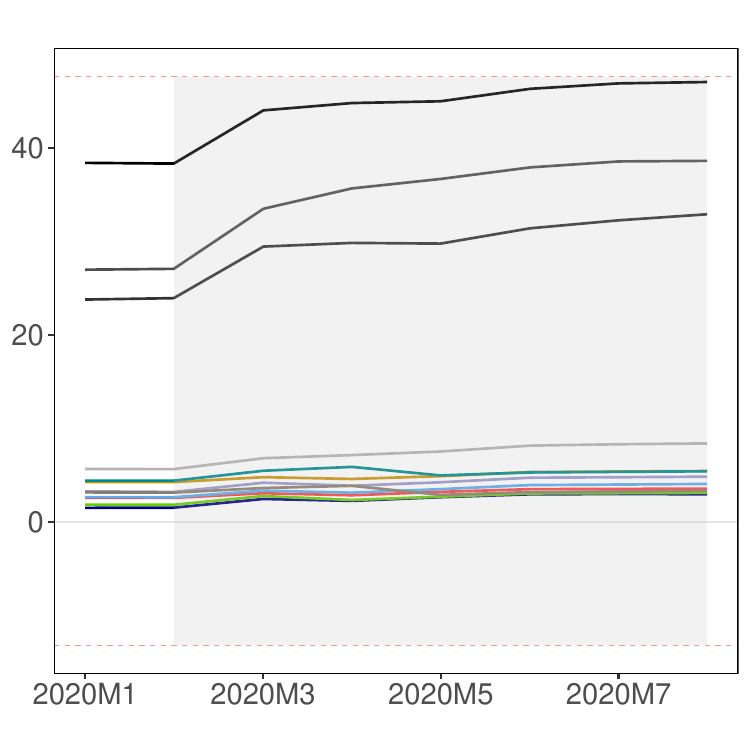}
\end{minipage}
\begin{minipage}{0.24\textwidth}
\centering
\includegraphics[scale=.3]{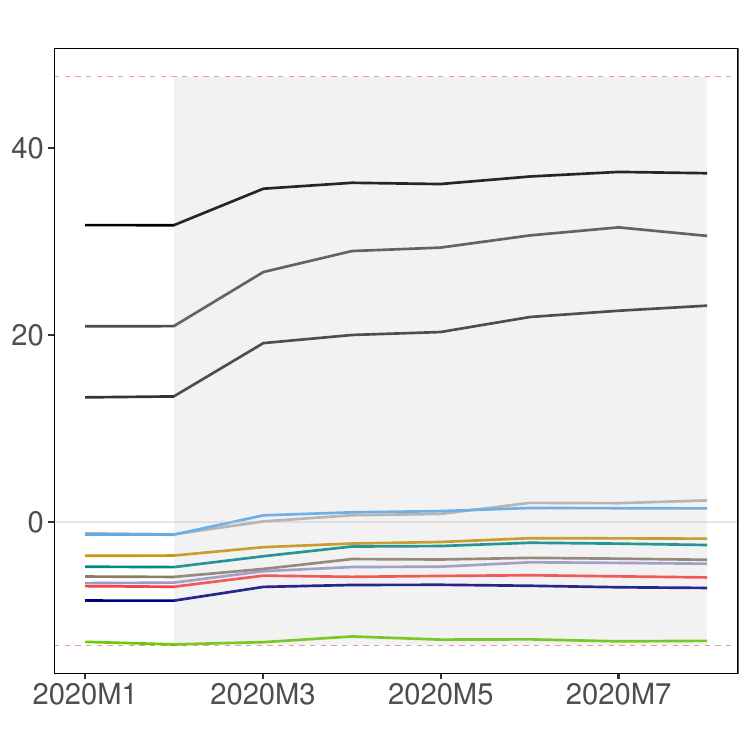}
\end{minipage}
\begin{minipage}{0.24\textwidth}
\centering
\includegraphics[scale=.3]{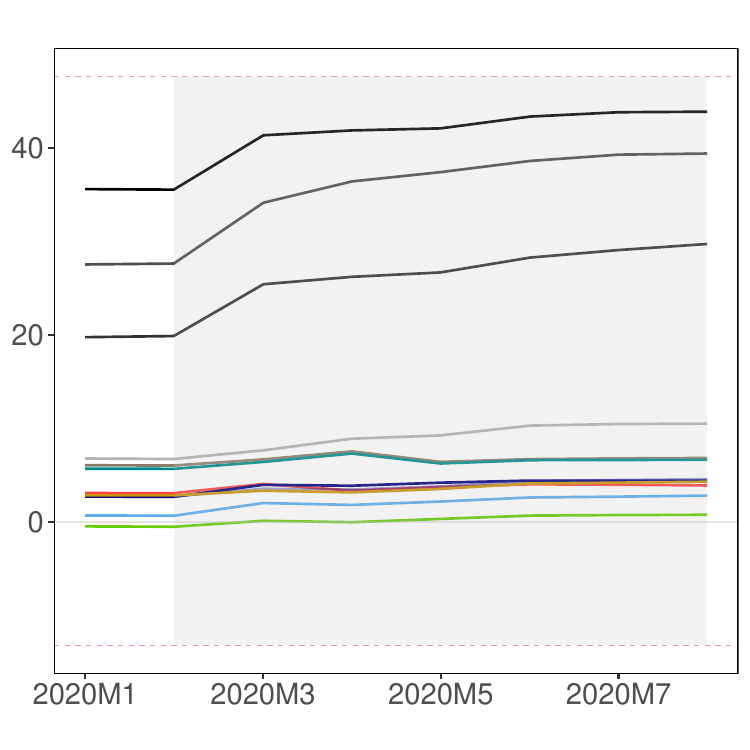}
\end{minipage}

\begin{minipage}{\textwidth}
\centering
\includegraphics[scale=.4]{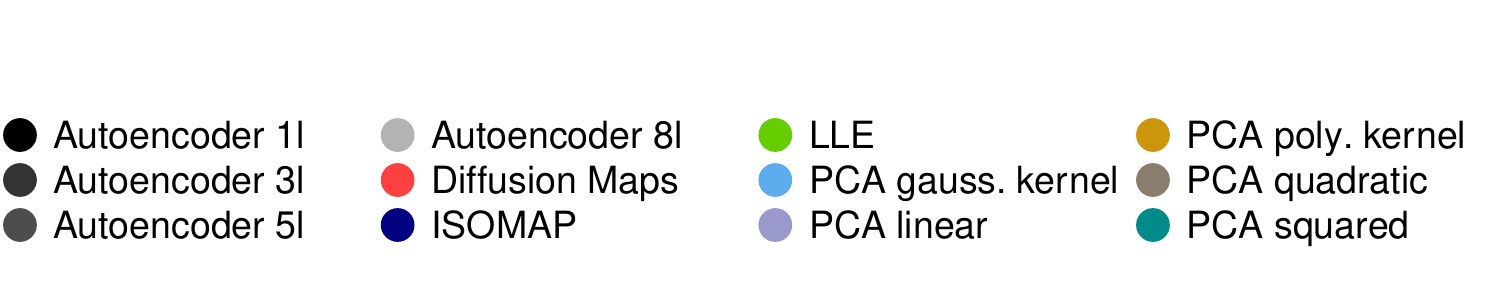}
\end{minipage}\vfill

\begin{minipage}{\textwidth}
\scriptsize \textit{Note:} The initial value in 2020:01 also takes into account the density forecast performance in the previous hold-out periods (ranging from 2002:01 to 2019:12).   The red dashed lines refer to the maximum/minimum Bayes factor over the full hold-out sample. The light gray shaded areas indicate the periods of the COVID-19 pandemic.
\end{minipage}
\end{figure}

\autoref{fig:lps1_pand} depicts the differences in LPLs for the period 2020:01 to 2020:08. For illustrative purposes, we only consider the models with 30 factors.\footnote{The findings for the other factors are very similar and available from the corresponding author upon request.} 

The figure provides a few interesting insights. First, we observe that in March 2020, models based on the Autoencoder improve upon the benchmark, irrespective of the prior and regression specification adopted. This finding is less pronounced for the other techniques in the constant parameter case. Comparing the performance of the constant parameter and the TVP regression models reveals that, irrespective of the prior, allowing for time variation in the parameters improves density forecasts during the pandemic. This finding is consistent with findings in, e.g., \cite{huber2020nowcasting}, who show that flexible models improve upon linear models during the pandemic due to increases in the predictive variance. 


\newpage

\subsection{Dynamic model learning based on density forecast performance}\label{sec:combi}
In the previous sub-section and Section \ref{sec:App A} of the Online Appendix we provide some evidence that model performance varies considerably over time (see \autoref{fig:lps1}). The key implication is that non-linear compression techniques (and time-varying parameters) might be useful during turbulent times whereas forecast evidence is less pronounced in normal times. In this sub-section, we ask whether combining models in a dynamic manner further improves predictive accuracy.

After having obtained the predictive densities of $y_{t+h}$ for the different dimension reduction techniques and model specifications, the goal is to exploit the advantages of both linear and non-linear approaches. This is achieved by combining models in a  model pool such that  better performing models over certain periods receive larger weights while inferior models are subsequently down-weighted. The literature on forecast combinations suggests several different weighting schemes, ranging from simply averaging over all models \citep[see, e.g.,][]{hendry2004pooling, Hall2007combi, clark2010averaging, berg2015combination} to estimating weights based on the models' performances according to the minimization of an objective or loss function \citep[see, e.g.,][]{timmermann2006combi, Hall2007combi, Geweke2011pools, conflitti2015combination, pettenuzzo2016combi} or according to the posterior probabilities of the predictive densities \citep[see, e.g.,][]{RafteryDMA, kk2012, beckmann2020exchange}. More recent approaches set up separate state space models which assume sophisticated law of motions for the weights associated with each predictive distribution \citep{billio2013combi, pettenuzzo2016combi, mcalinn2019bma}. These approaches, while being elegant and having the advantage of incorporating all available sources of uncertainty (i.e., also control for estimation uncertainty in the weights), are computationally cumbersome if the number of models to be combined is large.

Since our model space is huge, we use computationally efficient approximations to dynamically combine models. Our approach builds on combining predictive densities
according to their posterior probabilities. This is referred to as Bayesian model averaging (BMA). The resulting weights are capable of reflecting the predictive power of each model for the respective periods. Dynamic model averaging (DMA), as specified by \cite{RafteryDMA}, extends the approach by adding a discount (or \textit{forgetting}) factor to control for a model's forecasting performance in the recent past. The `recent past' is determined by the discount factor, with higher values attaching greater importance to past forecasting performances of the model and lower values gradually ignoring results of past predictive densities. Similar to \cite{RafteryDMA}, \cite{kk2012} and \cite{beckmann2020exchange}, we apply DMA to combine the predictive densities of our various models. These methods do not require computationally intensive MCMC or sequential Monte Carlo techniques and are thus fast and easy to implement.

DMA works as follows. Let $\bm \varrho_{t+h|t} = (\varrho_{t+h|t,1}, \dots, \varrho_{t+h|t,J})'$ denote a set of weights for $J$ competing models at time $t+h$ given all available information up to time $t$. These (horizon-specific) weights vary over time and depend on the recent predictive performance of the model according to:
\begin{align*}
\varrho_{t+h|t, j} &= \frac{\varrho_{t|t, j}^\delta}{\sum_{l=1}^J \varrho_{t|t, l}^\delta},\\
\varrho_{t+h|t+h, j} &= \frac{\varrho_{t+h|t, j} ~ p_j(y_{t+h}|y_{1:t})}{\sum_{l=1}^J \varrho_{t+h|t, l} ~ p_l(y_{t+h}|y_{1:t})}
\end{align*}
where $p_j(y_{t+h}|y_{1:t})$ denotes the $h$-month-ahead predictive distribution of model $j$ evaluated at $y_{t+h}$  and $\delta \in (0, 1]$ denotes a forgetting factor close to one. Intuitively speaking, the first equation is a prediction of the weights based on all available information up to time $t$ while the second equation shows how the weights get updated if new data flows in.

In our empirical work we set $\delta = 0.97$.\footnote{\cite{koop2013var} find robust results over the interval [0.95,1]. After optimizing over this set of parameter values we choose $\delta = 0.97$.}  Notice that if $\delta=1$, we obtain standard BMA weights while $\delta=0$ would imply that the weights depend exclusively on the forecasting performance in the last period.

\subsection{Forecasting performance of predictive combinations from dynamic model learning}\label{sec:combi_results}
Weights obtained by combining models according to their predictive power convey useful information about the adequacy of each model over time. In order to get a comprehensive picture of the effects of different model modifications, we combine our models and model specifications in various ways. 

\autoref{tab:lplcombi} presents the forecasting results when we use DMA to combine models. Again, all models are benchmarked to the AR model with constant parameters and the Minnesota prior. The first row depicts the relative performance of the single best performing model for the chosen time horizon. 

The table can be understood as follows. Each entry includes \textit{all} dimension reduction techniques. The rows define whether the model space includes all factors $q \in \{5, 15, 30\}$ or whether we combine models with a fixed number of factors exclusively. The columns refer to model spaces which include only constant parameter, time-varying parameter or both specifications in the respective model pool. Since we also discriminate between two competing priors we consider model weights conditioning on either the horseshoe or the Minnesota prior or average across both prior specifications (the first upper part of the table with $\{\text{HS, MIN}\}$).

Across the two forecast horizons considered, we find pronounced accuracy improvements for point and density forecasts relative to the AR model. When we benchmark the different combination strategies to the single best performing model we find no accuracy gains for both horizons. Differences in terms of LPLs 
are, however, rather small. This suggests that while the best performing model (i.e., a constant parameter regression with factors obtained through the Autoencoder) is hard to beat, one can effectively reduce model and specification uncertainty and thus obtain competitive forecasts without the need to rely on a single model. 

\begin{table*}[ht]
\caption{Forecast performance of predictive combinations.}\label{tab:lplcombi}
{\scriptsize
\begin{center}
\begin{tabular*}{\linewidth}{c @{\extracolsep{\fill}} llclllclll}
\toprule
\multicolumn{1}{l}{\bfseries }&\multicolumn{2}{c}{\bfseries Specification}&\multicolumn{1}{c}{\bfseries }&\multicolumn{3}{c}{\bfseries One-month-ahead}&\multicolumn{1}{c}{\bfseries }&\multicolumn{3}{c}{\bfseries One-quarter-ahead}\tabularnewline
\cline{2-3} \cline{5-7} \cline{9-11}
\multicolumn{1}{l}{}&\multicolumn{1}{c}{Prior}&\multicolumn{1}{c}{Combination}&\multicolumn{1}{c}{}&\multicolumn{1}{c}{const.}&\multicolumn{1}{c}{TVP}&\multicolumn{1}{c}{\{const., TVP\}}&\multicolumn{1}{c}{}&\multicolumn{1}{c}{const.}&\multicolumn{1}{c}{TVP}&\multicolumn{1}{c}{\{const., TVP\}}\tabularnewline
\midrule
{\scshape }&&&&&&&&&&\tabularnewline
  ~~& \multicolumn{2}{c}{Single best performing model}    &   &  \shadeRow  38.54&      &      &   & &  \shadeRow 48.50&            \tabularnewline
   ~~&   &   &   &  \shadeRow (0.80)&   &   &   & & \shadeRow (0.76)&      \tabularnewline
\midrule
{\scshape }&&&&&&&&&&\tabularnewline
   ~~&   \{HS, MIN\}&   q = \{05, 15, 30\}&   &   33.40&   31.53&   32.73&   &   45.88&   46.65&   46.32\tabularnewline
   ~~&   &   &   &   (0.83)&   (0.82)&   (0.82)&   &   (0.79)&   (0.79)&   (0.79)\tabularnewline
   ~~&   &   q = 05&   &    7.97&    8.68&    8.40&   &   46.49&   47.51&   47.09\tabularnewline
   ~~&   &   &   &   (0.93)&   (0.92)&   (0.92)&   &   (0.78)&   (0.78)&   (0.78)\tabularnewline
   ~~&   &   q = 15&   &    9.84&   14.81&   13.28&   &   40.45&   40.96&   40.83\tabularnewline
   ~~&   &   &   &   (0.94)&   (0.90)&   (0.91)&   &   (0.84)&   (0.83)&   (0.83)\tabularnewline
   ~~&   &   q = 30&   &   34.80&   33.45&   34.37&   &   33.25&   32.82&   33.05\tabularnewline
   ~~&   &   &   &   (0.83)&   (0.82)&   (0.82)&   &   (0.86)&   (0.87)&   (0.86)\tabularnewline
\midrule
{\scshape }&&&&&&&&&&\tabularnewline
   ~~&   HS&   q = \{05, 15, 30\}&   &   26.26&   29.17&   28.07&   &   47.29&   48.02&   47.75\tabularnewline
   ~~&   &   &   &   (0.85)&   (0.84)&   (0.84)&   &   (0.77)&   (0.77)&   (0.77)\tabularnewline
   ~~&   &   q = 05&   &    6.68&    6.74&    6.74&   &   \textbf{47.59}&   \textbf{48.68}&   \textbf{48.30}\tabularnewline
   ~~&   &   &   &   (0.94)&   (0.94)&   (0.94)&   &   (\textbf{0.76})&   (\textbf{0.76})&   (\textbf{0.76})\tabularnewline
   ~~&   &   q = 15&   &    8.03&    9.02&    8.68&   &   42.57&   42.65&   42.73\tabularnewline
   ~~&   &   &   &   (0.94)&   (0.94)&   (0.94)&   &   (0.83)&   (0.82)&   (0.82)\tabularnewline
   ~~&   &   q = 30&   &   28.05&   31.05&   29.89&   &   34.82&   33.95&   34.40\tabularnewline
   ~~&   &   &   &   (0.85)&   (0.84)&   (0.84)&   &   (0.85)&   (0.86)&   (0.85)\tabularnewline
\midrule
{\scshape }&&&&&&&&&&\tabularnewline
   ~~&   MIN&   q = \{05, 15, 30\}&   &   35.96&   32.23&   34.65&   &   43.37&   43.57&   43.50\tabularnewline
   ~~&   &   &   &   (0.82)&   (0.81)&   (0.81)&   &   (0.82)&   (0.82)&   (0.82)\tabularnewline
   ~~&   &   q = 05&   &    8.62&   10.10&    9.57&   &   44.80&   44.64&   44.74\tabularnewline
   ~~&   &   &   &   (0.92)&   (0.91)&   (0.91)&   &   (0.82)&   (0.82)&   (0.82)\tabularnewline
   ~~&   &   q = 15&   &   10.98&   17.02&   15.40&   &   26.16&   25.92&   26.01\tabularnewline
   ~~&   &   &   &   (0.94)&   (0.89)&   (0.90)&   &   (0.86)&   (0.86)&   (0.86)\tabularnewline
   ~~&   &   q = 30&   &   \textbf{37.27}&   \textbf{34.60}&   \textbf{36.46}&   &   23.63&   26.12&   25.27\tabularnewline
   ~~&   &   &   &   (\textbf{0.82})&   (\textbf{0.80})&   (\textbf{0.81})&   &   (0.86)&   (0.87)&   (0.87)\tabularnewline
\bottomrule
\end{tabular*}
\begin{tablenotes}[flushleft]
      \tiny
      \item \textit{Note:} The first (grey shaded) row states the results of the single best performing model as presented in the previous chapter for each forecast horizon benchmarked to the AR model with constant parameters and the Minnesota prior. All other rows show the relative results for the combinations of the different dimension reduction techniques according to the specification stated in the rows and columns headers. For example, the entry in row \{HS, MIN\}, $q = \{5,15,30\}$ and column const. combines all models estimated with constant parameters, the HS prior, the MIN prior, 5, 15 and 30 factors. Entries denote the differences in LPLs with relative RMSE in parantheses benchmarked against the AR model with constant parameters and the MIN prior.
    \end{tablenotes}
\end{center}}
\end{table*}

Comparing whether restricting the model a priori improves predictions yields mixed insights. For the one-month-ahead horizon we observe that pooling over models which use our variant of the Minnesota prior yields more favorable forecasts as compared to a pooling strategy which uses both priors or the horseshoe only. When we pool over constant and TVP regressions we find small decreases in predictive accuracy relative to a model pool which only includes constant parameter regressions.

Turning to one-quarter-ahead forecasts yields a similar picture. Using a large pool of models generally leads to slightly less precise forecasts. For higher-order forecasts our results suggest that pooling models which use the horseshoe yields higher LPLs. When we compare the different regression specifications we find that integrating out uncertainty with respect to whether parameters should be time-varying yields forecasts which are very similar to the strategy that only pools over constant parameter models. 

In general, the differences in predictive performance across the DMA-based averaging schemes are small. Hence, as a general suggestion we can recommend applying DMA and using the most exhaustive model space available (i.e., including both priors, the different number of factors and TVP and constant parameter regressions).  

\begin{figure}[h!]
\caption{Evolution of the weights determined by DMA for one-month-ahead cumulative LPLs.\label{fig:heatcombi1}}

\begin{minipage}{\textwidth}
\centering
(a) \textit{Parameter change}
\hspace{5pt}
\end{minipage}

\begin{minipage}{1.2\textwidth}
\centering \hspace*{-105pt}
\includegraphics[scale=.371]{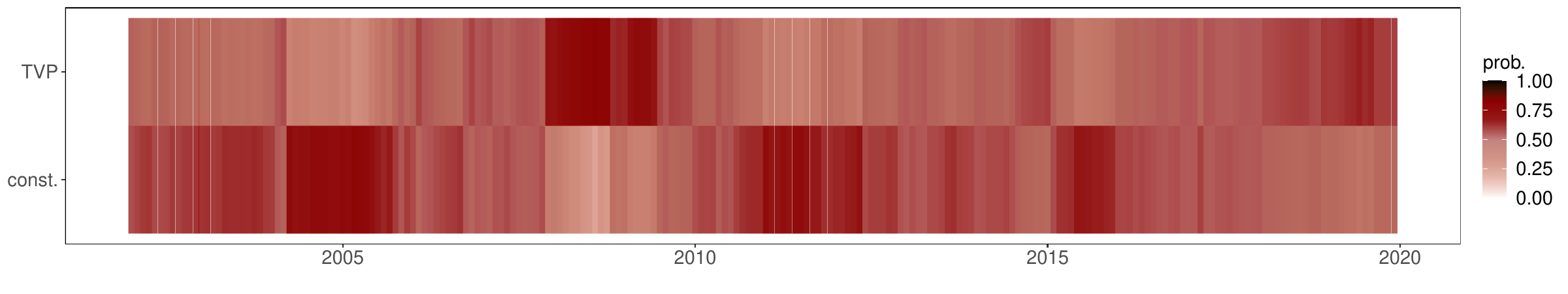}
\end{minipage}

\begin{minipage}{\textwidth}
\centering
(b) \textit{Number of factors}
\hspace{5pt}
\end{minipage}

\begin{minipage}{\textwidth}
\centering 
\includegraphics[scale=.362]{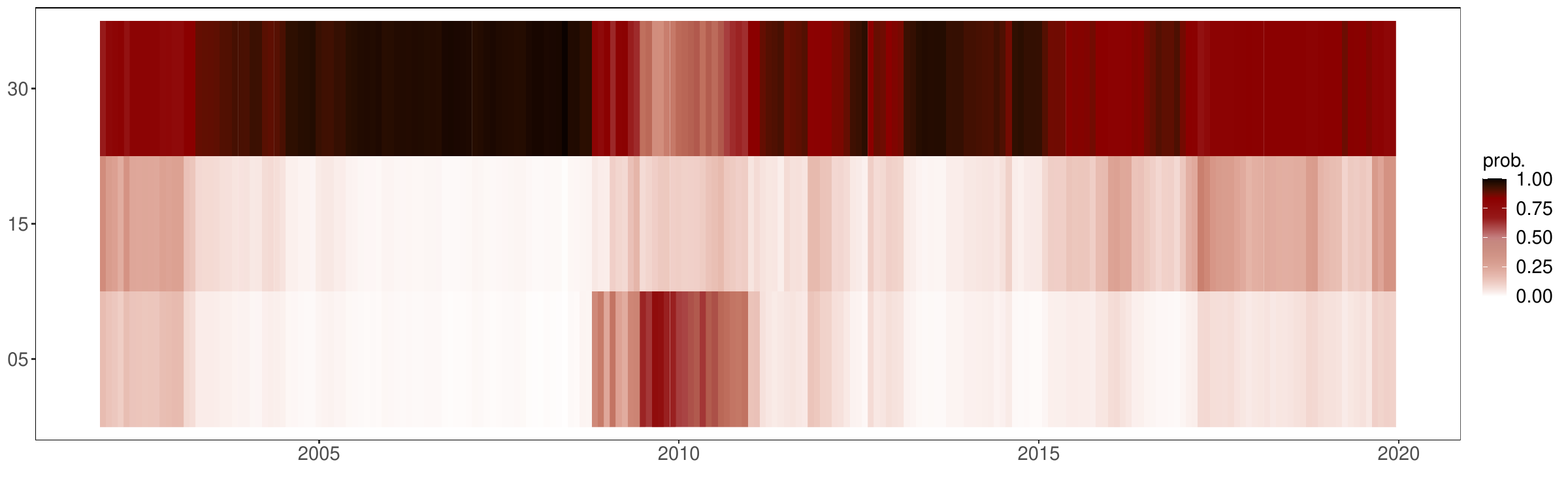}
\end{minipage}

\begin{minipage}{\textwidth}
\centering
(c) \textit{Model selection}
\hspace{5pt}
\end{minipage}

\begin{minipage}{1.2\textwidth}
\centering
\hspace{-132pt}
\includegraphics[scale=.4]{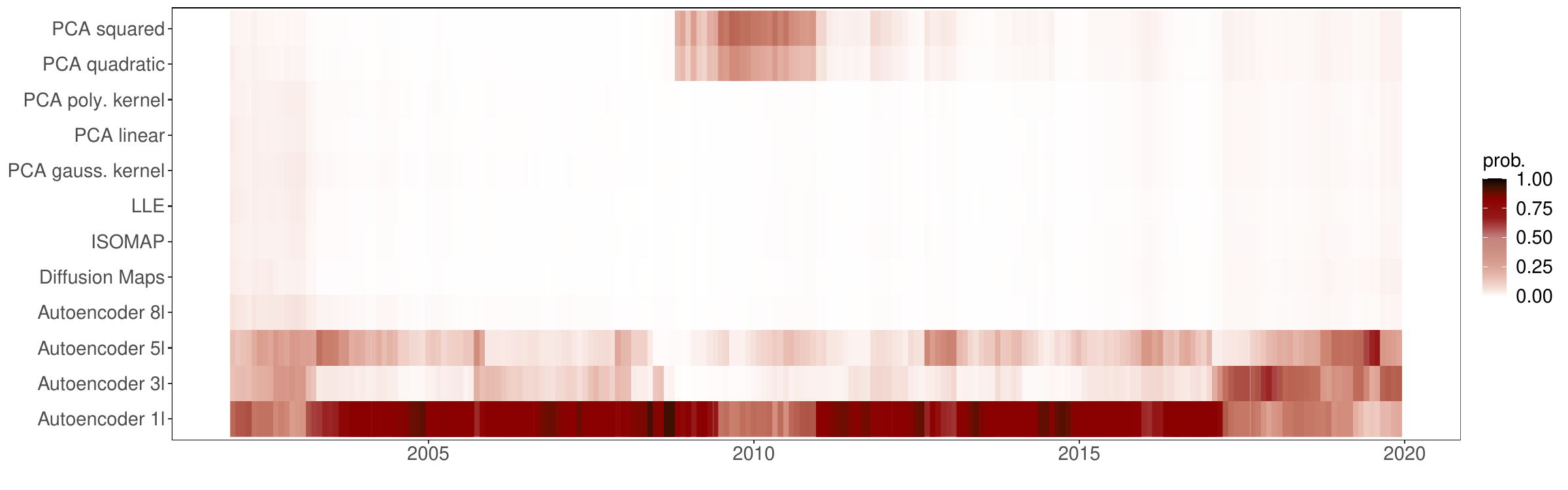}
\end{minipage}

\end{figure}

To investigate which model receives substantial posterior weight over time, \autoref{fig:heatcombi1} depicts the weights associated with the one-month-ahead LPLs over the hold-out period. Panel (a) displays the weight placed on models that allow for TVP, panel (b) shows the weight attached to the different number of factors and panel (c) shows the weight attached to each model. These weights are obtained by using the full model space (i.e., that includes both priors, TVP and constant parameter regressions and all number of factors). The weight placed on TVP specifications, for instance, is then simply obtained by summing up the weights associated with the different models that feature TVPs.


Starting with the top panel of the figure, we observe that during the beginning of the sample, appreciable model weight is placed on constant parameter models. In the mid of 2006, this changes and DMA places increasing posterior mass on models that allow for time-variation in the parameters. In the period from the beginning of 2007 to the onset of the financial crisis, we see that the weight on TVP models somewhat decreases. During the financial crisis, we again experience a pronounced increase in posterior weight towards TVP regression. In that period, constant parameter models only play a limited role in forming inflation forecasts. With some few exceptions, the remainder of the hold-out period is characterized by evenly distributed posterior mass across constant and TVP regressions.

The middle panel of \autoref{fig:heatcombi1} shows that DMA places increasing posterior mass on models with a large number of factors during the period prior to the global financial crisis. During the recession and the immediate period afterwards we observe that models with a different number of factors obtain substantial model weight. This suggests that if the number of factors is small, our dimension reduction techniques soak up information which might be useful during a recession (such as sharp changes in $\bm X$). If the number of factors becomes large this information is extracted as well but (potentially) reflected by more factors that display changes which are less pronounced. In the period after the global financial crisis we again find a large number of factors retrieving substantial posterior weight.

The bottom panel (panel (c)) of \autoref{fig:heatcombi1} provides information on how much weight is allocated to models that exploit non-linear dimension reduction techniques. This figure corroborates our full sample findings: the Autoencoder performs extremely well and dominates our pool of models. Notice, however, that this statement is not true during the global financial crisis. During that period we observe that models based on PCA squared and PCA quadratic feature large weights.  We also find that linear techniques (PCA linear) and other non-linear techniques (PCA with a Gaussian kernel, LLE, ISOMAP, diffusion maps) retreive almost no posterior weight over time.

Summing up this discussion we find that the single best performing model (the Autoencoder) is hard to beat when we dynamically combine models. However, this comparison is, to some extent, unfair since the researcher does not have this information at her disposal. Hence, combining models helps to integrate out this uncertainty by producing forecasts which are close to the single best performing model but, at the cost of higher computational costs, without the necessity of knowing the strongest single model specification.


\section{Closing remarks}\label{sec:conclusions}
In macroeconomics, the vast majority of researchers compress information using linear methods such as principal components to efficiently summarize information embodied in huge datasets in forecasting applications. Machine learning techniques describing large datasets with relatively few latent factors have gained relevance in the last years in various areas. In this paper, we have shown that using such approaches potentially improves real-time inflation forecasts for a wide range of competing model specifications. Our findings indicate that point forecasts of simpler models are hard to beat (especially at the one-month-ahead horizon). For density forecasts, however, we find that more sophisticated modeling techniques that rely on non-linear dimension reduction do particularly well. Among all the techniques considered, our results suggest that the Autoencoder, a particular form of a deep neural network, produces the most precise inflation forecasts (both in terms of point and density predictions). The large battery of competing models gives rise to substantial model uncertainty. We address this issue by using dynamic model averaging to dynamically weight different models, dimension reduction methods and priors. Doing so yields forecasts which are almost as accurate as the ones obtained from the single best performing models.

\clearpage
\small{\setstretch{0.85}
\addcontentsline{toc}{section}{References}
\bibliographystyle{cit_econometrica.bst}
\bibliography{bib}}

\newpage \normalsize
\setcounter{page}{1}
\begin{appendices}
\begin{center}
\LARGE\textbf{Online Appendix}\\[1em]
\Large\textbf{Real-time Inflation Forecasting \\ \vspace{-0.3em} Using Non-linear Dimension Reduction Techniques}\\[1em]
\large\MakeUppercase{Niko Hauzenberger}$^{1,2}$, \MakeUppercase{Florian Huber}$^1$, and \MakeUppercase{Karin Klieber}$^1$\\
$^1$\textit{University of Salzburg}\\
\vspace{-0.3em} $^2$\textit{Vienna University of Economics and Business} 
\end{center}

\setcounter{equation}{0}
\setcounter{table}{0}
\setcounter{figure}{0}
\renewcommand\theequation{A.\arabic{equation}}
\renewcommand\thetable{A.\arabic{table}}
\renewcommand\thefigure{A.\arabic{figure}}

\section{Empirical Appendix}\label{sec:App A}

\subsection{Properties of the factors cont.}

\begin{figure}[htb!]
\caption{Illustration of factors obtained from selected linear and non-linear dimension reduction techniques. \label{fig:factors}}

\begin{minipage}{0.49\textwidth}
\centering
a) PCA linear (q = 05)
\end{minipage}
\begin{minipage}{0.49\textwidth}
\centering
b) PCA squared (q = 05)
\end{minipage}

\begin{minipage}{0.49\textwidth}
\centering
\includegraphics[scale=.37]{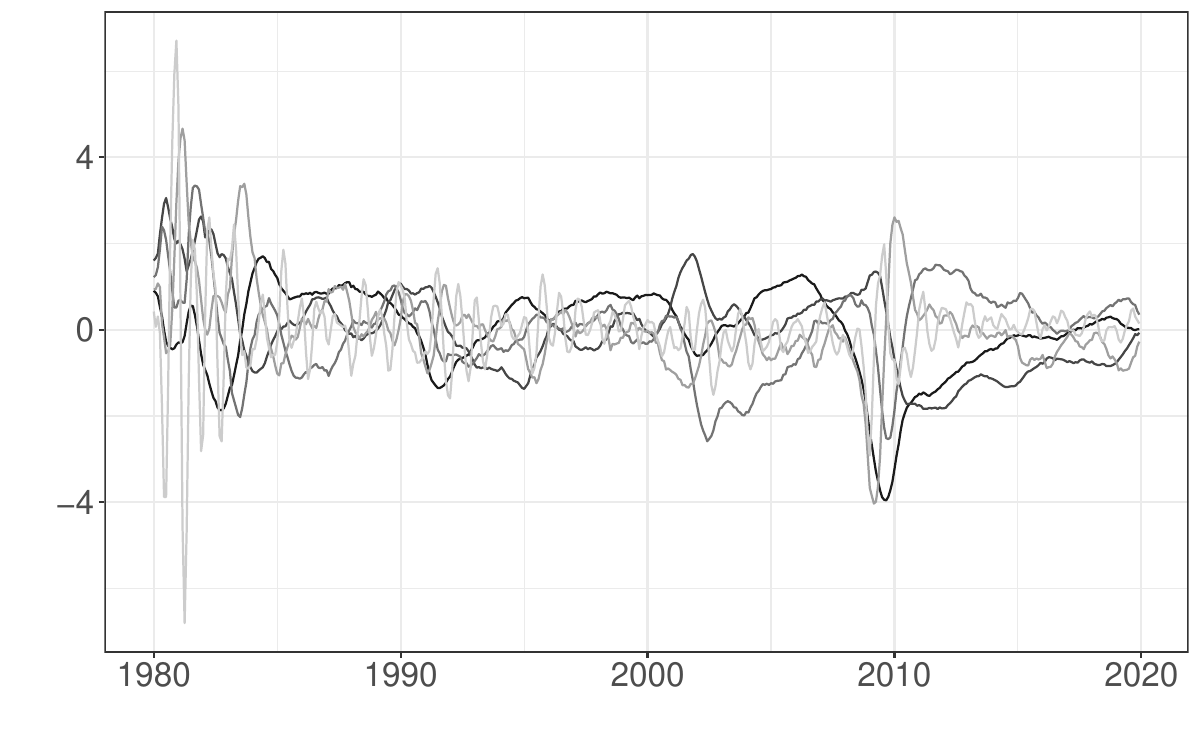}
\end{minipage}
\begin{minipage}{0.49\textwidth}
\centering
\includegraphics[scale=.37]{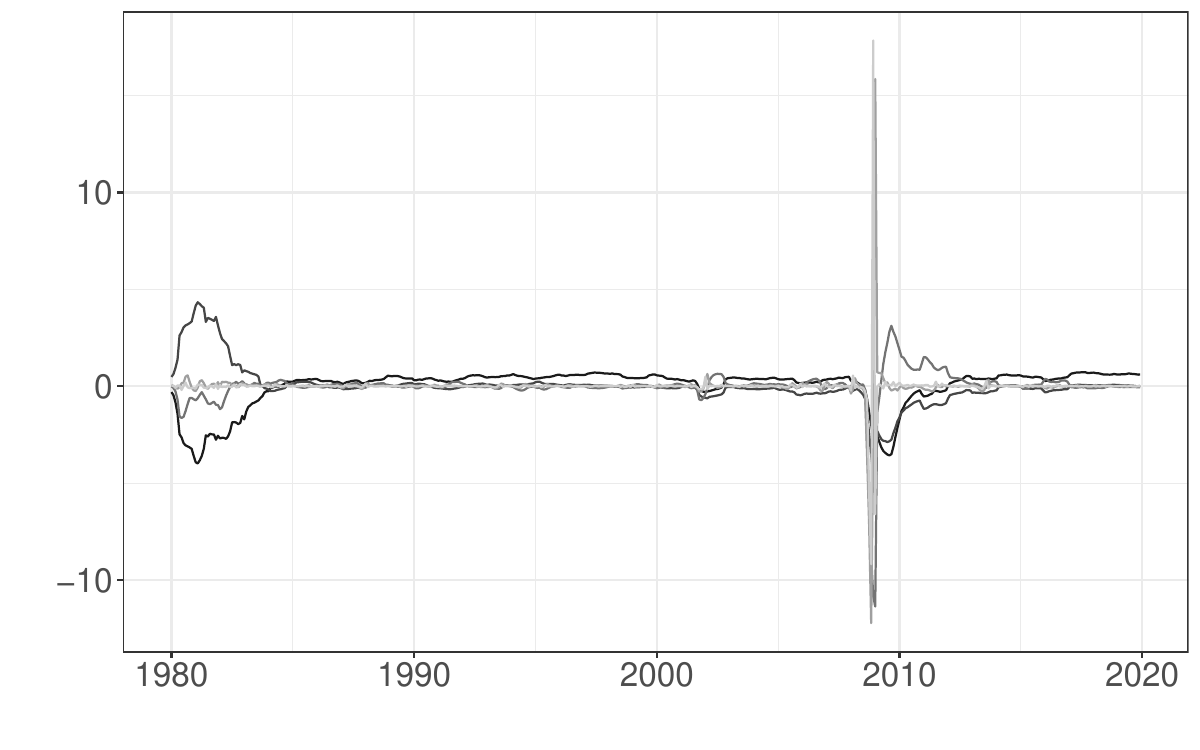}
\end{minipage}

\begin{minipage}{0.49\textwidth}
\centering
c) Autoencoder 1l (q = 05)
\end{minipage}
\begin{minipage}{0.49\textwidth}
\centering
d) Autoencoder 1l (q = 30)
\end{minipage}

\begin{minipage}{0.49\textwidth}
\centering
\includegraphics[scale=.37]{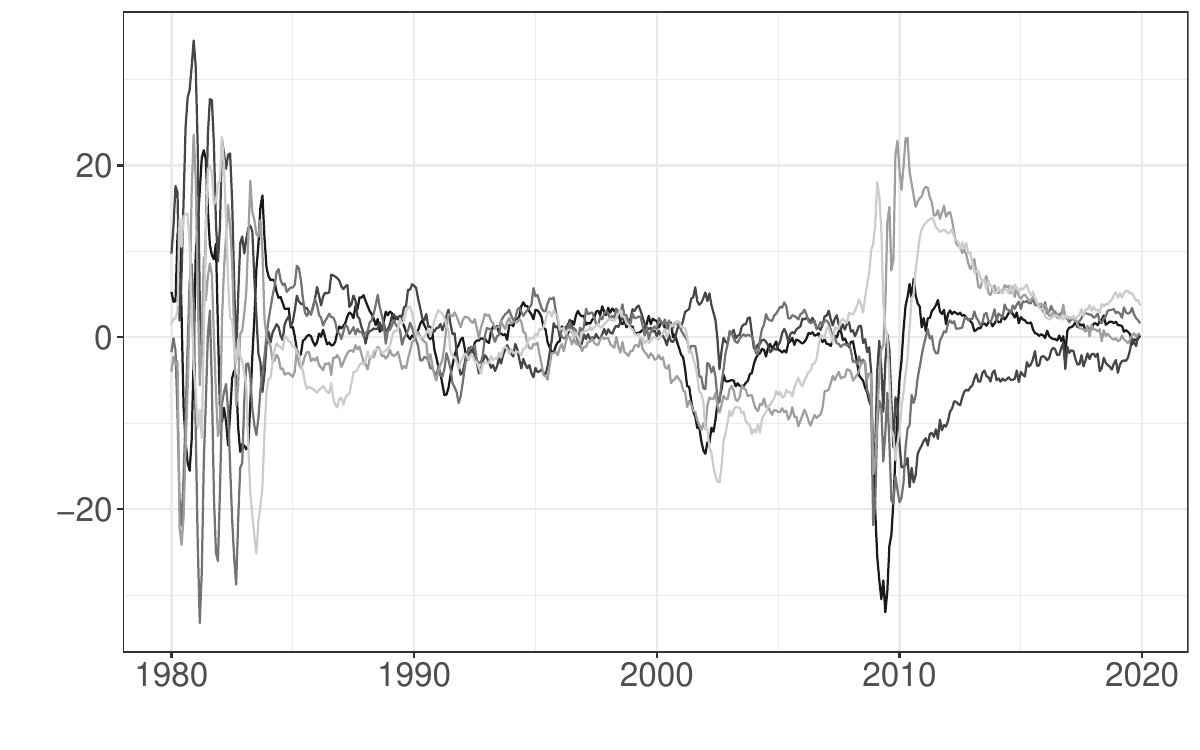}
\end{minipage}
\begin{minipage}{0.49\textwidth}
\centering
\includegraphics[scale=.37]{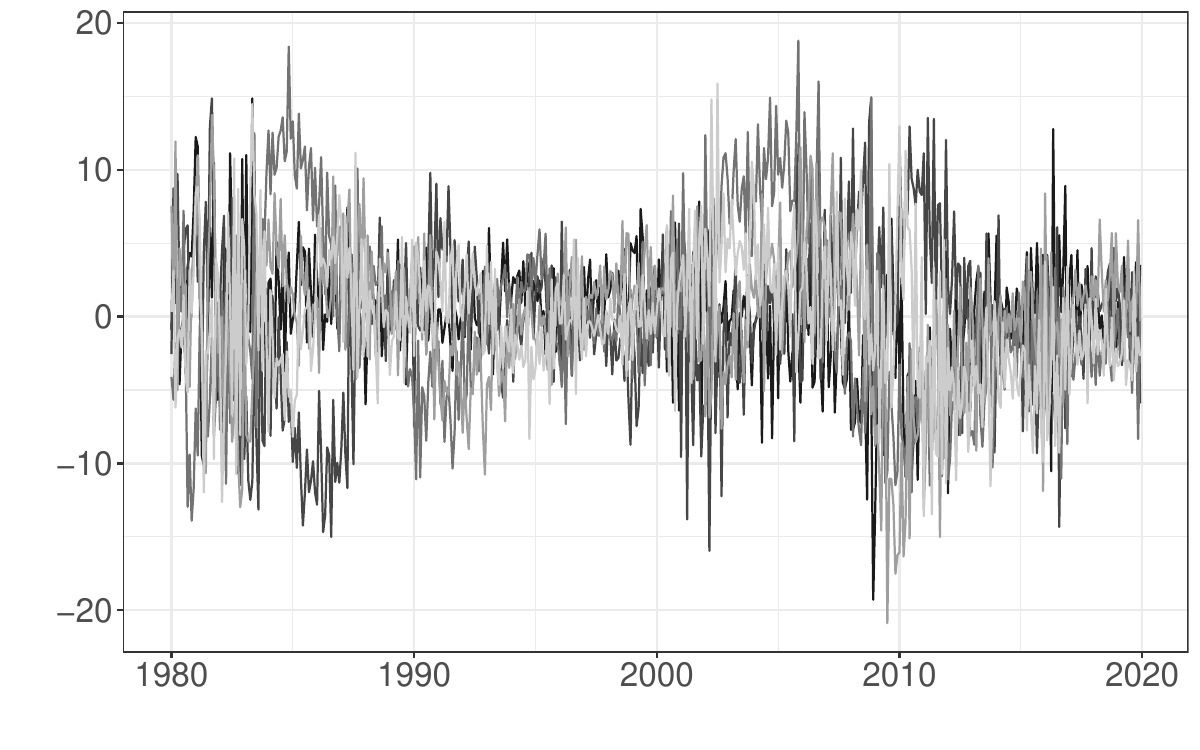}
\end{minipage}

\begin{minipage}{\textwidth}
\vspace{2pt}
\scriptsize \textit{Note:} This figure illustrates the factors obtained from linear and non-linear dimension reduction techniques applied to our US dataset with $K_0 = 104$ based on the last vintage (end of year $2019$). We choose the top-five factors according to their correlation with inflation for selected dimension reduction techniques. The factors are normalized with mean zero and variance one ranging from January $1980$ to December $2019$.
\end{minipage}
\end{figure}

\begin{figure}[!htpb]
\caption{Evolution of the one-month-ahead forecast metrics over the number of factors based on the Autoencoder 1l, const. (MIN), relative to the benchmark.
\label{fig:AUEN}}
\begin{minipage}{0.49\textwidth}
\centering
a) RMSE ratios
\vspace{2pt}
\end{minipage}
\begin{minipage}{0.49\textwidth}
\centering
b) LPLs
\vspace{2pt}
\end{minipage}
\begin{minipage}{0.49\textwidth}
\centering
\includegraphics[scale=.4]{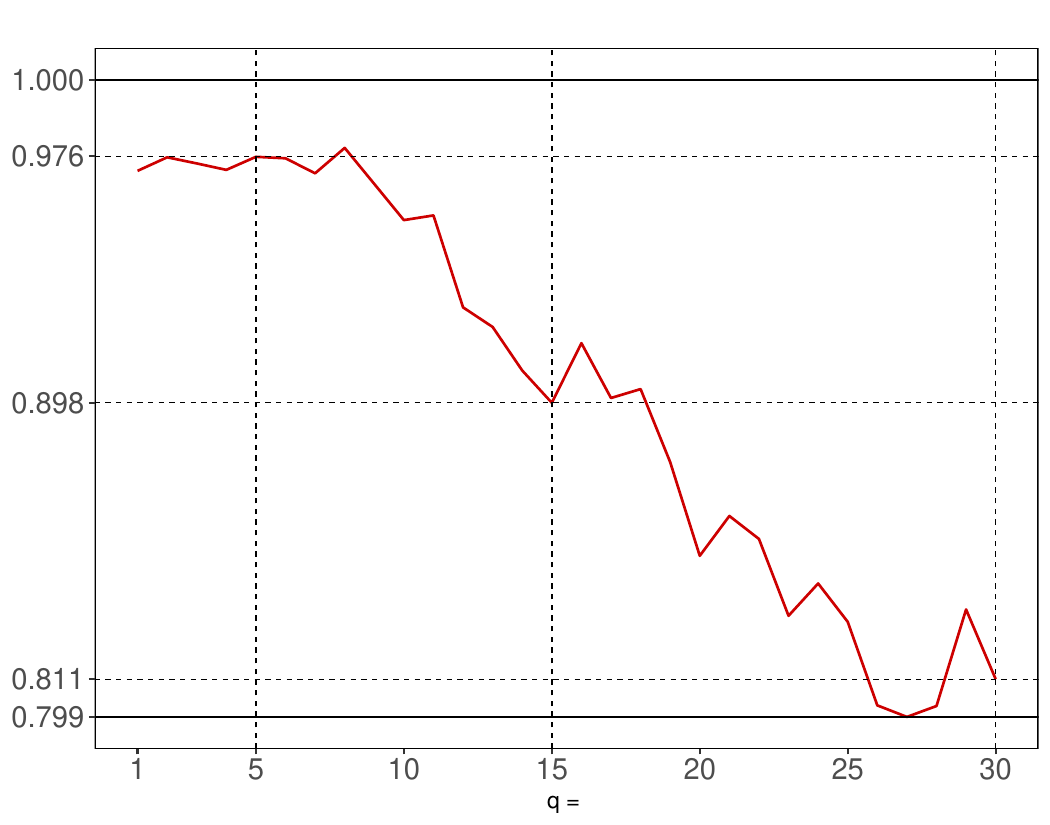}
\end{minipage}
\begin{minipage}{0.49\textwidth}
\centering
\includegraphics[scale=.4]{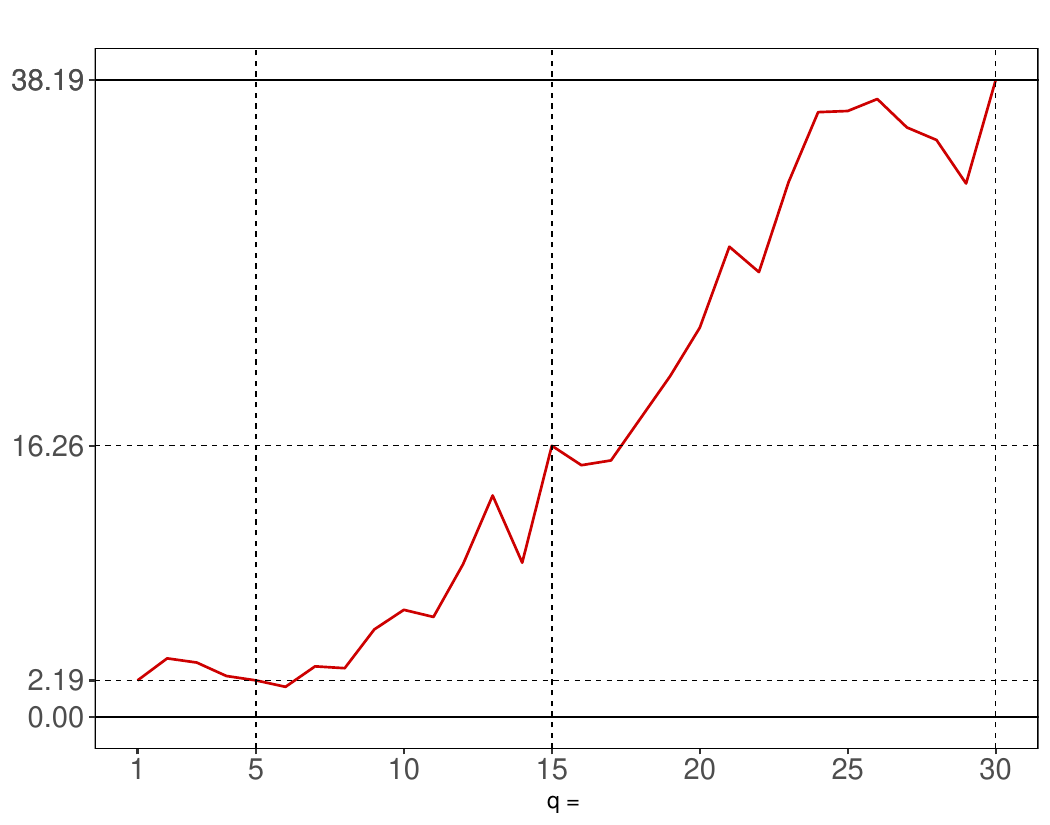}
\end{minipage}
\begin{minipage}{\textwidth}
\scriptsize \textit{Note:} The red solid lines indicate the evolution of the two forecast metrics. RMSE ratios below one (positive LPLs) indicate that the benchmark is outperformed in terms of point (density) forecasts. The black solid lines refer to the lowest (highest) RMSE ratio (LPL), while black dashed lines indicate the number of factors $q = \{5,15,30\}$ and the respective forecast metrics, which correspond to the entries in Table \ref{tab:main1}.
\end{minipage}
\end{figure}

\newpage

\subsection{Results based on extracting the factors from subgroups of the dataset}
\autoref{tab:main1_slwfst} and \autoref{tab:main3_slwfst} show the results of our forecasting exercise when dividing the dataset into fast- and slow-moving variables. The variables classified as real activity, housing, labor market and prices are assumed to respond with a certain delay to economic shocks and are thus considered as slow-moving variables. The quantities within the group of money stocks, reserves and loans, interest rates and stock market are referred to as fast-moving variables as they are assumed to show immediate reaction to economic shocks.

\begin{table*}[!tbp]
{\tiny
\begin{center}
\caption{One-month-ahead forecast performance based on factors from subgroups of the dataset. \label{tab:main1_slwfst}}
\begin{tabular*}{\textwidth}{l @{\extracolsep{\fill}} llcllll}
\toprule
\multicolumn{1}{l}{\bfseries }&\multicolumn{1}{c}{\bfseries Specification}&\multicolumn{1}{c}{\bfseries }&\multicolumn{1}{c}{const. (MIN)}&\multicolumn{1}{c}{const. (HS)}&\multicolumn{1}{c}{TVP (MIN)}&\multicolumn{1}{c}{TVP (HS)}\tabularnewline
\midrule
{\scshape }&&&&&&\tabularnewline
~~&   AR&   & \shadeBench  -329.28&    0.23&   -0.99&   -1.04\tabularnewline
 ~~&   &   &  \shadeBench (1.24)&   (0.98)&   (0.98)&   (0.98)\tabularnewline
   ~~&   Large ARX&   &      2.46&   -9.80***&      &   -8.95***\tabularnewline
   ~~&   &   &   (0.97)&   (1.06***)&   &   (1.04***)\tabularnewline
\midrule
   ~~&   Autoencoder 1l (q = 05)&   &      2.84&   -1.95&    2.47&   -1.60\tabularnewline
   ~~&   &   &   (0.97)&   (0.99)&   (0.98)&   (0.99)\tabularnewline
   ~~&   Autoencoder 1l (q = 15)&   &     13.19**&   17.72***&   16.24**&   19.57***\tabularnewline
   ~~&   &   &   (0.96*)&   (0.93**)&   (0.95*)&   (0.92**)\tabularnewline
   ~~&   Autoencoder 1l (q = 30)&   &   \textbf{30.27***}&   \textbf{41.90***}&   \textbf{36.90***}&   \textbf{38.79***}\tabularnewline
   ~~&   &   &   (\textbf{0.87***})&   (\textbf{0.81***})&   (\textbf{0.87***})&   (\textbf{0.83***})\tabularnewline
   ~~&   Autoencoder 3l (q = 05)&   &      2.42&   -2.17&    0.34&   -2.05\tabularnewline
   ~~&   &   &   (0.97)&   (0.99)&   (0.98)&   (0.99)\tabularnewline
   ~~&   Autoencoder 3l (q = 15)&   &      4.79&   -0.76&    3.97&    0.74\tabularnewline
   ~~&   &   &   (0.97)&   (1.00)&   (0.97)&   (0.99)\tabularnewline
   ~~&   Autoencoder 3l (q = 30)&   &     10.99*&   15.11**&   10.76&   13.35\tabularnewline
   ~~&   &   &   (0.95*)&   (0.94*)&   (0.94*)&   (0.93**)\tabularnewline
   ~~&   Autoencoder 5l (q = 05)&   &      2.86&   -3.12&    0.33&   -4.05\tabularnewline
   ~~&   &   &   (0.98)&   (1.00)&   (0.99)&   (0.99)\tabularnewline
   ~~&   Autoencoder 5l (q = 15)&   &      7.50*&    1.10&    6.53&   -2.33\tabularnewline
   ~~&   &   &   (0.96)&   (0.99)&   (0.97)&   (0.99)\tabularnewline
   ~~&   Autoencoder 5l (q = 30)&   &     12.92**&   19.91***&   15.60**&   20.94***\tabularnewline
   ~~&   &   &   (0.94**)&   (0.92***)&   (0.92**)&   (0.90***)\tabularnewline
   ~~&   Autoencoder 8l (q = 05)&   &      2.42&   -2.38&    2.22&   -2.59\tabularnewline
   ~~&   &   &   (0.98)&   (1.00)&   (0.99)&   (0.99)\tabularnewline
   ~~&   Autoencoder 8l (q = 15)&   &      1.91&   -3.60&    0.38&   -5.75\tabularnewline
   ~~&   &   &   (0.97)&   (1.00)&   (0.98)&   (1.00)\tabularnewline
   ~~&   Autoencoder 8l (q = 30)&   &      6.00&    7.66*&    8.49*&    7.82\tabularnewline
   ~~&   &   &   (0.97)&   (0.96*)&   (0.96)&   (0.95**)\tabularnewline
\midrule
   ~~&   Diffusion Maps (q = 05)&   &      3.17&   -1.52&    5.26&   -1.63\tabularnewline
   ~~&   &   &   (0.97)&   (0.99)&   (0.95)&   (0.99)\tabularnewline
   ~~&   Diffusion Maps (q = 15)&   &      4.76&   -5.84&    5.23&   -6.72\tabularnewline
   ~~&   &   &   (0.97)&   (1.02**)&   (0.97)&   (1.02)\tabularnewline
   ~~&   Diffusion Maps (q = 30)&   &      4.57&   -6.39*&    3.22&   -6.97\tabularnewline
   ~~&   &   &   (0.97)&   (1.02**)&   (0.97)&   (1.02)\tabularnewline
\midrule
{\scshape }&&&&&&\tabularnewline
   ~~&   ISOMAP (q = 05)&   &      1.84&   -2.54&    1.82&   -3.25\tabularnewline
   ~~&   &   &   (0.97)&   (1.00)&   (0.98)&   (0.99)\tabularnewline
   ~~&   ISOMAP (q = 15)&   &      1.00&   -7.00**&    1.50&   -8.73*\tabularnewline
   ~~&   &   &   (0.98)&   (1.02**)&   (0.98)&   (1.01)\tabularnewline
   ~~&   ISOMAP (q = 30)&   &      2.36&   -8.83***&    0.83&   -8.67**\tabularnewline
   ~~&   &   &   (0.97)&   (1.03***)&   (0.98)&   (1.02)\tabularnewline
\midrule
   ~~&   LLE (q = 05)&   &      1.94&   -2.82&    1.13&   -4.55\tabularnewline
   ~~&   &   &   (0.98)&   (0.99)&   (0.98)&   (1.00)\tabularnewline
   ~~&   LLE (q = 15)&   &      2.82&   -7.36**&    0.42&   -7.02\tabularnewline
   ~~&   &   &   (0.98)&   (1.02**)&   (0.98)&   (1.01)\tabularnewline
   ~~&   LLE (q = 30)&   &      1.07&   -6.55**&    0.10&   -6.46*\tabularnewline
   ~~&   &   &   (0.97)&   (1.02*)&   (0.98)&   (1.01)\tabularnewline
\midrule
   ~~&   PCA gauss. kernel (q = 05)&   &      3.07&   -2.30&    3.06&   -3.95\tabularnewline
   ~~&   &   &   (0.98)&   (0.99)&   (0.98)&   (0.99)\tabularnewline
   ~~&   PCA gauss. kernel (q = 15)&   &      3.38&   -2.29&    1.38&   -3.82\tabularnewline
   ~~&   &   &   (0.98)&   (1.00)&   (0.98)&   (1.00)\tabularnewline
   ~~&   PCA gauss. kernel (q = 30)&   &      1.31&   -2.86&    1.53&   -1.88\tabularnewline
   ~~&   &   &   (0.98)&   (0.99)&   (0.98)&   (0.99)\tabularnewline
\midrule
   ~~&   PCA linear (q = 05)&   &      3.34&   -1.75&    1.77&   -1.61\tabularnewline
   ~~&   &   &   (0.97)&   (0.99)&   (0.98)&   (0.99)\tabularnewline
   ~~&   PCA linear (q = 15)&   &      3.91&   -5.24*&    2.72&   -7.30\tabularnewline
   ~~&   &   &   (0.97)&   (1.01*)&   (0.98)&   (1.01)\tabularnewline
   ~~&   PCA linear (q = 30)&   &      2.37&   -6.33**&    1.85&   -6.30\tabularnewline
   ~~&   &   &   (0.97)&   (1.01**)&   (0.98)&   (1.00)\tabularnewline
\midrule
   ~~&   PCA poly. kernel (q = 05)&   &      3.86&   -1.00&    2.87&   -2.35\tabularnewline
   ~~&   &   &   (0.97)&   (0.99)&   (0.98)&   (0.98)\tabularnewline
   ~~&   PCA poly. kernel (q = 15)&   &      2.06&   -1.88&    0.88&   -3.37\tabularnewline
   ~~&   &   &   (0.98)&   (0.99)&   (0.98)&   (1.00)\tabularnewline
   ~~&   PCA poly. kernel (q = 30)&   &      3.30&   -1.22&    2.32&   -4.60\tabularnewline
   ~~&   &   &   (0.98)&   (0.99)&   (0.98)&   (1.00)\tabularnewline
\midrule
   ~~&   PCA quadratic (q = 05)&   &      6.45*&    2.31&    5.60&    1.53\tabularnewline
   ~~&   &   &   (0.96)&   (0.98)&   (0.97)&   (0.98)\tabularnewline
   ~~&   PCA quadratic (q = 15)&   &      5.29&    1.25&    6.70&   -1.02\tabularnewline
   ~~&   &   &   (0.97)&   (0.98)&   (0.95)&   (0.98)\tabularnewline
   ~~&   PCA quadratic (q = 30)&   &      4.24&   -7.79*&    3.30&   -8.45*\tabularnewline
   ~~&   &   &   (0.97)&   (1.02**)&   (0.97)&   (1.03)\tabularnewline
\midrule
   ~~&   PCA squared (q = 05)&   &      6.68*&    1.52&    3.85&    0.85\tabularnewline
   ~~&   &   &   (0.96)&   (0.96)&   (0.98)&   (0.98)\tabularnewline
   ~~&   PCA squared (q = 15)&   &      3.70&   -0.72&    6.04&   -1.14\tabularnewline
   ~~&   &   &   (0.97)&   (0.99)&   (0.95)&   (0.99)\tabularnewline
   ~~&   PCA squared (q = 30)&   &      3.97&   -4.69&    5.31&   -5.98\tabularnewline
   ~~&   &   &   (0.97)&   (1.02**)&   (0.94)&   (1.02)\tabularnewline
\bottomrule
\end{tabular*}
\begin{tablenotes}[flushleft]
\tiny
\item \textit{Note:} The table shows LPLs with RMSEs in parentheses below when we divide our dataset into slow- and fast-moving variables. The first (red shaded) entry gives the actual LPL and RMSE scores of our benchmark (an autoregressive (AR) model with constant parameters and a Minnesota prior). Asterisks indicate statistical significance for each model relative to the benchmark at the 1\% (***), 5\% (**) and 10\% (*) significance levels. Since the large ARX model with time-varying parameters would features $273$ period-specific coefficients and is computationally intractable, we assume that the TVPs feature a factor structure (with three factors) to reduce the dimension of the state space \citep[see Section \ref{sec:App B} of the Online Appendix and][]{chan2020reducing}.
\end{tablenotes}
\end{center}}
\end{table*}
\begin{table*}[!tbp]
{\tiny
\begin{center}
\caption{One-quarter-ahead forecast performance based on factors from subgroups of the dataset. \label{tab:main3_slwfst}}
\begin{tabular*}{\textwidth}{l @{\extracolsep{\fill}} llcllll}
\toprule
\multicolumn{1}{l}{\bfseries }&\multicolumn{1}{c}{\bfseries Specification}&\multicolumn{1}{c}{\bfseries }&\multicolumn{1}{c}{const. (MIN)}&\multicolumn{1}{c}{const. (HS)}&\multicolumn{1}{c}{TVP (MIN)}&\multicolumn{1}{c}{TVP (HS)}\tabularnewline
\midrule
{\scshape }&&&&&&\tabularnewline
 ~~&   AR&   &   \shadeBench  -322.66&     9.79&    20.02&    18.53\tabularnewline
   ~~&   &   &  \shadeBench (1.27)&   (0.94*)&   (0.89*)&   (0.89**)\tabularnewline
   ~~&   Large ARX&   &     19.68&    14.44&       &   11.03\tabularnewline
   ~~&   &   &   (0.89*)&   (0.94)&   &   (0.94)\tabularnewline
\midrule
   ~~&   Autoencoder 1l (q = 05)&   &     20.75&    22.80&    21.42&    22.79\tabularnewline
   ~~&   &   &   (0.90)&   (0.89)&   (0.90)&   (0.90)\tabularnewline
   ~~&   Autoencoder 1l (q = 15)&   &     21.62&    27.13&    19.14&    28.47\tabularnewline
   ~~&   &   &   (0.88**)&   (0.87**)&   (0.88*)&   (0.87**)\tabularnewline
   ~~&   Autoencoder 1l (q = 30)&   &     26.67&    32.02*&    23.59&    34.50**\tabularnewline
   ~~&   &   &   (0.86**)&   (0.84**)&   (0.87**)&   (0.83**)\tabularnewline
   ~~&   Autoencoder 3l (q = 05)&   &     13.35&     7.90&    11.89&     9.35\tabularnewline
   ~~&   &   &   (0.88**)&   (0.89**)&   (0.88**)&   (0.89**)\tabularnewline
   ~~&   Autoencoder 3l (q = 15)&   &     17.99&    16.21&    17.48&    16.48\tabularnewline
   ~~&   &   &   (0.87**)&   (0.87**)&   (0.88**)&   (0.87**)\tabularnewline
   ~~&   Autoencoder 3l (q = 30)&   &     25.22&    22.89&    21.76&    22.21\tabularnewline
   ~~&   &   &   (0.86**)&   (0.85**)&   (0.86**)&   (0.85**)\tabularnewline
   ~~&   Autoencoder 5l (q = 05)&   &     13.35&     9.44&    13.31&     9.64\tabularnewline
   ~~&   &   &   (0.88**)&   (0.90**)&   (0.88**)&   (0.90**)\tabularnewline
   ~~&   Autoencoder 5l (q = 15)&   &     17.35&    16.19&    19.28&    17.79\tabularnewline
   ~~&   &   &   (0.87**)&   (0.87**)&   (0.87**)&   (0.87**)\tabularnewline
   ~~&   Autoencoder 5l (q = 30)&   &     22.07&    22.93&    19.89&    22.27\tabularnewline
   ~~&   &   &   (0.86**)&   (0.85**)&   (0.86**)&   (0.85**)\tabularnewline
   ~~&   Autoencoder 8l (q = 05)&   &     14.69&    10.49&    13.31&     8.05\tabularnewline
   ~~&   &   &   (0.87**)&   (0.89**)&   (0.88**)&   (0.88**)\tabularnewline
   ~~&   Autoencoder 8l (q = 15)&   &     18.81&    14.15&    12.66&    14.37\tabularnewline
   ~~&   &   &   (0.87**)&   (0.87**)&   (0.87**)&   (0.87**)\tabularnewline
   ~~&   Autoencoder 8l (q = 30)&   &     16.73&    17.04&    13.88&    17.67\tabularnewline
   ~~&   &   &   (0.87**)&   (0.86**)&   (0.87**)&   (0.86**)\tabularnewline
\midrule
   ~~&   Diffusion Maps (q = 05)&   &     23.30&    24.21&    21.04&    21.18\tabularnewline
   ~~&   &   &   (\textbf{0.85**})&   (0.84**)&   (\textbf{0.85***})&   (0.84**)\tabularnewline
   ~~&   Diffusion Maps (q = 15)&   &     24.54&    24.61&    23.08&    23.34\tabularnewline
   ~~&   &   &   (0.86**)&   (0.85**)&   (0.85***)&   (0.85**)\tabularnewline
   ~~&   Diffusion Maps (q = 30)&   &     22.62&    19.15&    18.59&    15.70\tabularnewline
   ~~&   &   &   (0.85**)&   (0.86**)&   (0.85**)&   (0.86**)\tabularnewline
\midrule
   ~~&   ISOMAP (q = 05)&   &     19.64&    23.67&    17.33&    22.84\tabularnewline
   ~~&   &   &   (0.86**)&   (0.85**)&   (0.87**)&   (0.85**)\tabularnewline
   ~~&   ISOMAP (q = 15)&   &     16.16&     5.90&     6.23&     1.56\tabularnewline
   ~~&   &   &   (0.87**)&   (0.88*)&   (0.89*)&   (0.88*)\tabularnewline
   ~~&   ISOMAP (q = 30)&   &     14.74&     7.15&     5.99&     9.99\tabularnewline
   ~~&   &   &   (0.88**)&   (0.88*)&   (0.89*)&   (0.88*)\tabularnewline
\midrule
   ~~&   LLE (q = 05)&   &     20.22&    18.90&    17.84&    19.08\tabularnewline
   ~~&   &   &   (0.86**)&   (0.87**)&   (0.87**)&   (0.87**)\tabularnewline
   ~~&   LLE (q = 15)&   &     13.71&   -19.34&    -7.33&   -16.46\tabularnewline
   ~~&   &   &   (0.87**)&   (0.91)&   (0.89)&   (0.91)\tabularnewline
   ~~&   LLE (q = 30)&   &     -1.75&   -25.40&   -14.52&   -24.81\tabularnewline
   ~~&   &   &   (0.88**)&   (0.90)&   (0.89*)&   (0.90)\tabularnewline
\midrule
   ~~&   PCA gauss. kernel (q = 05)&   &     22.26&    19.58&    20.70&    21.06\tabularnewline
   ~~&   &   &   (0.87**)&   (0.88**)&   (0.88**)&   (0.88**)\tabularnewline
   ~~&   PCA gauss. kernel (q = 15)&   &     18.35&    11.90&    19.22&    12.83\tabularnewline
   ~~&   &   &   (0.88**)&   (0.89*)&   (0.88**)&   (0.89*)\tabularnewline
   ~~&   PCA gauss. kernel (q = 30)&   &     20.06&    14.91&    20.25&    13.55\tabularnewline
   ~~&   &   &   (0.88**)&   (0.87**)&   (0.87**)&   (0.87*)\tabularnewline
\midrule
   ~~&   PCA linear (q = 05)&   &     20.93&    19.89&    17.90&    20.71\tabularnewline
   ~~&   &   &   (0.88**)&   (0.90)&   (0.88*)&   (0.89*)\tabularnewline
   ~~&   PCA linear (q = 15)&   &     21.04&    17.39&    15.61&    19.46\tabularnewline
   ~~&   &   &   (0.88*)&   (0.89*)&   (0.89*)&   (0.89*)\tabularnewline
   ~~&   PCA linear (q = 30)&   &     20.39&    19.62&    17.91&    19.18\tabularnewline
   ~~&   &   &   (0.88**)&   (0.87*)&   (0.88*)&   (0.87*)\tabularnewline
\midrule
   ~~&   PCA poly. kernel (q = 05)&   &     18.07&    18.57&    20.22&    19.27\tabularnewline
   ~~&   &   &   (0.88**)&   (0.89*)&   (0.88**)&   (0.89*)\tabularnewline
   ~~&   PCA poly. kernel (q = 15)&   &     19.06&    18.08&    18.79&    17.68\tabularnewline
   ~~&   &   &   (0.88*)&   (0.88*)&   (0.89*)&   (0.88*)\tabularnewline
   ~~&   PCA poly. kernel (q = 30)&   &     21.74&    20.98&    20.81&    22.47\tabularnewline
   ~~&   &   &   (0.88**)&   (0.87**)&   (0.88**)&   (0.87**)\tabularnewline
\midrule
   ~~&   PCA quadratic (q = 05)&   &     25.67&    23.56&    27.81*&    24.94\tabularnewline
   ~~&   &   &   (0.89*)&   (0.89)&   (0.89*)&   (0.89)\tabularnewline
   ~~&   PCA quadratic (q = 15)&   &     30.48**&   \textbf{37.88**}&    26.86*&   \textbf{38.19**}\tabularnewline
   ~~&   &   &   (0.87**)&   (0.88**)&   (0.88*)&   (0.87**)\tabularnewline
   ~~&   PCA quadratic (q = 30)&   &     10.62&    16.05&     6.92&    15.24\tabularnewline
   ~~&   &   &   (0.89)&   (0.88)&   (0.89)&   (0.89)\tabularnewline
\midrule
   ~~&   PCA squared (q = 05)&   &   \textbf{30.57**}&    29.17*&   \textbf{30.16**}&    30.22*\tabularnewline
   ~~&   &   &   (0.87*)&   (0.88)&   (0.87*)&   (0.86)\tabularnewline
   ~~&   PCA squared (q = 15)&   &     21.21&    37.54**&    17.82&    36.69**\tabularnewline
   ~~&   &   &   (0.87**)&   (\textbf{0.83**})&   (0.87*)&   (\textbf{0.83**})\tabularnewline
   ~~&   PCA squared (q = 30)&   &     14.14&    23.64&    10.74&    24.69\tabularnewline
   ~~&   &   &   (0.89)&   (0.84**)&   (0.89)&   (0.85**)\tabularnewline
\bottomrule
\end{tabular*}
\begin{tablenotes}[flushleft]
\tiny
\item \textit{Note:} The table shows LPLs with RMSEs in parentheses below when we divide our dataset into slow- and fast-moving variables. The first (red shaded) entry gives the actual LPL and RMSE scores of our benchmark (an autoregressive (AR) model with constant parameters and a Minnesota prior). Asterisks indicate statistical significance for each model relative to the benchmark at the 1\% (***), 5\% (**) and 10\% (*) significance levels. Since the large ARX model with time-varying parameters would features $273$ period-specific coefficients and is computationally intractable, we assume that the TVPs feature a factor structure (with three factors) to reduce the dimension of the state space \citep[see Section \ref{sec:App B} of the Online Appendix and][]{chan2020reducing}.
\end{tablenotes}
\end{center}}
\end{table*}

\clearpage

\subsection{Forecast performance over time}
\begin{figure}[!htbp]
\caption{Evolution of one-month-ahead cumulative LPLs against the benchmark. \label{fig:lps1}}
\begin{minipage}{\textwidth}
\centering \vspace*{-5pt}
\small (a) \textit{No dimension reduction}
\hspace{5pt}
\end{minipage}
\begin{minipage}{\textwidth}
\centering
\includegraphics[scale=.3]{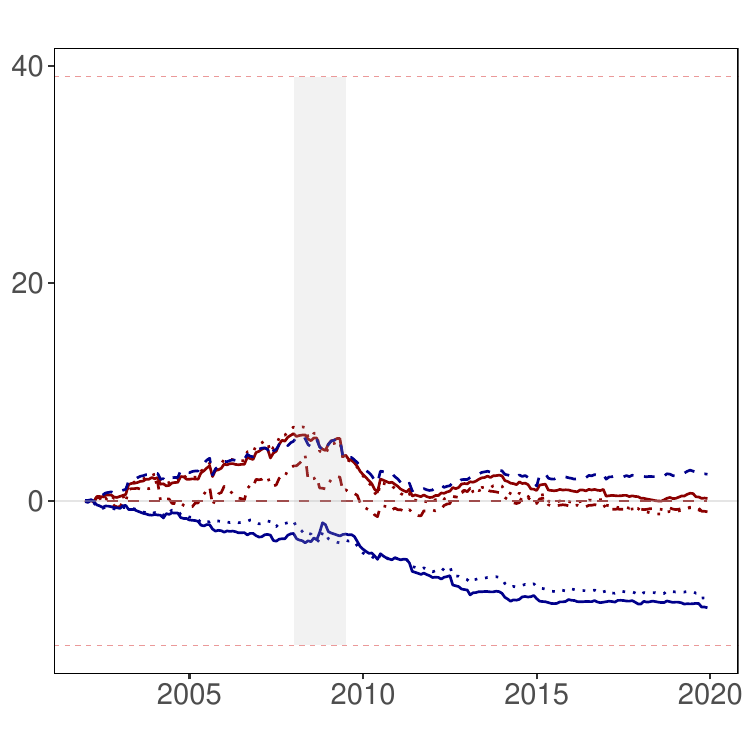}
\end{minipage}
\begin{minipage}{\textwidth}
\centering \hspace{2cm}
\includegraphics[scale=.4]{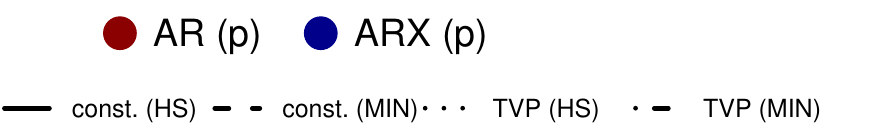}
\end{minipage}\vfill

\begin{minipage}{\textwidth}
\centering
\vspace{5pt}
\small (b) \textit{q = 5}
\vspace{2pt}
\end{minipage}

\begin{minipage}{0.24\textwidth}
\centering
\scriptsize \textit{const. (HS)}
\end{minipage}
\begin{minipage}{0.24\textwidth}
\centering
\scriptsize \textit{const. (MIN)}
\end{minipage}
\begin{minipage}{0.24\textwidth}
\centering
\scriptsize \textit{TVP (HS)}
\end{minipage}
\begin{minipage}{0.24\textwidth}
\centering
\scriptsize \textit{TVP (MIN)}
\end{minipage}

\begin{minipage}{0.24\textwidth}
\centering
\includegraphics[scale=.3]{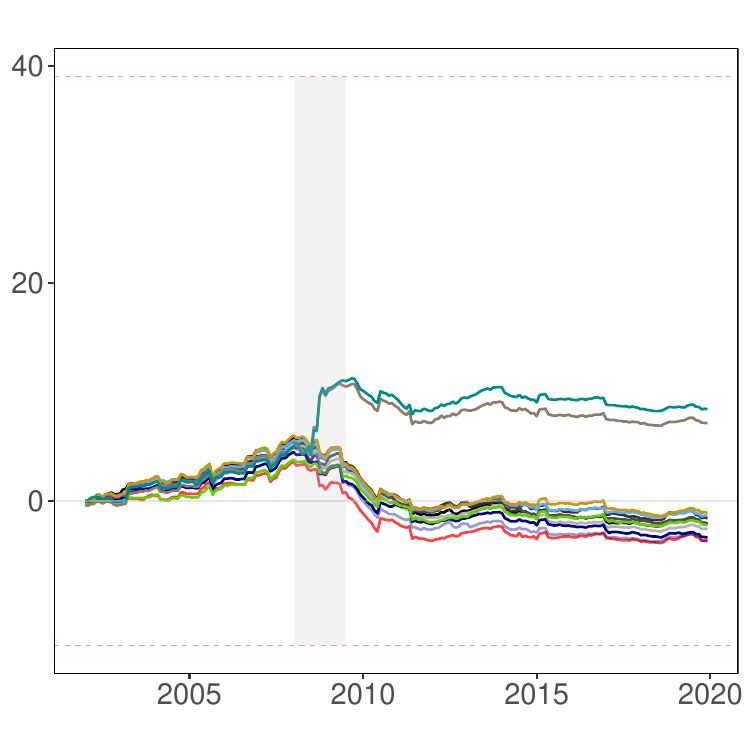}
\end{minipage}
\begin{minipage}{0.24\textwidth}
\centering
\includegraphics[scale=.3]{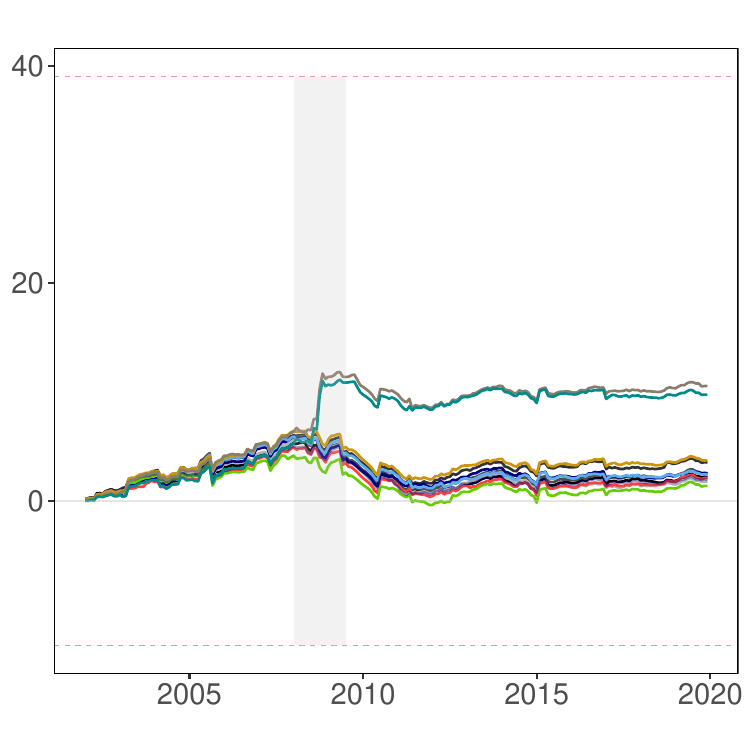}
\end{minipage}
\begin{minipage}{0.24\textwidth}
\centering
\includegraphics[scale=.3]{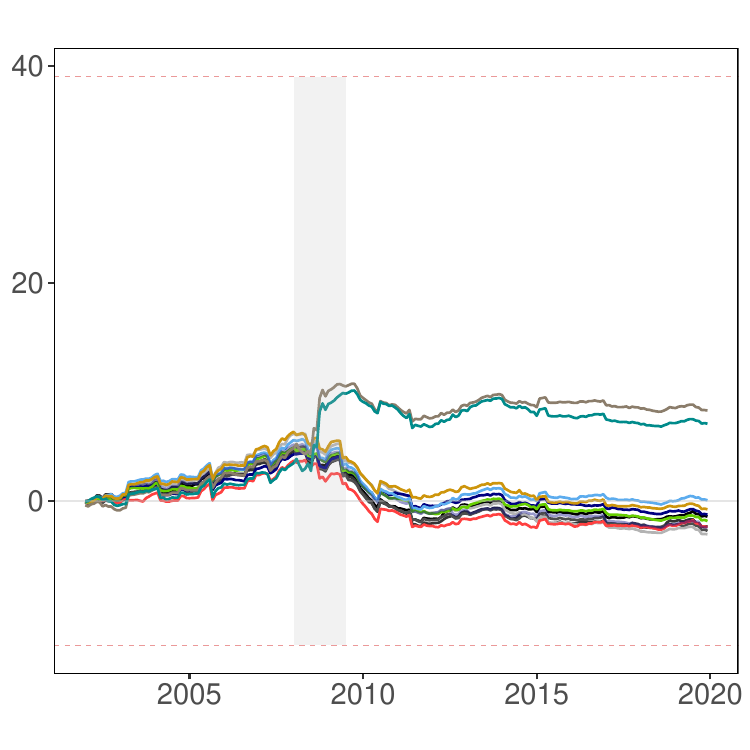}
\end{minipage}
\begin{minipage}{0.24\textwidth}
\centering
\includegraphics[scale=.3]{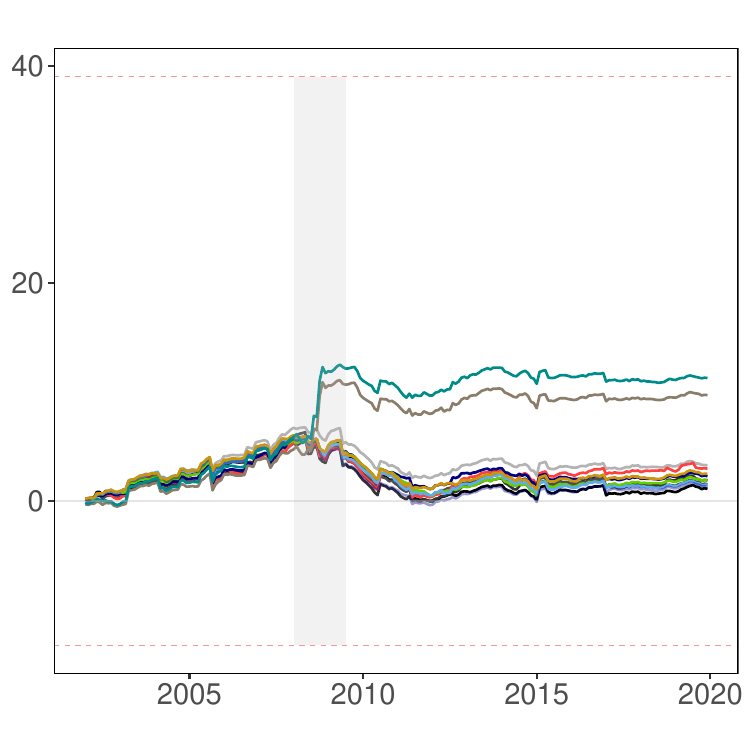}
\end{minipage}

\begin{minipage}{\textwidth}
\centering
\small (c) \textit{q = 15}
\end{minipage}

\begin{minipage}{0.24\textwidth}
\centering
\includegraphics[scale=.3]{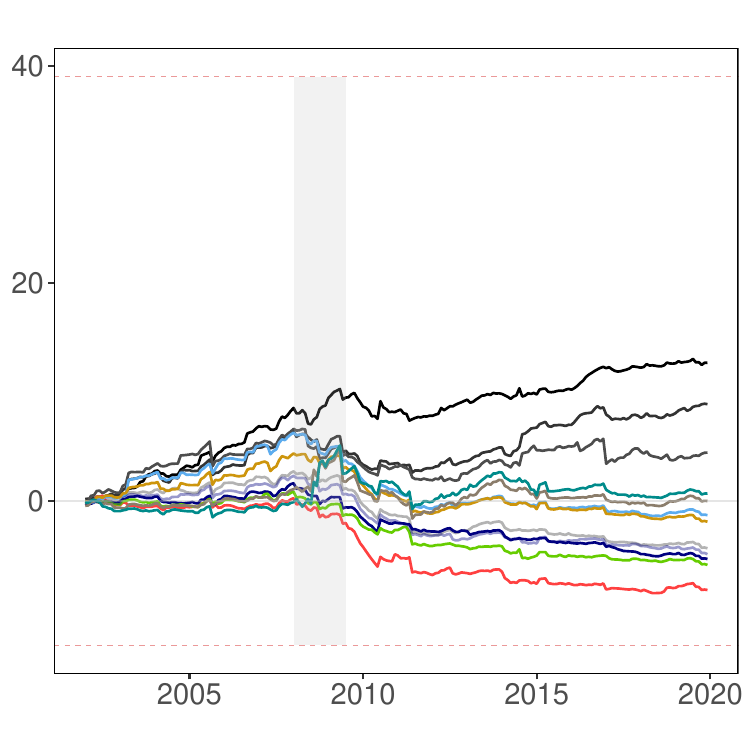}
\end{minipage}
\begin{minipage}{0.24\textwidth}
\centering
\includegraphics[scale=.3]{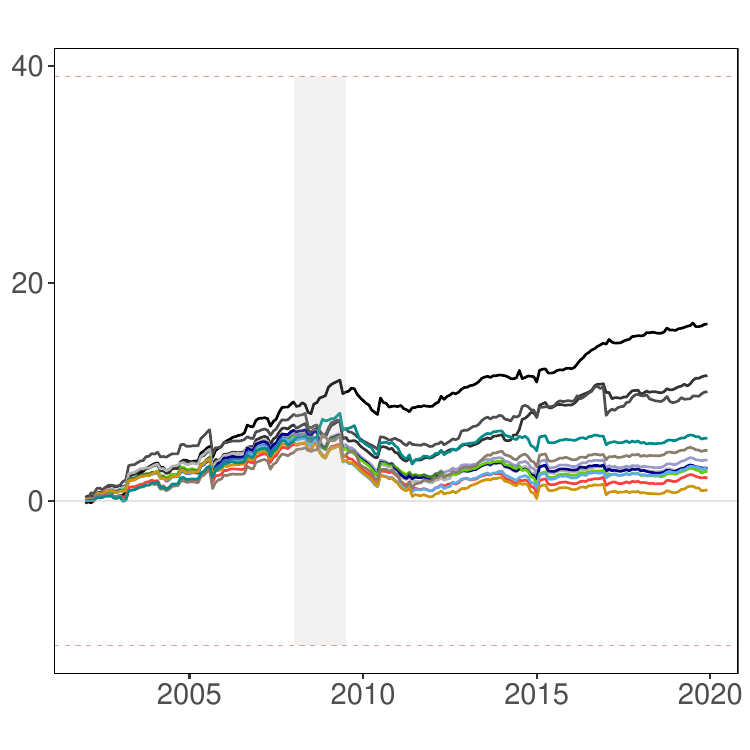}
\end{minipage}
\begin{minipage}{0.24\textwidth}
\centering
\includegraphics[scale=.3]{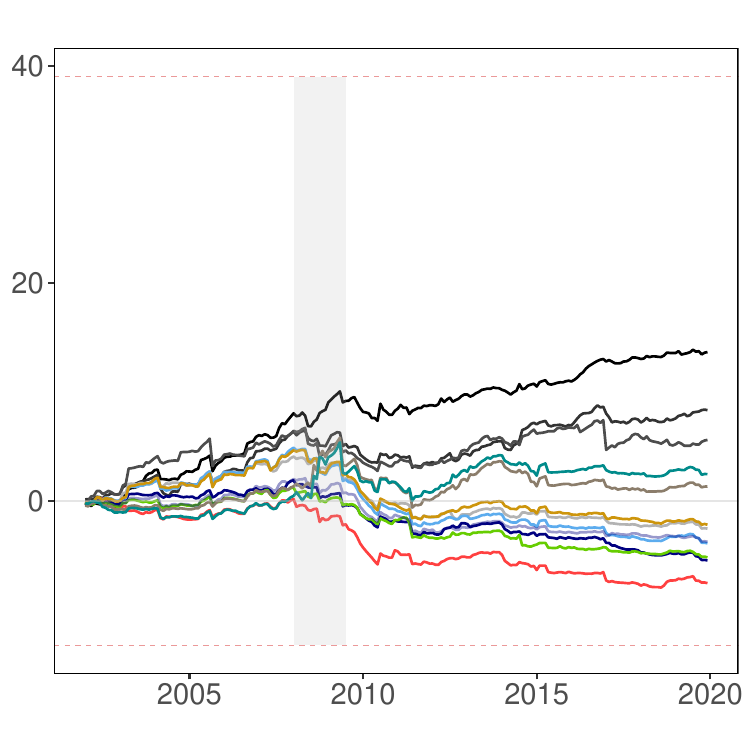}
\end{minipage}
\begin{minipage}{0.24\textwidth}
\centering
\includegraphics[scale=.3]{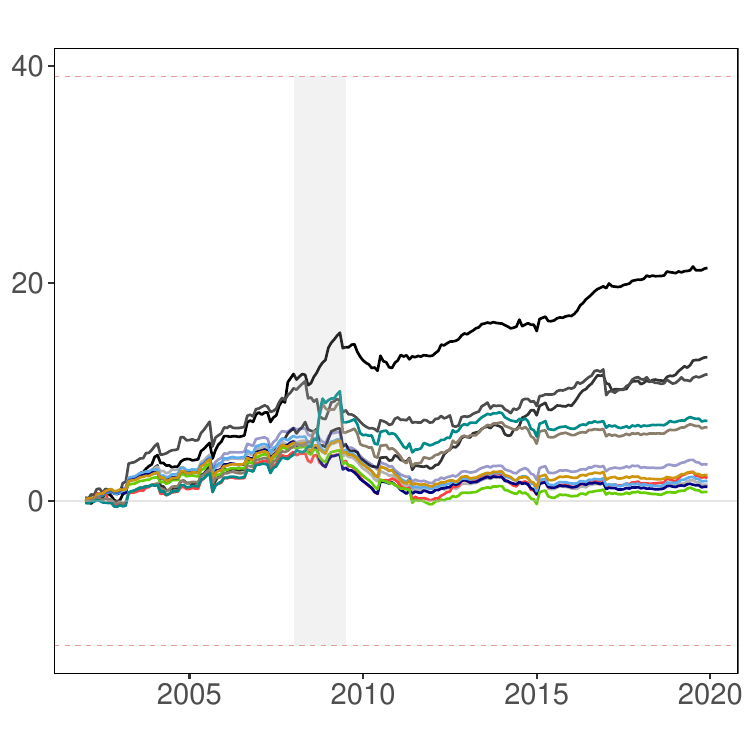}
\end{minipage}

\begin{minipage}{\textwidth}
\centering
\small (d) \textit{q = 30}
\end{minipage}

\begin{minipage}{0.24\textwidth}
\centering
\includegraphics[scale=.3]{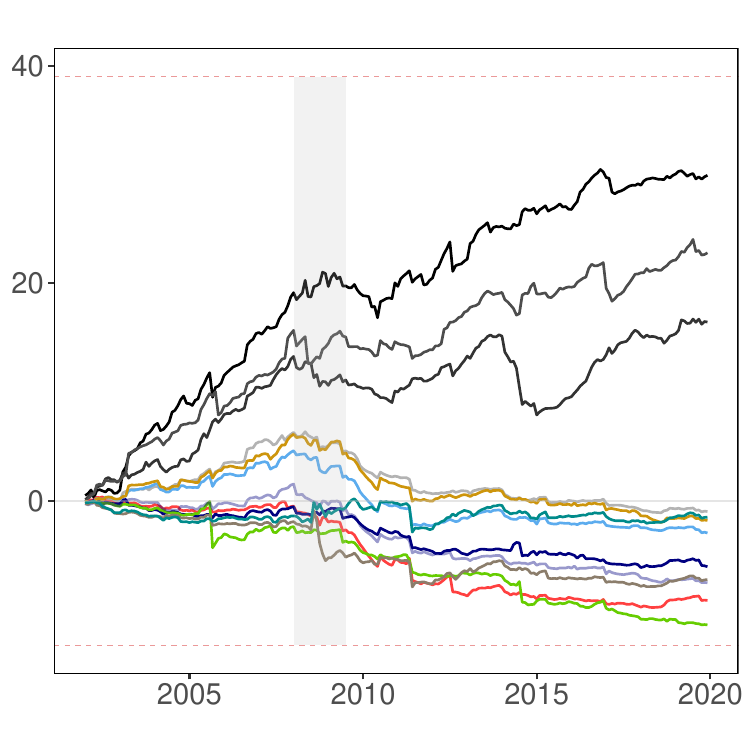}
\end{minipage}
\begin{minipage}{0.24\textwidth}
\centering
\includegraphics[scale=.3]{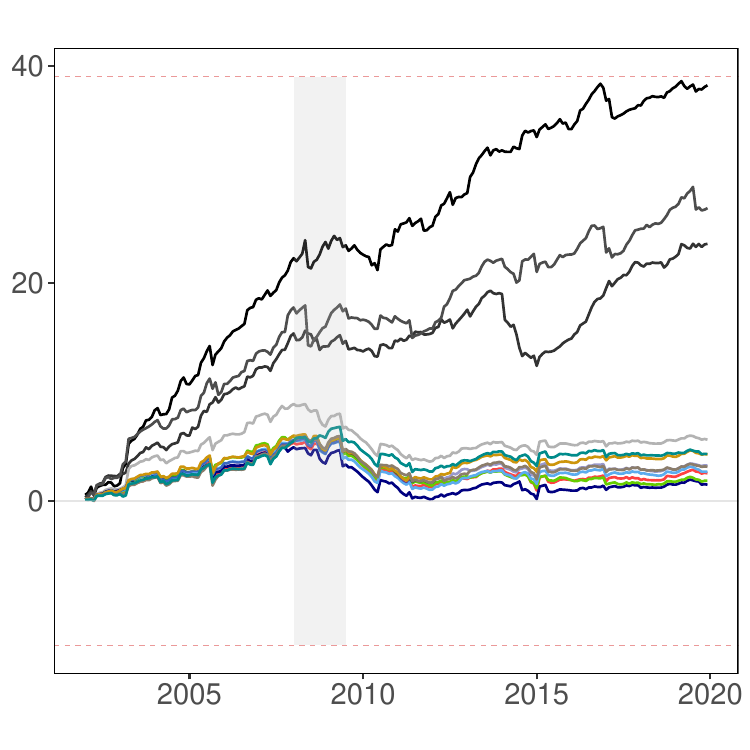}
\end{minipage}
\begin{minipage}{0.24\textwidth}
\centering
\includegraphics[scale=.3]{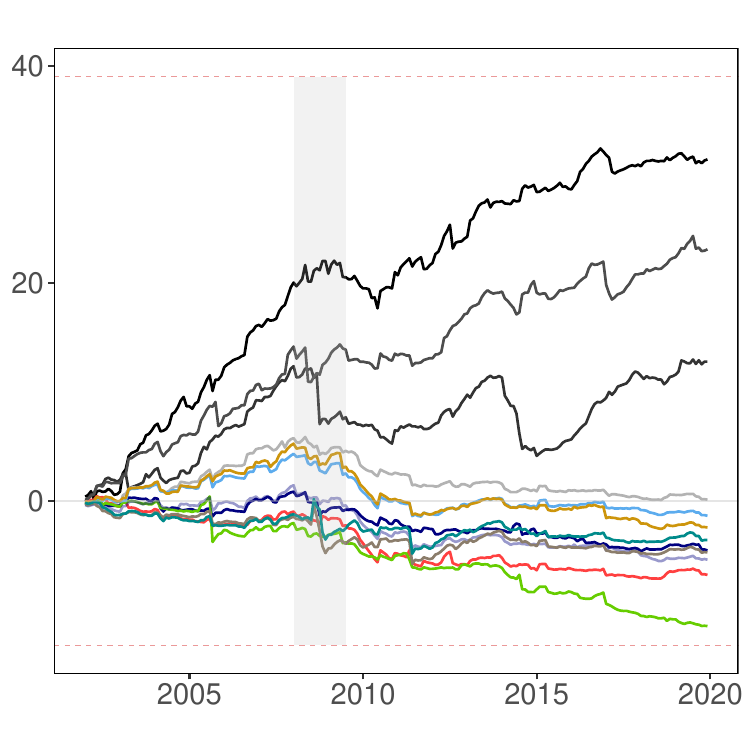}
\end{minipage}
\begin{minipage}{0.24\textwidth}
\centering
\includegraphics[scale=.3]{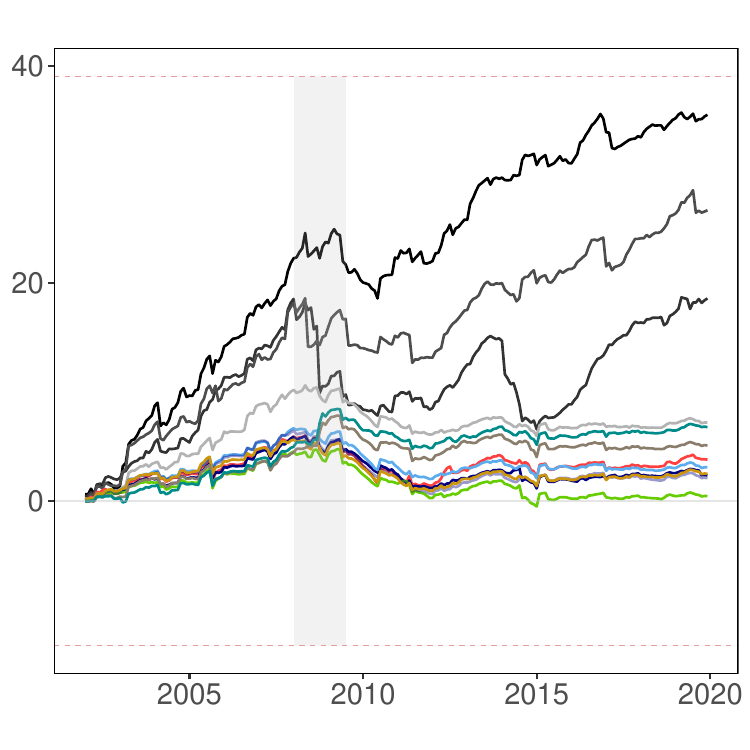}
\end{minipage}

\begin{minipage}{\textwidth}
\centering
\includegraphics[scale=.4]{legend_col.pdf}
\end{minipage}\vfill
\begin{minipage}{\textwidth}
\vspace{2pt}
\scriptsize \textit{Note:} The red dashed lines denote the max./min. LPLs at the end of the hold-out sample, while the gray shaded areas indicate the NBER recessions.
\end{minipage}

\end{figure}

\autoref{fig:lps1} depicts the cumulative LPLs  along the hold-out sample for the one-month-ahead forecasts. The figure suggests a great deal of performance variation over time. This figure shows that the different variants of the Autoencoder display a consistently stronger performance throughout the hold-out period. In some periods (i.e., the global financial crisis and early 2015) we observe declines in the relative model performance vis-\'{a}-vis the AR benchmark.
\vspace*{-5pt}
\clearpage

\begin{landscape}
\begin{table*}[ht]
{\tiny
\caption*{\subsection{Probability integral transforms of higher order forecasts}}
\caption{Test statistics of one-quarter-ahead probability integral transformations \label{tab:PITs3}}
\begin{center}
\begin{tabular*}{\linewidth}{l @{\extracolsep{\fill}} lclllclllclllclll}
\toprule
\multicolumn{1}{l}{\bfseries }&\multicolumn{1}{c}{\bfseries Specification}&\multicolumn{1}{c}{\bfseries }&\multicolumn{3}{c}{\bfseries const. (MIN)}&\multicolumn{1}{c}{\bfseries }&\multicolumn{3}{c}{\bfseries const. (HS)}&\multicolumn{1}{c}{\bfseries }&\multicolumn{3}{c}{\bfseries TVP (MIN)}&\multicolumn{1}{c}{\bfseries }&\multicolumn{3}{c}{\bfseries TVP (HS)}\tabularnewline
\cmidrule{4-6} \cmidrule{8-10} \cmidrule{12-14} \cmidrule{16-18}
\multicolumn{1}{l}{}&\multicolumn{1}{c}{}&\multicolumn{1}{c}{}&\multicolumn{1}{c}{Mean}&\multicolumn{1}{c}{Variance}&\multicolumn{1}{c}{AR(1) coef.}&\multicolumn{1}{c}{}&\multicolumn{1}{c}{Mean}&\multicolumn{1}{c}{Variance}&\multicolumn{1}{c}{AR(1) coef.}&\multicolumn{1}{c}{}&\multicolumn{1}{c}{Mean}&\multicolumn{1}{c}{Variance}&\multicolumn{1}{c}{AR(1) coef.}&\multicolumn{1}{c}{}&\multicolumn{1}{c}{Mean}&\multicolumn{1}{c}{Variance}&\multicolumn{1}{c}{AR(1) coef.}\tabularnewline
\midrule
{\scshape }&&&&&&&&&&&&&&&&&\tabularnewline
   ~~&   AR&   &   -0.027&   1.240&   0.547***&   &   -0.021&   1.272&   0.583***&   &   -0.017&   1.201&   0.648***&   &   -0.025&   1.297&   0.639***\tabularnewline
   ~~&   Large ARX&   &   -0.050&   1.342&   0.693***&   &   0.082&   1.347&   0.707***&   &   &  &  &   &   0.067&   1.384&   0.712***\tabularnewline
\midrule
   ~~&   Autoencoder 1l (q = 05)&   &   -0.056&   1.402&   0.700***&   &   -0.029&   1.466&   0.708***&   &   -0.054&   1.448&   0.703***&   &   -0.029&   1.462&   0.708***\tabularnewline
   ~~&   Autoencoder 1l (q = 15)&   &   -0.032&   1.357&   0.701***&   &   -0.008&   1.374&   0.697***&   &   -0.029&   1.380&   0.702***&   &   -0.008&   1.384&   0.702***\tabularnewline
   ~~&   Autoencoder 1l (q = 30)&   &   -0.034&   1.400&   0.702***&   &   0.028&   1.383&   0.701***&   &   -0.028&   1.378&   0.705***&   &   0.028&   1.396&   0.704***\tabularnewline
   ~~&   Autoencoder 3l (q = 05)&   &   -0.054&   1.402&   0.700***&   &   -0.045&   1.411&   0.705***&   &   -0.048&   1.383&   0.701***&   &   -0.045&   1.439&   0.707***\tabularnewline
   ~~&   Autoencoder 3l (q = 15)&   &   -0.027&   1.402&   0.701***&   &   0.022&   1.430&   0.707***&   &   -0.016&   1.389&   0.698***&   &   0.017&   1.432&   0.698***\tabularnewline
   ~~&   Autoencoder 3l (q = 30)&   &   -0.023&   1.374&   0.702***&   &   0.034&   1.440&   0.701***&   &   -0.024&   1.389&   0.704***&   &   0.030&   1.440&   0.703***\tabularnewline
   ~~&   Autoencoder 5l (q = 05)&   &   -0.060&   1.393&   0.696***&   &   -0.061&   1.479&   0.705***&   &   -0.052&   1.409&   0.700***&   &   -0.054&   1.451&   0.704***\tabularnewline
   ~~&   Autoencoder 5l (q = 15)&   &   -0.031&   1.436&   0.703***&   &   -0.008&   1.487&   0.702***&   &   -0.032&   1.442&   0.700***&   &   -0.006&   1.487&   0.703***\tabularnewline
   ~~&   Autoencoder 5l (q = 30)&   &   -0.022&   1.395&   0.707***&   &   0.040&   1.418&   0.702***&   &   -0.019&   1.420&   0.703***&   &   0.031&   1.417&   0.700***\tabularnewline
   ~~&   Autoencoder 8l (q = 05)&   &   -0.051&   1.403&   0.700***&   &   -0.029&   1.456&   0.712***&   &   -0.056&   1.428&   0.698***&   &   -0.039&   1.463&   0.712***\tabularnewline
   ~~&   Autoencoder 8l (q = 15)&   &   -0.054&   1.452&   0.704***&   &   -0.035&   1.524&   0.692***&   &   -0.064&   1.496&   0.702***&   &   -0.038&   1.526&   0.693***\tabularnewline
   ~~&   Autoencoder 8l (q = 30)&   &   -0.070&   1.624&   0.704***&   &   -0.065&   1.794*&   0.699***&   &   -0.085&   1.707*&   0.698***&   &   -0.070&   1.777*&   0.702***\tabularnewline
\midrule
   ~~&   Diffusion Maps (q = 05)&   &   -0.079&   1.467&   0.701***&   &   -0.073&   1.464&   0.704***&   &   -0.090&   1.535&   0.710***&   &   -0.078&   1.478&   0.707***\tabularnewline
   ~~&   Diffusion Maps (q = 15)&   &   -0.048&   1.416&   0.696***&   &   -0.031&   1.458&   0.698***&   &   -0.065&   1.475&   0.705***&   &   -0.038&   1.495&   0.699***\tabularnewline
   ~~&   Diffusion Maps (q = 30)&   &   -0.052&   1.467&   0.703***&   &   -0.013&   1.496&   0.699***&   &   -0.057&   1.489&   0.708***&   &   -0.015&   1.513&   0.700***\tabularnewline
\midrule
   ~~&   ISOMAP (q = 05)&   &   -0.060&   1.411&   0.704***&   &   -0.065&   1.449&   0.704***&   &   -0.069&   1.426&   0.703***&   &   -0.066&   1.428&   0.704***\tabularnewline
   ~~&   ISOMAP (q = 15)&   &   -0.060&   1.409&   0.702***&   &   -0.048&   1.489&   0.701***&   &   -0.069&   1.461&   0.707***&   &   -0.049&   1.498&   0.703***\tabularnewline
   ~~&   ISOMAP (q = 30)&   &   -0.039&   1.491&   0.702***&   &   -0.008&   1.538&   0.703***&   &   -0.012&   1.562&   0.705***&   &   -0.004&   1.548&   0.703***\tabularnewline
\midrule
   ~~&   LLE (q = 05)&   &   -0.069&   1.451&   0.704***&   &   -0.097&   1.541&   0.721***&   &   -0.081&   1.492&   0.707***&   &   -0.101&   1.545&   0.720***\tabularnewline
   ~~&   LLE (q = 15)&   &   -0.108&   1.698*&   0.688***&   &   -0.099&   1.774*&   0.726***&   &   -0.128&   1.773*&   0.711***&   &   -0.093&   1.750*&   0.724***\tabularnewline
   ~~&   LLE (q = 30)&   &   -0.094&   1.612*&   0.680***&   &   -0.071&   2.110*&   0.691***&   &   -0.119&   1.711&   0.713***&   &   -0.069&   2.037*&   0.704***\tabularnewline
\midrule
   ~~&   PCA gauss. kernel (q = 05)&   &   -0.050&   1.415&   0.704***&   &   -0.024&   1.435&   0.713***&   &   -0.045&   1.388&   0.703***&   &   -0.032&   1.457&   0.711***\tabularnewline
   ~~&   PCA gauss. kernel (q = 15)&   &   -0.041&   1.412&   0.706***&   &   -0.022&   1.458&   0.715***&   &   -0.035&   1.408&   0.703***&   &   -0.021&   1.430&   0.716***\tabularnewline
   ~~&   PCA gauss. kernel (q = 30)&   &   -0.038&   1.395&   0.704***&   &   -0.009&   1.412&   0.714***&   &   -0.039&   1.407&   0.708***&   &   -0.006&   1.419&   0.712***\tabularnewline
\midrule
   ~~&   PCA linear (q = 05)&   &   -0.052&   1.382&   0.698***&   &   -0.025&   1.435&   0.713***&   &   -0.048&   1.402&   0.703***&   &   -0.024&   1.447&   0.711***\tabularnewline
   ~~&   PCA linear (q = 15)&   &   -0.039&   1.391&   0.701***&   &   -0.003&   1.397&   0.712***&   &   -0.038&   1.426&   0.707***&   &   -0.008&   1.434&   0.708***\tabularnewline
   ~~&   PCA linear (q = 30)&   &   -0.042&   1.378&   0.699***&   &   0.003&   1.396&   0.705***&   &   -0.044&   1.400&   0.703***&   &   0.003&   1.393&   0.704***\tabularnewline
\midrule
   ~~&   PCA poly. kernel (q = 05)&   &   -0.050&   1.378&   0.701***&   &   -0.038&   1.451&   0.711***&   &   -0.055&   1.413&   0.702***&   &   -0.043&   1.456&   0.717***\tabularnewline
   ~~&   PCA poly. kernel (q = 15)&   &   -0.033&   1.392&   0.699***&   &   -0.021&   1.428&   0.712***&   &   -0.036&   1.404&   0.705***&   &   -0.020&   1.413&   0.711***\tabularnewline
   ~~&   PCA poly. kernel (q = 30)&   &   -0.041&   1.357&   0.700***&   &   -0.012&   1.398&   0.706***&   &   -0.041&   1.384&   0.702***&   &   -0.012&   1.394&   0.707***\tabularnewline
\midrule
   ~~&   PCA quadratic (q = 05)&   &   -0.077&   1.216&   0.695***&   &   -0.084&   1.169&   0.690***&   &   -0.075&   1.193&   0.696***&   &   -0.081&   1.153&   0.691***\tabularnewline
   ~~&   PCA quadratic (q = 15)&   &   -0.143&   1.468&   0.716***&   &   -0.129&   1.347&   0.701***&   &   -0.160&   1.477&   0.713***&   &   -0.131&   1.339&   0.704***\tabularnewline
   ~~&   PCA quadratic (q = 30)&   &   -0.071&   1.327&   0.700***&   &   -0.053&   1.413&   0.705***&   &   -0.072&   1.297&   0.708***&   &   -0.053&   1.408&   0.708***\tabularnewline
\midrule
   ~~&   PCA squared (q = 05)&   &   -0.065&   1.129&   0.682***&   &   -0.082&   1.154&   0.684***&   &   -0.063&   1.132&   0.683***&   &   -0.080&   1.150&   0.686***\tabularnewline
   ~~&   PCA squared (q = 15)&   &   -0.116&   1.417&   0.714***&   &   -0.119&   1.336&   0.688***&   &   -0.121&   1.421&   0.710***&   &   -0.123&   1.337&   0.690***\tabularnewline
   ~~&   PCA squared (q = 30)&   &   -0.051&   1.289&   0.698***&   &   -0.071&   1.402&   0.700***&   &   -0.062&   1.317&   0.695***&   &   -0.071&   1.413&   0.699***\tabularnewline
\bottomrule
\end{tabular*}
\begin{tablenotes}[flushleft]
      \tiny
      \item \textit{Note:} This table summarizes the normalized forecast errors, which are obtained with probability integral transformations (PIT). 
      Similar to \cite{clark2011} we show the mean, the variance and the AR($1$) coefficient of the normalized forecast errors. Given a well-calibrated model (i.e. the null-hypothesis), normalized forecast errors should have zero mean, a variance of one and experience no autocorrelation. These conditions are tested separately: 1) To test for a zero mean we compute the p-values with a Newey–West variance (with five lags). 2) To test for a unit variance we regress the squared normalized forecast errors on an intercept and allow for a Newey–West variance (with three lags). 3) To test for no autocorrelation we obtain the p-values with an AR(1) model that features an unconditional mean and heteroskedasticity-robust standard errors. Asterisks indicate statistical significance for each model at the 1\% (***), 5\% (**) and 10\% (*) significance levels. 
    \end{tablenotes}
\end{center}}
\end{table*}

\end{landscape}

\setcounter{equation}{0}
\setcounter{table}{0}
\setcounter{figure}{0}
\renewcommand\theequation{B.\arabic{equation}}
\renewcommand\thetable{B.\arabic{table}}
\renewcommand\thefigure{B.\arabic{figure}}

\section{Technical Appendix}\label{sec:App B}

\subsection{Non-centered parameterization}
To implement the Bayesian priors to achieve shrinkage in the TVP regression defined by Eq. (\ref{eq: TVP_regression}) and Eq. (\ref{eq: state}), we use the non-centered parameterization proposed in  \cite{fs_wagner}. Intuitively speaking, this allows us to move the process innovation variances into the observation equation and discriminate between a time-invariant and a time-varying part of the model. The non-centered parameterization of the model is given by:
\begin{align}
y_{t+h} &= \bm d_{t+h}'\bm \beta_0 + \bm d_{t+h}'\sqrt{\bm V}\tilde{\bm \beta}_{t+h} + \epsilon_{t+h}, \quad \epsilon_{t+h} \sim \mathcal{N} (0,\sigma^2_{t+h}) \label{eq:ncn} \\
\tilde{\bm \beta}_{t+h} &= \tilde{\bm \beta}_{t+h-1} + \varepsilon_{t+h}, \quad \varepsilon_{t+h} \sim \mathcal{N} (0,\bm I_M),\quad  \tilde{\bm \beta}_0 = \bm 0_M,
\end{align}\label{eq: ncp}
where the $j$th element in $\bm \tilde{\bm \beta}_{t+h}$ is given by $\tilde{\beta}_{jt+h} = \frac{\beta_{jt+h}-\beta_{j0}}{v_j}$  for $j=1, \dots, M$.

Conditional on the normalized states $\tilde{\bm \beta}$, Eq. (\ref{eq:ncn}) can be written  as a linear regression model as follows:
\begin{equation}
y_{t+h} = \bm D'_{t+h} \bm \alpha + \epsilon_{t+h},
\end{equation}
with $\bm D_{t+h} = [ \bm d'_{t+h}, (\bm \tilde{\bm \beta}_{t+h} \odot \bm d_{t+h})']'$ denoting a $2M$-dimensional vector of regressors and $\bm \alpha = (\bm \beta'_0, v_1, \dots, v_M')$ is  a $2M$-dimensional coefficient vector. This parameterization implies that the state innovation variances (or more precisely the square roots) are moved into the observation equation and we can estimate them alongside $\bm \beta_0$ (conditional on the states $\bm \tilde{\bm \beta}_{t+h}$). 

Since the large ARX model with time-varying parameters would features $273$ period-specific coefficients it is computationally intractable. To reduce the dimension of the state space, we assume a factor structure for the normalized TVPs, $\tilde{\bm \beta}_{t+h}$, with three factors. This implies that $\bm V$ is a full $M \times M$ matrix but of reduced-rank. We leave the remaining setup (prior, etc.) unchanged and follow \cite{chan2020reducing} to perform posterior simulation. Conditional on the prior the posterior sampling steps remain quite similar. For further details, see \cite{chan2020reducing}.

\subsection{Prior setup}\label{sec:App Prior}
\subsubsection{Priors on the regression coefficients}
We use a multivariate Gaussian prior on $\bm \alpha$:
\begin{equation}
\bm \alpha|\underline{\bm V}  \sim \mathcal{N} (\bm 0, \underline{\bm V}).
\end{equation}
Since our dependent variable is transformed to be approximately stationary, we set the prior mean equal to zero. This reflects the notion that inflation, as defined in Eq. (\ref{eq:inflation}), follows a white noise process a priori.
Moreover, $\underline{\bm V}$ denotes a $2M$-dimensional prior variance-covariance matrix, $\underline{\bm V} = \text{diag}\left(\tau^2_1, \dots, \tau^2_{2M} \right)$. This matrix collects the prior shrinkage parameters $\tau_j$ associated with the time-invariant regression coefficients and the process innovation standard deviations.

In the empirical work, we consider two prior variants that differ in the specification of the prior variance-covariance matrix $\underline{\bm V}$. The first is the horseshoe \citep[HS,][]{carvalho2010horseshoe} prior and the second is an adaptive Minnesota \citep[MIN, see][]{carriero2015bayesian, giannone2015prior} prior.

\begin{enumerate}



\item \textbf{Horseshoe prior:}\\
The horseshoe prior of \cite{carvalho2010horseshoe} achieves shrinkage by introducing local  and global shrinkage parameters \citep[see][]{polson2010shrink}. These follow a standard half-Cauchy distribution restricted to the positive real numbers. That is:
\begin{equation}\label{eq:HS}
\tau^2_j = \varsigma^2\zeta^2_j,  \quad \varsigma \sim \mathcal{C}^{+}(0,1), \quad \zeta_j \sim \mathcal{C}^{+}(0,1)
\end{equation}
 While the global component $\varsigma$ strongly pushes all coefficients in $\bm \alpha$ towards the prior mean (i.e., zero), the local scalings $\{\zeta_j\}_{j = 1}^{2M}$ allow for variable-specific departures from zero in light of a global scaling parameter close to zero. This flexibility leads to heavy tails in the marginal prior (obtained after integrating out $\zeta_j$) which turns out to be useful for forecasting.

\item \textbf{Adaptive Minnesota prior:}\\
Inspired by \cite{chan2021minnesota}, we consider a simplified version of the adaptive hierarchical Minnesota prior. This setup is also closely related to the one of \cite{carriero2015bayesian} and \cite{giannone2015prior}. That is, we allow for different treatment of own lags of inflation, the exogenous factors as well as the square root of the state innovation variances (governing the time variation). To capture this notion of the adaptive Minnesota prior and to be consistent with notation of the horseshoe prior, we impose the following structure on the diagonal element of the prior variance-covariance matrix: 
\begin{equation}
\tau_j^2 = \varsigma_j^2\zeta^2_j,    
\end{equation}
with 
\begin{equation*}
\varsigma^2_j =
\begin{cases}
\varsigma_1^2 & \text{for time-invariant coefficients related the own lags of inflation } (j =1,\dots,p),\\
\varsigma_2^2 & \text{for time-invariant coefficients related to the $k$th factor } (j = p+1,\dots,M), \\
\varsigma_3^2 & \text{for all state innovation standard deviations } (j =M+1,\dots,2M)  
\end{cases} 
\end{equation*}
and 
\begin{equation*}
\zeta_j^2 =
\begin{cases}
\frac{1}{l^2} & \text{for coefficients associated with the own lags of inflation } (l =1,\dots,p),\\
\frac{\hat{\sigma}^2_\pi}{\hat{\sigma}^2_k} & \text{for coefficients associated with the $k$th exogenous factor } (k =1,\dots,q). \\
\end{cases}
\end{equation*}
Here, $\varsigma_1$, $\varsigma_2$ and $\varsigma_3$ are global shrinkage parameters, $\{\zeta_j\}_{j = 1}^{2M}$ denote local scalings and $\hat{\sigma}_\pi^2$ as well as $\{\hat{\sigma}_k^2\}_{k =1}^{q}$ refer to OLS variances of an AR($1$) model on inflation and the $k$th exogenous factor, respectively. We assume that the global scaling parameters $\varsigma_1$, $\varsigma_2$ and $\varsigma_3$ feature a hierarchical prior structure and are standard half-Cauchy distributed \citep[see][]{polson2012half}, while the local scalings are set to fixed and known values. This structure contrasts with the horseshoe prior in Eq. (\ref{eq:HS}), where we do not discriminate between certain groups of coefficients and where both hyperparameters (i.e., global and local scalings) are stochastic quantities.
\end{enumerate}

\subsection{Full conditional posterior simulation}
We carry out posterior inference by using a Markov chain Monte Carlo (MCMC) algorithm to simulate from the joint posterior of the parameters, the log-volatilities and the TVPs. This MCMC algorithm consists of the following steps:

\begin{enumerate}
\item Conditional on the time-varying part of the coefficients and the stochastic volatilities,  we draw $(\bm \beta_0, v_1, \dots, v_M)'$ from $\mathcal{N}(\overline{\bm \beta}, \overline{\bm V})$ with $\overline{\bm V}=(\tilde{\bm D}' \tilde{\bm D} + \underline{\bm V}^{-1})^{-1}$ and $\underline{\bm \beta} = \overline{\bm V} (\tilde{\bm D} \tilde{\bm y})$.  $\tilde{\bm y}$ is a $T-$dimensional vector with typical element $y_t/\sigma_t$ and $\tilde{\bm D}$ is a $T \times (2M)$ matrix with typical row $\bm D_t/\sigma_t$. 

\item Controlling for all other model parameters, the full history of $\tilde{\bm \beta}_{t+h}$ is sampled using the forward-filtering backward-sampling (FFBS) algorithm proposed by \cite{carterkohn, fs1994}. For constant parameter models this step is skipped.

\item The stochastic volatilities $\log \sigma^2_{t+h}$ are drawn by employing the algorithm of \cite{kastner2014ancillarity} implemented in the \texttt{stochvol} R-package of \cite{kastner2016dealing}.

\item Sampling the diagonal elements of $\underline{\bm V}$ depends on the specific prior setup chosen.
\begin{itemize}
\item In case we adopt the horseshoe prior, we rely on the hierarchical representation of \cite{makalic2015simple}.  Introducing auxiliary random quantities which follow an inverse Gamma distribution we can draw $\zeta_j$ and $\varsigma$ as follows:
\begin{align*}
\zeta^2_j | \beta_j, \varsigma, \eta &\sim \mathcal{G}^{-1} \left(1,  \eta_j^{-1} + \frac{\beta_j^2}{2 \varsigma^2} \right),\\
\varsigma^2 | \beta_j, \zeta_j, \varphi &\sim \mathcal{G}^{-1} \left( \frac{2M+1}{2}, \varphi^{-1} +  \sum_{j=1}^{2M} \frac{\beta_j^2}{2 \zeta_j^2} \right),\\
\eta_j|\zeta_j &\sim \mathcal{G}^{-1}\left(1, 1 + \zeta_j^{-2}\right) \quad \text{and} \\
\varphi | \varsigma &\sim \mathcal{G}^{-1}\left(1, 1 + \varsigma^{-2}\right).
\end{align*}
\item If the Minnesota prior is used, updating of $\varsigma_1$, $\varsigma_2$ and $\varsigma_3$ can be done in a similar fashion as we do with the horseshoe prior. The main difference, however, is that we only need to update the global hyperparameters and the respective auxiliary quantities: 

\begin{align*}
\varsigma^2_1 | \beta_j, \zeta_j, \varphi_1 &\sim \mathcal{G}^{-1} \left( \frac{p+1}{2}, \varphi_1^{-1} + \sum_{j=1}^{p} \frac{\beta_j^2}{ 2\zeta_j^2} \right),\\
\varsigma^2_2 | \beta_j, \zeta_j, \varphi_2 &\sim \mathcal{G}^{-1} \left( \frac{M-p+1}{2}, \varphi_2^{-1} +  \sum_{j=p+1}^{M} \frac{\beta_j^2}{2 \zeta_j^2} \right),\\
\varsigma^2_3 | \beta_j, \zeta_j, \varphi_3 &\sim \mathcal{G}^{-1} \left( \frac{M+1}{2}, \varphi_3^{-1} +  \sum_{j=M+1}^{2M} \frac{\beta_j^2}{2 \zeta_j^2} \right) \quad \text{and}\\
\varphi_i | \varsigma_i &\sim \mathcal{G}^{-1}\left(1, 1 + \varsigma_i^{-2}\right), \quad \text{for } i = 1,2,3.\\
\end{align*}
\end{itemize}
\end{enumerate}
We sample from the relevant full conditional posterior distributions iteratively. This is repeated $10,000$ times and the first $2,000$ draws are discarded as burn-in.

\setcounter{equation}{0}
\setcounter{table}{0}
\setcounter{figure}{0}
\renewcommand\theequation{C.\arabic{equation}}
\renewcommand\thetable{C.\arabic{table}}
\renewcommand\thefigure{C.\arabic{figure}}

\clearpage
\section{Data Appendix}\label{sec:App C}

\begin{figure}
\caption{Variable set for the small ARX models.} \label{fig:corrfreq}

\begin{minipage}{\textwidth}
\centering
a) Autoencoder 1l (q = 30)
\hspace{5pt}
\end{minipage}

\begin{minipage}{\textwidth}
\centering
\includegraphics[scale=.39]{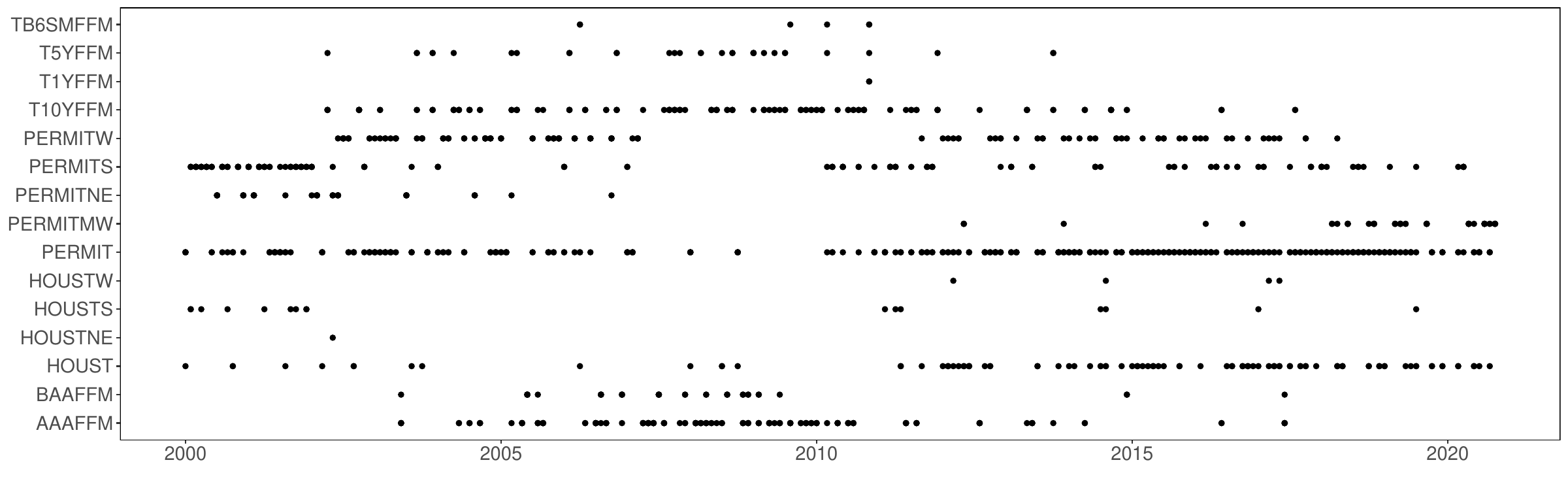}
\end{minipage}

\begin{minipage}{\textwidth}
\centering
b) PCA quadratic (q = 05)
\hspace{5pt}
\end{minipage}

\begin{minipage}{\textwidth}
\centering
\includegraphics[scale=.4]{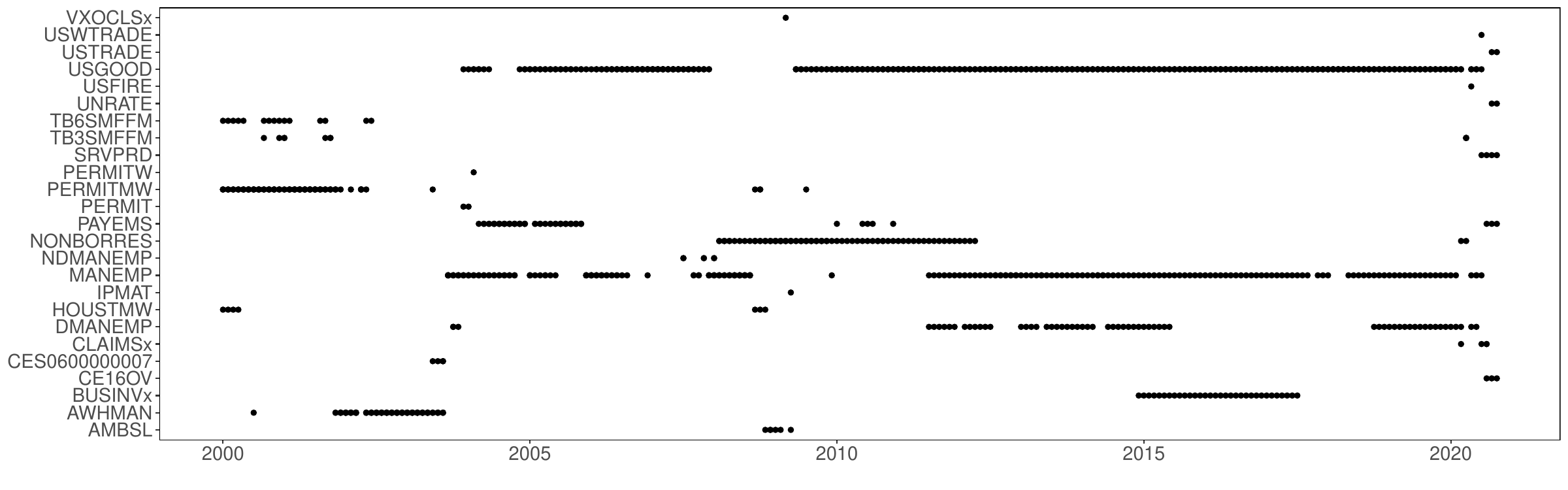}
\end{minipage}

\begin{minipage}{\textwidth}
\centering
c) PCA squared (q = 05)
\hspace{5pt}
\end{minipage}

\begin{minipage}{\textwidth}
\centering
\includegraphics[scale=.4]{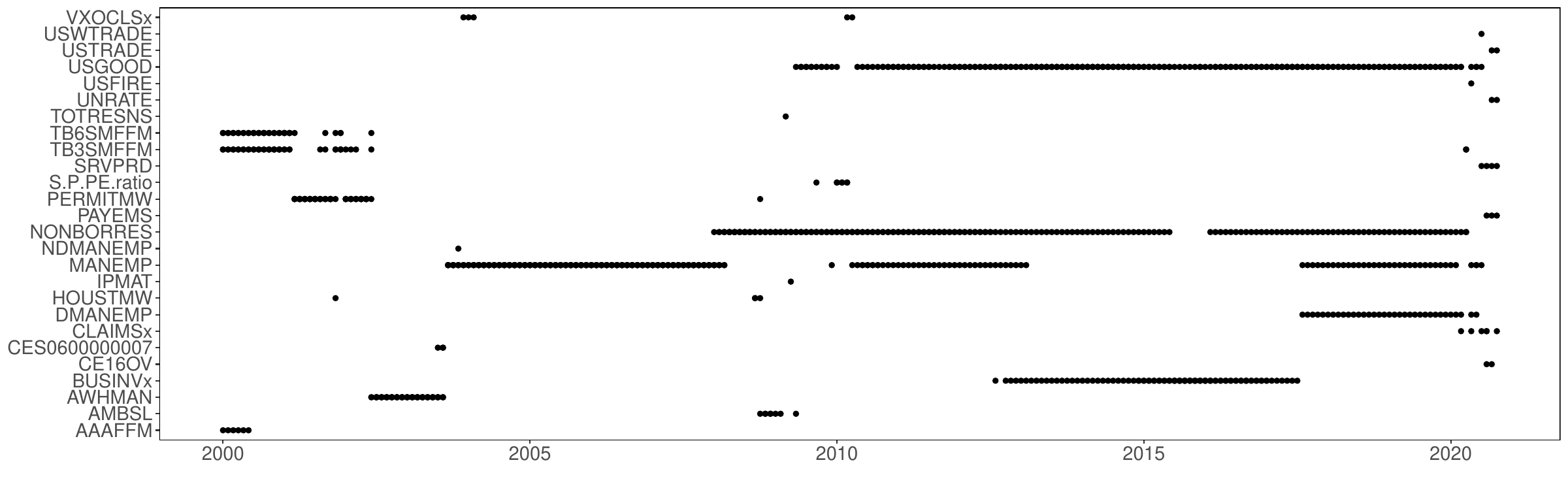}
\end{minipage}

\begin{minipage}{\textwidth}
\scriptsize \textit{Note:} The plots show the top-five correlated variables in each vintage and hence the variable set for the best performing models as shown in \autoref{tab:mainARX}. Note that each variable may occur more than once per vintage as we consider the lagged dataset.
\end{minipage}

\end{figure}

The Federal Reserve Economic Data (FRED) contains monthly observations of macroeconomic variables for the US and is available for download at \url{https://research.stlouisfed.org}. Details on the dataset can be found in \cite{mccracken2016fred}. For each data vintage (available from 1999:08), the time series start in January 1959. Due to missing values in some of the series, we preselect 105 variables and transform them according to \autoref{tab:data1_us}. We select all variables for our models except for the ARX model. In this case, we apply our variable selection approach described in Sub-section \ref{sec:varselect} and choose the variables according to their correlation with the factors obtained from the different dimension reduction techniques. For the small ARX models, we determine the top-five correlated variables in each vintage for each dimension reduction technique. As an example, \autoref{fig:corrfreq} shows the outcome of the first step for the best performing models (i.e., the Autoencoder with one layer and $30$ factors, PCA quadratic and PCA squared with five factors). \autoref{tab:corr_variables} shows the top-five correlated variables for each model when aggregated over all vintages. For the large ARX model, we aggregate the counts over all models and choose the most frequently occurring variables (top-twenty). Those are indicated by column \textit{PART}. 

\begin{table*}[ht]
\caption{Data description of slow moving variables \label{tab:data1_us}}
{\tiny
\begin{center}
\begin{tabular*}{\textwidth}{l @{\extracolsep{\fill}} llcccl}
\toprule
\multicolumn{1}{l}{Classification}&\multicolumn{1}{c}{\ FRED.Mnemonic}&\multicolumn{1}{c}{\ Description}&\multicolumn{1}{c}{\ Trans I(0)}&\multicolumn{1}{c}{\ PART}&\multicolumn{1}{c}{\ FULL}\tabularnewline
\midrule
\textbf{Real activity} &RPI&Real personal income  &$5$&&x\tabularnewline
&W875RX1&Real personal income ex transfer receipts&$5$&&x\tabularnewline
&INDPRO&IP Index &$5$&x&x\tabularnewline
&IPFPNSS&IP: Final Products&$5$&&x\tabularnewline
&IPFINAL&IP: Final Products (Market Group)&$5$&&x\tabularnewline
&IPCONGD&IP: Consumer Goods&$5$&&x\tabularnewline
&IPMAT&IP: Materials &$5$&&x\tabularnewline
&IPMANSICS&IP: Manufacturing (SIC)&$5$&x&x\tabularnewline
&CUMFNS&Capacity Utilization: Manufacturing&$2$&x&x\tabularnewline
&RETAILx&Retail and Food Services Sales &$5$&&x\tabularnewline
&AMDMNOx&New Orders for Durable goods&$5$&&x\tabularnewline
&ANDENOx&New Orders for Nondefense Capital goods&$5$&&x\tabularnewline
&AMDMUOx&Unfilled Orders for Durable goods&$5$&&x\tabularnewline
&BUSINVx&Total Business Inventories&$5$&&x\tabularnewline
&ISRATIOx&Total Business: Inventories to Sales Ratio&$2$&&x\tabularnewline
&UMCSENTx&Consumer Sentiment Index&$2$&&x\tabularnewline
&CMRMTSPLx&Real Manu. and TradeIndustries Sales&$5$&&x\tabularnewline
\midrule
\textbf{Housing} &HOUST&Housing Starts: Total New Privately Owned&$4$&x&x\tabularnewline
&HOUSTNE&Housing Starts, Northeast&$4$&&x\tabularnewline
&HOUSTMW&Housing Starts, Midwest&$4$&&x\tabularnewline
&HOUSTS&Housing Starts, South&$4$&&x\tabularnewline
&HOUSTW&Housing Starts, West&$4$&&x\tabularnewline
&PERMIT&New Private Housing Permits (SAAR)&$4$&x&x\tabularnewline
&PERMITNE&New Private Housing Permits, Northeast (SAAR)&$4$&&x\tabularnewline
&PERMITMW&New Private Housing Permits, Midwest (SAAR)&$4$&&x\tabularnewline
&PERMITS&New Private Housing Permits, South (SAAR)&$4$&&x\tabularnewline
&PERMITW&New Private Housing Permits, West (SAAR)&$4$&&x\tabularnewline
\midrule
\textbf{Labor market} &CLF16OV&Civilian Labor Force&$5$&&x\tabularnewline
&CE16OV&Civilian Employment &$5$&&x\tabularnewline
&UNRATE&Civilian Unemployment Rate&$2$&&x\tabularnewline
&UEMPMEAN&Average Duration of Unemployment (Weeks)&$2$&&x\tabularnewline
&UEMPLT5&Civilians Unemployed : Less Than 5 Weeks&$5$&&x\tabularnewline
&UEMP5TO14&Civilians Unemployed for 5-14 Weeks&$5$&&x\tabularnewline
&UEMP15OV&Civilians Unemployed : 15 Weeks \& Over&$5$&&x\tabularnewline
&UEMP15T26&Civilians Unemployed for 15-26 Weeks&$5$&&x\tabularnewline
&UEMP27OV&Civilians Unemployed for 27 Weeks and Over&$5$&&x\tabularnewline
&CLAIMSx&Initial Claims &$5$&&x\tabularnewline
&PAYEMS&All Employees: Total nonfarm&$5$&&x\tabularnewline
&USGOOD&All Employees: Goods-Producing Industries&$5$&x&x\tabularnewline
&CES1021000001&All Employees: Mining and Logging: Mining&$5$&&x\tabularnewline
&USCONS&All Employees: Construction&$5$&&x\tabularnewline
&MANEMP&All Employees: Manufacturing&$5$&x&x\tabularnewline
&DMANEMP&All Employees: Durable goods&$5$&&x\tabularnewline
&NDMANEMP&All Employees: Nondurable goods&$5$&&x\tabularnewline
&SRVPRD&All Employees: Service-Providing Industries&$5$&&x\tabularnewline
&USWTRADE&All Employees: Wholesale Trade&$5$&&x\tabularnewline
&USTRADE&All Employees: Retail Trade&$5$&&x\tabularnewline
&USFIRE&All Employees: Financial Activities&$5$&&x\tabularnewline
&USGOVT&All Employees: Government&$5$&&x\tabularnewline
&CES0600000007&Avg Weekly Hours: Goods-Producing&$1$&x&x\tabularnewline
&AWOTMAN&Avg Weekly Overtime Hourse: Manufacturing&$2$&&x\tabularnewline
&AWHMAN&Avg Weekly Hours: Manufacturing&$1$&x&x\tabularnewline
&CES0600000008&Avg Hourly Earnings: Goods-Producing&$6$&&x\tabularnewline
&CES2000000008&Avg Hourly Earnings: Construction&$6$&&x\tabularnewline
&CES3000000008&Avg Hourly Earnings: Manufacturing&$6$&&x\tabularnewline
\midrule
\textbf{Prices} &OILPRICEx&Crude Oil, , spliced WTI and Cushing&$6$&&x\tabularnewline
&PPICMM&PPI: Metals and metal products&$6$&&x\tabularnewline
&CPIAUCSL&CPI : All Items&$6$&&x\tabularnewline
&CPIAPPSL&CPI : Apparel&$6$&&x\tabularnewline
&CPITRNSL&CPI : Transportation&$6$&&x\tabularnewline
&CPIMEDSL&CPI : Medical Care&$6$&&x\tabularnewline
&CUSR0000SAC&CPI : Commodities&$6$&x&x\tabularnewline
&CUSR0000SAS&CPI : Services&$6$&&x\tabularnewline
&CPIULFSL&CPI : All Items Less Food&$6$&&x\tabularnewline
&CUSR0000SA0L5&CPI : All Items Less Medical Care&$6$&&x\tabularnewline
\bottomrule
\end{tabular*}
\end{center}}
\end{table*}

\begin{table*}[ht]
\captionsetup{labelformat=empty}
\caption{Data description of fast moving variables (cont.)}
{\tiny
\begin{center}
\begin{tabular*}{\textwidth}{l @{\extracolsep{\fill}} llcccl}
\toprule
\multicolumn{1}{l}{Classification}&\multicolumn{1}{c}{\ FRED.Mnemonic}&\multicolumn{1}{c}{\ Description}&\multicolumn{1}{c}{\ Trans I(0)}&\multicolumn{1}{c}{\ PART}&\multicolumn{1}{c}{\ FULL}\tabularnewline
\midrule
\textbf{Money stocks} &M1SL&M1 Money Stock&$6$&&x\tabularnewline
&M2SL&M2 Money Stock&$6$&&x\tabularnewline
&M2REAL&Real M2 Money Stock&$5$&&x\tabularnewline
&AMBSL&St. Louis Adjusted Monetary Base&$6$&x&x\tabularnewline
\midrule
\textbf{Reserves \& loans} &TOTRESNS&Total Reserves of Depository Institutions&$6$&&x\tabularnewline
&NONBORRES&Reserves of Depository Institutions &$7$&x&x\tabularnewline
&BUSLOANS&Commercial and Industrial Loans &$6$&&x\tabularnewline
&REALLN&Real Estate Loans at All Commerical Banks&$6$&x&x\tabularnewline
&NONREVSL&Total Nonrevolving Credit&$6$&&x\tabularnewline
&CONSPI&Nonrevolving consumer credit to Personal Income&$2$&&x\tabularnewline
&MZMSL&MZM Money Stock&$6$&&x\tabularnewline
&DTCOLNVHFNM&Consumer Motor Vehicle Loans Outstanding&$6$&&x\tabularnewline
&DTCTHFNM& Total Consumer Loans and Leases Outstanding&$6$&&x\tabularnewline
&INVEST&Securities in Bank Credit at All Commercial Banks&$6$&&x\tabularnewline
&CP3Mx&3-Month AA Financial Commercial Paper Rate&$2$&&x\tabularnewline
\midrule
\textbf{Interest rates} &FEDFUNDS&Effective Federal Funds Rate&$2$&&x\tabularnewline
&TB3MS&3-Month Treasury Bill&$2$&&x\tabularnewline
&TB6MS&6-Month Treasury Bill&$2$&&x\tabularnewline
&GS1&1-Year Treasury Rate&$2$&&x\tabularnewline
&GS5&5-Year Treasury Rate&$2$&&x\tabularnewline
&GS10&10-Year Treasury Rate&$2$&&x\tabularnewline
&AAA&Moody's Seasoned Aaa Corporate Bond Yield&$2$&&x\tabularnewline
&BAA&Moody's Seasoned Baa Corporate Bond Yield &$2$&&x\tabularnewline
&COMPAPFFx&3-Month Commercial Paper Minus FEDFUNDS&$1$&x&x\tabularnewline
&TB3SMFFM&3-Month Treasury C  Minus FEDFUNDS&$1$&x&x\tabularnewline
&TB6SMFFM&6-Month Treasury C  Minus FEDFUNDS&$1$&x&x\tabularnewline
&T1YFFM&1-Year Treasury C  Minus FEDFUNDS&$1$&x&x\tabularnewline
&T5YFFM&5-Year Treasury C  Minus FEDFUNDS&$1$&&x\tabularnewline
&T10YFFM&10-Year Treasury C  Minus FEDFUNDS&$1$&x&x\tabularnewline
&AAAFFM&Moody's Aaa Corporate Bond  Minus FEDFUNDS&$1$&x&x\tabularnewline
&BAAFFM&Moody's Baa Corporate Bond  Minus FEDFUNDS&$1$&x&x\tabularnewline
\midrule
\textbf{Stock market} &TWEXMMTH& Trade Weighted U.S. Dollar Index: Major Currencies&$5$&&x\tabularnewline
&EXSZUSx&Switzerland / U.S. Foreign Exchange Rate&$5$&&x\tabularnewline
&EXJPUSx&Japan / U.S. Foreign Exchange Rate&$5$&&x\tabularnewline
&EXUSUKx&U.S. / UK Foreign Exchange Rate&$5$&&x\tabularnewline
&EXCAUSx&Canada / U.S. Foreign Exchange Rate&$5$&&x\tabularnewline
&S.P.500& S\&Ps Common Stock Price Index: Composite&$5$&&x\tabularnewline
&S.P..indust& S\&Ps Common Stock Price Index: Industrials&$5$&&x\tabularnewline
&S.P.div.yield& S\&Ps Composite Common Stock: Dividend Yield&$2$&&x\tabularnewline
&S.P.PE.ratio& S\&Ps Composite Common Stock: Price-Earnings Ratio&$5$&&x\tabularnewline
\bottomrule
\end{tabular*}
\begin{tablenotes}[flushleft]
      \tiny
      \item \textit{Note:} Column \texttt{Trans I(0)} denotes the transformation of each time series to achieve approximate stationarity: (1) no transformation, (2) $\Delta x_t$, (4) $\log(x_t)$, (5) $\Delta \log(x_t)$, (6) $\Delta^2 \log(x_t)$, (7) $\Delta (x_t/x_{t-1}-1.0)$
    \end{tablenotes}
\end{center}}
\end{table*}

\begin{landscape}\begin{table*}
{\tiny
\caption{Top-five correlated variables with the factors of each specification \label{tab:corr_variables}}
\begin{center}
\begin{tabular*}{\linewidth}{l @{\extracolsep{\fill}} llllllllllll}
\toprule
\multicolumn{1}{l}{}&\multicolumn{1}{c}{Specification}&\multicolumn{1}{c}{}&\multicolumn{9}{c}{The five variables with the highest prevalence among the group of the top-five correlated}\tabularnewline
\cmidrule{4-12}
\multicolumn{1}{l}{}&\multicolumn{1}{c}{}&\multicolumn{1}{c}{}&\multicolumn{1}{c}{1.}&\multicolumn{1}{c}{}&\multicolumn{1}{c}{2.}&\multicolumn{1}{c}{}&\multicolumn{1}{c}{3.}&\multicolumn{1}{c}{}&\multicolumn{1}{c}{4.}&\multicolumn{1}{c}{}&\multicolumn{1}{c}{5.}\tabularnewline
\midrule
   ~~&   Autoencoder 1l (q = 05)&   &   PERMIT   (22\%)&   &   HOUST   (17\%)&   &   PERMITW   (13\%)&   &   T10YFFM   (6\%)&   &   PERMITMW   (6\%)\tabularnewline
   ~~&   Autoencoder 1l (q = 15)&   &   PERMIT   (25\%)&   &   PERMITS   (12\%)&   &   T10YFFM   (10\%)&   &   PERMITW   (10\%)&   &   HOUST   (8\%)\tabularnewline
   ~~&   Autoencoder 1l (q = 30)&   &   PERMIT   (30\%)&   &   PERMITS   (13\%)&   &   T10YFFM   (12\%)&   &   AAAFFM   (12\%)&   &   PERMITW   (10\%)\tabularnewline
   ~~&   Autoencoder 3l (q = 05)&   &   PERMIT   (20\%)&   &   PERMITS   (14\%)&   &   HOUST   (9\%)&   &   PERMITW   (9\%)&   &   BAAFFM   (7\%)\tabularnewline
   ~~&   Autoencoder 3l (q = 15)&   &   CES0600000007   (20\%)&   &   T10YFFM   (12\%)&   &   AAAFFM   (9\%)&   &   PERMITNE   (9\%)&   &   AWHMAN   (8\%)\tabularnewline
   ~~&   Autoencoder 3l (q = 30)&   &   CES0600000007  (20\%)&   &   T10YFFM   (12\%)&   &   PERMIT   (11\%)&   &   PERMITS   (10\%)&   &   AAAFFM   (10\%)\tabularnewline
   ~~&   Autoencoder 5l (q = 05)&   &   PERMIT   (24\%)&   &   HOUST   (12\%)&   &   PERMITS   (11\%)&   &   PERMITW   (11\%)&   &   PERMITMW   (7\%)\tabularnewline
   ~~&   Autoencoder 5l (q = 15)&   &   CES0600000007   (26\%)&   &   PERMITNE   (11\%)&   &   T10YFFM   (11\%)&   &   AAAFFM   (10\%)&   &   AWHMAN   (8\%)\tabularnewline
   ~~&   Autoencoder 5l (q = 30)&   &   CES0600000007   (27\%)&   &   T10YFFM   (11\%)&   &   AAAFFM   (11\%)&   &   PERMITMW  (8\%)&   &   PERMITNE   (8\%)\tabularnewline
   ~~&   Autoencoder 8l (q = 05)&   &   PERMIT   (25\%)&   &   HOUST   (14\%)&   &   PERMITW   (9\%)&   &   T10YFFM   (7\%)&   &   PERMITS   (7\%)\tabularnewline
   ~~&   Autoencoder 8l (q = 15)&   &   T10YFFM   (12\%)&   &   PERMITNE   (12\%)&   &   PERMITMW   (10\%)&   &   PERMIT   (10\%)&   &   BAAFFM   (10\%)\tabularnewline
   ~~&   Autoencoder 8l (q = 30)&   &   PERMIT  (17\%)&   &   T10YFFM   (14\%)&   &   PERMITNE   (14\%)&   &   BAAFFM   (11\%)&   &   CES0600000007   (10\%)\tabularnewline
\midrule
   ~~&   Diffusion Maps (q = 05)&   &   S.P.div.yield   (31\%)&   &   TB3MS   (10\%)&   &   COMPAPFFx   (7\%)&   &   NONBORRES   (7\%)&   &   HOUSTNE   (6\%)\tabularnewline
   ~~&   Diffusion Maps (q = 15)&   &   NONBORRES   (40\%)&   &   AMBSL   (22\%)&   &   TOTRESNS   (7\%)&   &   IPMAT   (7\%)&   &   MZMSL   (5\%)\tabularnewline
   ~~&   Diffusion Maps (q = 30)&   &   NONBORRES   (33\%)&   &   AMBSL   (27\%)&   &   TOTRESNS   (9\%)&   &   MZMSL  (9\%)&   &   UMCSENTx   (5\%)\tabularnewline
\midrule
   ~~&   ISOMAP (q = 05)&   &   BAAFFM   (20\%)&   &   AAAFFM   (18\%)&   &   PERMITNE   (11\%)&   &   PERMITMW   (9\%)&   &   AWHMAN   (8\%)\tabularnewline
   ~~&   ISOMAP (q = 15)&   &   TB6SMFFM   (17\%)&   &   T5YFFM   (12\%)&   &   TB3SMFFM   (11\%)&   &   T10YFFM   (8\%)&   &   BAAFFM   (6\%)\tabularnewline
   ~~&   ISOMAP (q = 30)&   &   TB6SMFFM   (18\%)&   &   T1YFFM   (16\%)&   &   VXOCLSx   (7\%)&   &   T5YFFM   (6\%)&   &   TB3SMFFM   (6\%)\tabularnewline
\midrule
   ~~&   LLE (q = 05)&   &   PERMITMW   (20\%)&   &   BAAFFM   (16\%)&   &   HOUST   (11\%)&   &   PERMITS   (10\%)&   &   AWHMAN   (7\%)\tabularnewline
   ~~&   LLE (q = 15)&   &   PERMITMW   (10\%)&   &   TB6SMFFM   (10\%)&   &   TB3SMFFM   (9\%)&   &   AWHMAN   (9\%)&   &   MANEMP   (8\%)\tabularnewline
   ~~&   LLE (q = 30)&   &   COMPAPFFx   (18\%)&   &   TB6SMFFM   (13\%)&   &   TB3SMFFM   (10\%)&   &   T1YFFM   (7\%)&   &   PERMITMW   (7\%)\tabularnewline
\midrule
   ~~&   PCA gauss. kernel (q = 05)&   &   AAAFFM   (35\%)&   &   BAAFFM   (15\%)&   &   CES0600000007   (13\%)&   &   AWHMAN   (9\%)&   &   T10YFFM   (5\%)\tabularnewline
   ~~&   PCA gauss. kernel (q = 15)&   &   VXOCLSx   (16\%)&   &   COMPAPFFx   (15\%)&   &   TB6SMFFM   (11\%)&   &   USGOOD   (9\%)&   &   MANEMP   (9\%)\tabularnewline
   ~~&   PCA gauss. kernel (q = 30)&   &   INDPRO   (27\%)&   &   IPMANSICS   (26\%)&   &   CUMFNS   (18\%)&   &   IPFPNSS   (6\%)&   &   GS5   (6\%)\tabularnewline
\midrule
   ~~&   PCA linear (q = 05)&   &   T10YFFM   (38\%)&   &   PERMITNE   (14\%)&   &   AAAFFM   (12\%)&   &   AWHMAN   (9\%)&   &   PERMITS   (5\%)\tabularnewline
   ~~&   PCA linear (q = 15)&   &   T1YFFM   (16\%)&   &   TB6SMFFM   (15\%)&   &   TB6MS   (9\%)&   &   GS1   (9\%)&   &   FEDFUNDS   (5\%)\tabularnewline
   ~~&   PCA linear (q = 30)&   &   INDPRO   (20\%)&   &   CUMFNS   (16\%)&   &   IPMANSICS   (12\%)&   &   M2REAL   (7\%)&   &   CUSR0000SA0L5   (6\%)\tabularnewline
\midrule
   ~~&   PCA poly. kernel (q = 05)&   &   T10YFFM   (29\%)&   &   AWHMAN   (15\%)&   &   CES0600000007   (6\%)&   &   AAAFFM   (6\%)&   &   PERMITS   (6\%)\tabularnewline
   ~~&   PCA poly. kernel (q = 15)&   &   TB6SMFFM   (20\%)&   &   T1YFFM   (16\%)&   &   COMPAPFFx   (10\%)&   &   GS1   (8\%)&   &   TB6MS   (7\%)\tabularnewline
   ~~&   PCA poly. kernel (q = 30)&   &   INDPRO   (22\%)&   &   IPMANSICS   (13\%)&   &   CUMFNS   (13\%)&   &   M2REAL   (9\%)&   &   CUSR0000SA0L5   (6\%)\tabularnewline
\midrule
   ~~&   PCA quadratic (q = 05)&   &   USGOOD   (42\%)&   &   MANEMP   (17\%)&   &   PERMITMW   (8\%)&   &   AWHMAN  (7\%)&   &   NONBORRES   (7\%)\tabularnewline
   ~~&   PCA quadratic (q = 15)&   &   NONBORRES   (33\%)&   &   AWHMAN   (18\%)&   &   HOUSTMW   (8\%)&   &   HOUSTW   (7\%)&   &   PERMITMW   (6\%)\tabularnewline
   ~~&   PCA quadratic (q = 30)&   &   PERMITW   (16\%)&   &   AWHMAN   (14\%)&   &   HOUST   (11\%)&   &   CES0600000007  (11\%)&   &   HOUSTW   (8\%)\tabularnewline
\midrule
   ~~&   PCA squared (q = 05)&   &   MANEMP   (27\%)&   &   USGOOD   (21\%)&   &   NONBORRES   (19\%)&   &   BUSINVx  (8\%)&   &   AWHMAN  (6\%)\tabularnewline
   ~~&   PCA squared (q = 15)&   &   NONBORRES   (56\%)&   &   PAYEMS   (10\%)&   &   AWHMAN   (9\%)&   &   TB3SMFFM   (6\%)&   &   VXOCLSx  (5\%)\tabularnewline
   ~~&   PCA squared (q = 30)&   &   NONBORRES  (27\%)&   &   CP3Mx   (11\%)&   &   VXOCLSx  (10\%)&   &   AWHMAN   (10\%)&   &   PAYEMS   (8\%)\tabularnewline
\bottomrule
\end{tabular*}
\begin{tablenotes}[flushleft]
      \tiny
      \item \textit{Note:} The table shows the first five variables with the highest average correlation among the factors obtained from the different dimension reduction techniques. We determine the top-five correlated variables for each vintage and present those most frequently occurring. This overview serves mostly as a simple illustration of our unsupervised variable selection approach used in Table \ref{tab:mainARX}. In parentheses, we present the frequency of occurrence over these vintages. For example, when we obtain five factors with the Autoencoder (1 layer), the variable PERMIT is in $22\%$ of all vintages among the top-five correlated variables.
    \end{tablenotes}
\end{center}}
\end{table*}\end{landscape}

\end{appendices}
\end{document}